\documentclass[a4paper,fleqn,usenatbib]{mnras}

\usepackage{newtxtext,newtxmath}

\usepackage[T1]{fontenc}
\usepackage{ae,aecompl}

\usepackage{graphicx}	
\usepackage{amsmath}	
\usepackage{amssymb}	

\usepackage[usenames,dvipsnames]{color}
\usepackage{latexsym}
\usepackage{soul}

\defcitealias{2015MNRAS.453.2830M}{Paper~I}

\title[On the relevance of chaos II]{On the relevance of chaos for halo stars in the solar
neighbourhood II}

\author[N.P. Maffione et al.]{Nicolas P. Maffione$^{1,2}$,\thanks{E-mail: npmaffione@unrn.edu.ar (NPM)}
Facundo A. G\'omez$^{3,4,5}$,
Pablo M. Cincotta$^{6,7}$, \newauthor
Claudia M. Giordano$^{6,7}$,
Robert J. J. Grand$^{8,9}$,
Federico Marinacci$^{10}$,
R{\"u}diger Pakmor$^{8}$,\newauthor
Christine M. Simpson$^{8}$,
Volker Springel$^{8,9,5}$
and Carlos S. Frenk$^{11}$\\
$^{1}$Laboratorio de Procesamiento de Se\~nales Aplicado y Computaci\'on de Alto Rendimiento, Sede Andina, \\Universidad Nacional de R\'io Negro, Mitre 630, San Carlos de Bariloche, R8400AHN, R\'io Negro, Argentina\\
$^{2}$CONICET, Mitre 630, San Carlos de Bariloche, R8400AHN, R\'io Negro, Argentina\\
$^{3}$Departamento de F\'isica y Astronom\'ia, Universidad de La Serena, Av. Juan Cisternas 1200 N, La Serena, Chile\\
$^{4}$ Instituto de Investigaci\'on Multidisciplinar en Ciencia y Tecnolog\'ia, Universidad de La Serena, Ra\'ul Bitr\'an 1305, La Serena, Chile\\
$^{5}$Max-Planck-Institut f\"ur Astrophysik, Karl-Schwarzschild-Str. 1, D-85748, Garching, Germany\\
$^{6}$Grupo de Caos en Sistemas Hamiltonianos, Facultad de Ciencias Astron\'omicas y Geof\'isicas, \\Universidad Nacional de La Plata, Paseo del Bosque s/n, La Plata, B1900FWA, Buenos Aires, Argentina\\
$^{7}$Instituto de Astrof\'isica de La Plata, Universidad Nacional de La Plata, CONICET, Paseo del Bosque s/n, \\La Plata, B1900FWA, Buenos Aires, Argentina\\
$^{8}$Heidelberger Institut f\"ur Theoretische Studien, Schloss-Wolfsbrunnenweg 35, 69118 Heidelberg, Germany\\
$^{9}$Zentrum f\"ur Astronomie der Universitat Heidelberg, Astronomisches Recheninstitut, M\"onchhofstrasse 12-14, 69120 Heidelberg, Germany\\
$^{10}$Department of Physics, Kavli Institute for Astrophysics and Space Research, MIT, Cambridge, MA 02139, USA\\
$^{11}$Institute for Computational Cosmology, Department of Physics, Durham University, South Road, Durham, DH1 3LE, UK\\
}

\date{Accepted XXX. Received YYY; in original form ZZZ}

\pubyear{2017}

\begin{document}
\label{firstpage}
\pagerange{\pageref{firstpage}--\pageref{lastpage}}
\maketitle

\begin{abstract}
In a previous paper based on dark matter only simulations we show that, in the approximation of an analytic and static potential  
describing the strongly triaxial and cuspy shape of Milky Way-sized haloes, diffusion due to chaotic mixing in the
neighbourhood of the Sun does not efficiently erase phase space signatures of past accretion events. In this second paper we further explore the 
effect of chaotic mixing using multicomponent Galactic potential models and solar neighbourhood-like volumes extracted from
fully cosmological hydrodynamic simulations, thus naturally accounting for the gravitational potential associated with  baryonic 
components, such as the bulge and disc. Despite the strong change in the global Galactic potentials with respect to those obtained in dark 
matter only simulations, our  results confirm that a large fraction of  halo particles evolving on chaotic orbits exhibit their chaotic 
behaviour after periods of time significantly larger than a Hubble time. In addition, significant diffusion in phase space is not observed on those particles that do 
 exhibit chaotic behaviour within a Hubble time.

\end{abstract}




\section{Introduction}
\label{sec:introduction}

We are crossing the gates into a new era in astronomy research, the era of multi-messenger observations, big data and extremely detailed numerical simulations. In this promising scenario, the field of galactic archaeology is in an extraordinary position thanks to the arrival of the satellite {\it Gaia} \citep[see][]{2001A&A...369..339P,2008IAUS..248..529L}. A first glimpse at the extraordinary quality of the full 6-dimensional phase space catalog that {\it Gaia} will provide throughout its lifetime   has already been publicly released \citep{2015A&A...574A.115M,2016A&A...595A...4L}. Several studies based on these early-stage data have already pushed the boundaries on the characterization of the extended solar neighbourhood phase space structure and its relation to the  Galaxy's formation history \citep{2017ApJ...845..101B,2017MNRAS.470.1360B,2017arXiv170703833I,2017MNRAS.466L.113M, 2017arXiv171206616S}.

According to the $\Lambda$ cold dark matter ($\Lambda$CDM) cosmological model, galaxies increase their mass through merger and accretion of  smaller systems. These accretion events are expected to play a very important role in shaping the present-day chemical distribution \citep{2014SAAS...37..145M} and morphological and 
kinematical structure of the host galaxy \citep[see][for a complete review on near-field cosmology]{2014SAAS...37....1B}. Stellar haloes of large galaxies such as our own are believed to be primarily formed  as  a  result  of  the  accumulation  of  tidal  debris  associated  with  ancient  as  well  as  recent  and ongoing accretion events \citep{2008A&ARv..15..145H}. Galactic discs are also strongly affected by such interactions. In addition to heating and thickening pre-existing discs, satellites can generate substructure in both chemical abundance and phase space, as well as induce secular phenomena such as bars, spiral arms, and warps \citep[see, e.g.][]{2009MNRAS.397.1599Q,2012MNRAS.419.2163G,2012MNRAS.423.3727G,2013MNRAS.429..159G,2014ApJ...781L..20M,2015MNRAS.450..266W,2016MNRAS.456.2779G,2017MNRAS.465.3446G,2017arXiv171002538L,2018MNRAS.473.1218L}. 

Signatures of these accretion events can provide strong constraints on the formation history of galaxies \citep[see][and references therein, for a recent and comprehensive discussion on the subject]{2016ASSL..420.....N}, such as our own Milky Way \citep[MW,][]{2002ARA&A..40..487F}. Thus, much effort has been devoted to develop methods and tools to efficiently identify and quantify substructure in different Galactic distributions. To first order, the stellar halo can be approximated as a collisionless component \citep{1987gady.book.....B} and, thus, retains its dynamical memory providing an ideal place to search for signatures of accretion events. Studies based on numerical models have predicted that a few hundred kinematically cold stellar streams should be currently crossing our solar neighbourhood \citep{1999MNRAS.307..495H,2003MNRAS.339..834H,2008A&ARv..15..145H}. However, due to limitations  of studies based on pre-{\it Gaia} astrometric catalogues \citep[e.g. with][]{1997ESASP1200.....E,2000AJ....120.1579Y,2006AJ....131.1163S,2006ChJAA...6..265Z,2008AJ....136..421Z}, only a handful of stellar streams were identified \citep[e.g.][]{1999MNRAS.307..495H,1999Natur.402...53H,2006MNRAS.365.1309H}. Furthermore, studies based on {\it Gaia} first data release, DR1, have already revealed more substructures in our stellar halo \citep[e.g.][]{2017A&A...598A..58H,2017A&A...608A..73K}, but the number of identified substructures is still far from the few hundred streams predicted by the models. Only with the arrival of {\it Gaia} second data release, DR2, we will be able to provide a more robust quantification of the ammount of substructure in the solar neighbourhood. 

A valid concern regarding our ability to identify signatures from old accretion events  relates to the longevity of cold kinematical structures. It is well known that  dark matter (DM) haloes of MW-like galaxies are expected to be triaxial \citep{2002ApJ...574..538J,2006MNRAS.367.1781A,2011MNRAS.416.1377V}, and that a fraction of the orbits hosted by the corresponding triaxial potentials will exhibit chaotic behaviour \citep[see e.g.][]{1993ApJ...409..563S,1996ApJ...460..136M,1996ApJ...471...82M,2000MNRAS.319...43S,2002MNRAS.337..619V,2003MNRAS.345..727K,2004A&A...428..905K,2005CeMDA..91..173M,2007LNP...729..297E}. As shown by \citet{1999MNRAS.307..495H,2008MNRAS.385..236V,2013MNRAS.436.3602G} and \citet[][hereinafter \citetalias{2015MNRAS.453.2830M}]{2015MNRAS.453.2830M}, the density of a stellar stream on a chaotic orbit decays exponentially with time, as opposed to the power-law decay associated with regular orbits. As a consequence, the identification of stellar streams on chaotic orbits is extremely challenging as they can quickly blend with the background stellar distribution, even in velocity space.  More importantly,  within relevant time-scales, strong chaotic behaviour can lead to diffusion in the space of pseudo-integrals of motion, such as energy and angular momentum \citep{1992IAUS..149..471P,1997A&A...319..796S,2013ApJ...767...93V}. As a result, signatures of stellar streams can be effectively erased, hindering our hopes of constraining our Galactic accretion history through the identification and quantification of substructures in phase space \citep[for a comprehensive review on chaos in galaxies, see the book by][and references therein]{2002ocda.book.....C}. 

In \citetalias{2015MNRAS.453.2830M} we tackled this problem by characterizing the orbital distribution of star particles located 
within solar neighbourhood-like  volumes extracted from  stellar halo models based on DM-only simulations \citep[][]{2008Natur.456...73S,2008MNRAS.391.1685S,2010MNRAS.406..744C}. Our results showed that $\sim70$~per cent of these orbits, evolving within  strongly triaxial potentials, could be classified as chaotic.  However, only $\la20$~per cent of these particles 
 revealed their chaotic nature within a Hubble time.  The remaining orbits classified as chaotic ($\sim50$~per cent of the total), revealed their chaotic behaviour only after a Hubble time. These orbits, dubbed as ``sticky'' \citep[see][and \citetalias{2015MNRAS.453.2830M} and references therein for further details]{2000CSF....11.2281T}, are particularly important. They have an intrinsically chaotic nature. However, for halo stars moving on such orbits, chaotic mixing will not have enough time to act. Furthermore, an analysis based on first order expansions of the underlying potentials demonstrated that diffusion in phase space is not significant on any realistic time-scale 
\citep[in agreement with previous works, see for instance:][]{2004A&A...423..745G,2006A&A...455..499C,2014PhyD..266...49C}, even for those orbits that revealed their chaotic nature within a Hubble time.

Although chaotic mixing is non-negligible \citep[e.g.][]{2015ApJ...799...28P,2016MNRAS.460..497H,2017MNRAS.470...60E} and might be strongly relevant for the morphological structure of very cold streams \citep[see for instance][]{2016MNRAS.455.1079P,2016ApJ...824..104P}, our results suggested that it  
is not efficient at erasing  signatures of accretion events. However, as mentioned before, this study was based on stellar halo models extracted from DM-only simulations. Due to the lack of a baryon component, the overall galactic potentials were 
clearly a poor representation of the true Galactic potential, especially within the inner Galactic regions. The addition of baryons  does not only  modify the potential through their additional mass distribution, but also significantly alters the density profile of the DM halo within which the baryons are embedded \citep[e.g.][]{2004ApJ...616...16G,2016MNRAS.457.1931S,2016MNRAS.458.1559Z}. Previous studies based on cosmological hydrodynamical simulations have shown that, when baryons are taken into account, DM haloes present a significantly more oblate distribution in the inner regions \citep[for instance:][]{1994ApJ...431..617D,2006PhRvD..74l3522G,2008ApJ...681.1076D,2010MNRAS.407..435A}, thus enhancing the asymmetry within the inner and the outer galactic regions. 

In this work, we take a  step forward on this matter by characterizing the effects of chaotic mixing in solar neighbourhood-like volumes 
extracted from fully cosmological hydrodynamical simulations of the formation of MW-like galaxies \citep[][]{2014MNRAS.437.1750M,2017MNRAS.467..179G}. These simulations naturally account for the effects associated with the gravitational potential of the baryonic components (such as the bulge and disc), and thus can be used  to characterize the efficiency of chaotic mixing in a more realistic scenario.

The paper is organized as follows: in Section~\ref{sec:methodology} we briefly introduce the simulations, models and techniques used in the present study. Our results on the actual relevance of chaos in erasing kinematic signatures of accretion events in the local stellar halo are presented in Sections~\ref{sec:relevance} and~\ref{sec:diffusion} and, finally, we discuss and summarise our results in Section~\ref{sec:discussion}.
 
\section{Methodology}
\label{sec:methodology}

In this section we briefly describe the simulations and numerical tools used to characterize and quantify chaotic behaviour within solar neighbourhood-like phase space volumes.

\subsection{Simulations}
\label{subsec:meth-sim} 
In this study we focus on a set of seven fully cosmological hydrodynamic zoom-in simulations of MW-like galaxies, extracted from \citet{2014MNRAS.437.1750M} and \citet{2017MNRAS.467..179G}.

The simulations were carried out using the  $N$-body +  moving-mesh, magnetohydrodynamic code {\sc arepo} \citep{2010MNRAS.401..791S,2016MNRAS.455.1134P}. A standard $\Lambda$CDM cosmology was adopted in both cases. The values chosen for the different cosmological parameters are very similar and can be 
found on the corresponding papers. 


The baryonic physics model implemented in {\sc arepo} follows a number of processes that play a key role in the
formation of late-type galaxies, such as gas cooling/heating, star formation, mass return and metal enrichment from  stellar evolution, the  growth of  supermassive black  holes, magnetic fields \citep{2013MNRAS.432..176P,2017MNRAS.469.3185P} and  feedback both from stellar sources and from black hole  accretion. The parameters that regulate the efficiency of each physical process were chosen by comparing the results obtained in simulations of cosmologically representative volumes to a wide range of observations of the galaxy population \citep{2013MNRAS.436.3031V,2014MNRAS.437.1750M,2016MNRAS.459..199G}. 

In order to contrast our results with those presented in \citetalias{2015MNRAS.453.2830M}, first we use a hydrodynamic re-simulation of one of the haloes from the {\it Aquarius Project} \citep{2008Natur.456...73S,2008MNRAS.391.1685S}, also run with the code {\sc arepo}. This simulation, namely Aq-C4 (for simulation Aq-C at the resolution level 4), was first introduced in \citet{2014MNRAS.437.1750M}. 

However, most of the simulations used in this work are taken from the {\it Auriga Project}. This suite is composed of 30 high-resolution cosmological zoom-in simulations of the formation of late-type galaxies within  MW-sized DM haloes. The haloes were selected from a lower resolution DM-only simulation from the {\it Eagle Project} \citep{2015MNRAS.446..521S}, a periodic box of side 100 Mpc. Each halo was chosen to have, at $z=0$, a virial mass in the range of $10^{12}$ -- $2 \times 10^{12}~ {\rm M}_{\odot}$ and to be more distant than nine times the virial radius from any other halo of mass more than $3$~per cent of its own mass. The typical DM particle and gas cell mass resolutions for the simulations used in this work 
(Aq-C4 and {\it Auriga}, also resolution level 4) are $\sim 3\times 10^{5}~{\rm M}_{\odot}$ and $\sim 6 \times  10^{4}~{\rm M}_{\odot}$, respectively. The gravitational softening length used for DM and stars grows with scale factor up to a  maximum of 369 pc, after which it  is kept  constant in physical units. The  softening length  of gas cells  scales with the  mean radius  of the  cell, but is never allowed to drop below the stellar softening length. A resolution  study  across  three resolution levels \citep{2017MNRAS.467..179G} shows that many galaxy properties, such  as   surface  density   profiles,  orbital circularity  distributions, star formation histories  and  disc  vertical  structures  are already well converged at the resolution level used in this work. We will refer to the {\it Auriga} simulations as \textquotedblleft Au-i'', with \textquotedblleft i'' enumerating the different initial conditions, as in \citet{2017MNRAS.467..179G}. The main properties of each simulation at $z=0$ are listed in Table~\ref{table:t1}. A detailed description of how these parameters were obtained is given in \citet{2014MNRAS.437.1750M} and \citet{2017MNRAS.467..179G}. 

\begin{table}
\centering
\caption{Main properties of the Aq-C4 and the six {\it Auriga} simulations (resolution level 4) at $z=0$ from \citet{2014MNRAS.437.1750M} and \citet{2017MNRAS.467..179G}, respectively. The first column labels the simulation. From left to right, the columns give the virial radius, $r_{200}$; the concentration parameter  $c_{\rm NFW}$ of the underlying DM haloes; the DM mass  $M_{200}$,  and the stellar mass  $M_\star$ inside the virial radius.}
\label{table:t1}
\begin{tabular}{@{}lcccc} \hline Name & $r_{200}$ & $c_{{\rm NFW}}$ & $M_{200}$ & $M_\star$\\
& [kpc] & & [$10^{10}$~M$_{\odot}$] & [$10^{10}$~M$_{\odot}$] \\
\hline \\
Aq-C4 & $234.4$ & 16.03 & $145.71$ & $5.31$\\
Au-3 & $239.02$ & 15.6 & $145.78$ & $7.75$\\
Au-6 & $213.83$ & 11 & $104.39$ & $4.75$\\
Au-15 & $225.4$ & 7.9 & $122.25$ & $3.93$\\
Au-16 & $241.48$ & 9.3 & $150.33$ & $5.41$\\
Au-19 & $224.57$ & 8.3 & $120.9$ & $5.32$\\
Au-21 & $238.65$ & 14.2 & $145.09$ & $7.72$\\
\hline
\end{tabular}
\end{table}

Finally, it is important to highlight that, as discussed below, our analytic Galactic potentials do not account for the effect of Galactic bars. Thus, the simulations used in this work were chosen so that they do not present strong Galactic bars at $z=0$. Note that, even though the time varying potential associated with a bar can enhance the strength of chaotic diffusion in the very inner galactic region \citep{2001A&A...373..511F,2003AJ....125..785Q,2007A&A...467..145C,2008A&A...488..161C,2011ApJ...733...39S}, it is unlikely to play a significant role in erasing signatures of past and ongoing Galactic accretion events within the solar neighbourhood and beyond. Though, we defer the detailed study of this aspect to future work (Maffione et al., in preparation).

\subsection{The galactic potential}
\label{subsec:meth-pot}

In order to characterize the dynamics of MW-like stellar haloes by recourse to the chaos indicator (hereinafter CI) used in the present effort, the 
high-precision numerical integration of both the equations of motion and their first variational equations is required (see Section~\ref{subsec:meth-ind-OFLI} for further details). Indeed, the variational equations are needed to track the temporal evolution of the separation between initially nearby orbits in phase space (see the Appendix in \citetalias{2015MNRAS.453.2830M} for details). Therefore, as an analytic representation of the  underlying galactic potential is in order,  we describe the potential of each simulated galaxy by a superposition of suitable analytic and static models representing the different galactic components. Other approximations, based on series expansion of the underlying potential can be very accurate \citep[i.e.][]{1973Ap&SS..23...55C,1992ApJ...386..375H,1999AJ....117..629W,2008MNRAS.383..971K,2011MNRAS.416.2697L,2013MNRAS.434.3174V,2014ApJ...792...98M}. However, a rather large number of terms should be considered, thus rendering unfeasible the derivation of the first variational equations (for further discussion see \citetalias{2015MNRAS.453.2830M}).

Our analytic Galactic potential contains four different contributions corresponding to the central nuclear region, the bulge, the disc and the DM halo:
\begin{equation}
\Phi_{\rm MW}=\Phi_{\rm nuc}+\Phi_{\rm bul}+\Phi_{\rm disc}+\Phi_{\rm DM2}.
\label{eq:mw}
\end{equation}

\begin{figure}
\begin{center}
\begin{tabular}{c}
\hspace{-5mm}\includegraphics[width=1\linewidth]{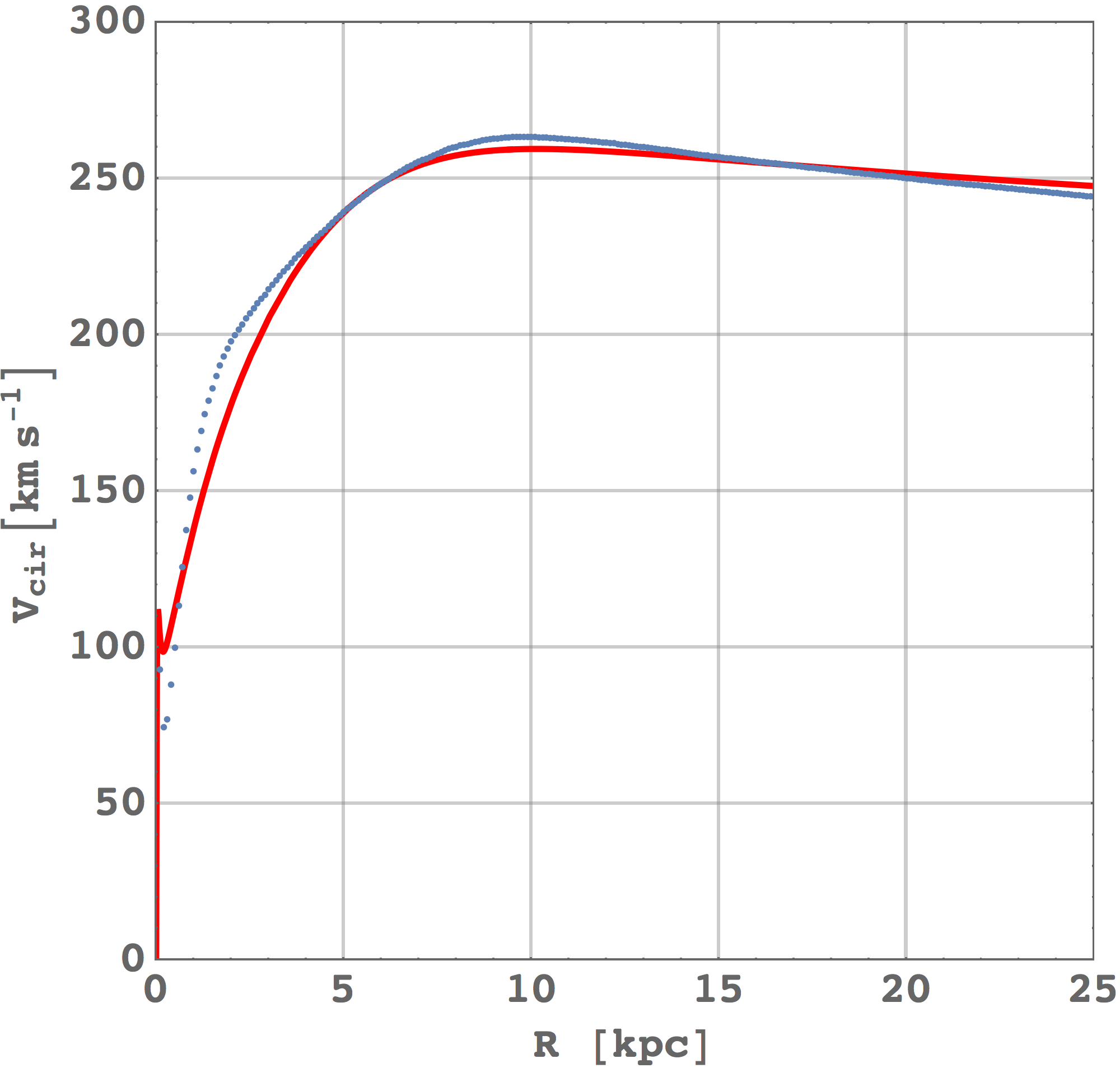}
\end{tabular}
\end{center}
\caption{Total circular velocity curves $V_{\rm cir}$ as a function of galactocentric distance $R$ for one of our analytic and static representations of the MW (red solid line) and its simulated MW-like galaxy counterpart taken from a cosmological hydrodynamic simulation (blue points). The resemblance, particularly in the solar neighbourhood-like volume (i.e. $5\la {\rm R [kpc]}\la11$), is reassuring.}
\label{fig:3c}
\end{figure}

The values of the parameters that describe each Galactic component are directly extracted from the numerical simulations, presented in last section, as described in \citet{2014MNRAS.437.1750M} and \citet{2017MNRAS.467..179G}. The concentration  and virial radius that describe each DM halo are obtained by fitting a Navarro, Frenk \& White profile \citep{1996ApJ...462..563N,1997ApJ...490..493N} to the corresponding DM particle distribution, and are listed on Table~\ref{table:t1}. For the stellar component, a decomposition of the  surface density profile into an exponential disc and a S{\'e}rsic profile is 
 performed to obtain disc scalelengths and bulge effective radii, as well as their relative mass contributions. Masses are slightly re-calibrated by fitting our analytic models to the total circular velocity curves extracted from the simulations, as illustrated in Fig.~\ref{fig:3c}.  Note that, through this process, the total stellar mass is kept constant. Resulting values are listed on Table \ref{table:t2}.

We highlight that, as in \citetalias{2015MNRAS.453.2830M}, the cosmological simulations considered in this work are only used to extract the parameters that characterize the underlying potentials, and to obtain realistic models of the phase space distributions of solar neighbourhood-like volumes (see next section).  Our goal is not to accurately characterize the impact of chaos in the Aq-C4 or {\it Auriga} local stellar haloes themselves. Instead, we aim to obtain reasonable descriptions of these numerically simulated galaxies to reflect in our results the expected variations in the galactic potential associated with the galaxy formation process. Sampling  initial conditions from a self-consistent model and later evolving them in a slightly different potential should increase the fraction of chaotic orbits in the sample \citep{2012MNRAS.419.1951V}. Thus, under the approximation of our static potentials, the fraction of chaotic orbits within the phase space volumes analyzed here will likely be overestimated with respect to that associated to the best possible analytical representation of the underlying potential.

We acknowledge that, despite the benefits of dealing with analytic and static representations of the galactic potentials, these models have their own strong limitations. For example, we are not accounting for how substructure as well as time-dependance could enhance the efficiency of diffusion in phase space \citep[see][and Section~\ref{sec:discussion} for further discussion]{2013MNRAS.433.2576P}. These issues will be tackle in a forthcoming paper.

In what follows, we describe each galactic component.

\subsubsection{The central nuclear region}

The presence of a supermassive black hole and a nuclear star cluster in the inner galactic regions \citep{2002A&A...384..112L} can significantly amplify the amount of chaos as their mass profiles contribute to a cuspy shape \citep[see for instance][]{1998ApJ...506..686V,2001PhRvE..64e6209K}. In our analysis, this component is particularly important for box orbits that are currently crossing our simulated solar neighbourhood-like volumes.
Therefore, to model such a component, which dominates the mass distribution within the inner $\sim30$ pc, we use a Plummer sphere \citep{1911MNRAS..71..460P}:
\begin{equation}
\Phi_{\rm nuc}=-\frac{B}{\sqrt{r^2+\left(\epsilon^{\rm s}_{\rm nuc}\right)^2}}\ ,
\end{equation}
where the constant $B$ is defined as $B=G\,M_{\rm nuc}$, with $G$ the gravitational constant, $M_{\rm nuc}$ the estimated mass enclosed in the central region, $r=\sqrt{x^2+y^2+z^2}$ the usual galactocentric distance and $\epsilon^{\rm s}_{\rm nuc}$ the radial scalelength. All the values of the parameters for MW model C4 are obtained from simulation Aq-C4 \citep{2014MNRAS.437.1750M}, except for $\epsilon^{\rm s}_{{\rm nuc}}$ which is taken from \citet{2002A&A...384..112L}. The nuclear region has the same values of the parameters for all of our MW models and it is not included in Table~\ref{table:t2} for the sake of brevity: the mass being $M_{\rm nuc}=2\times10^{8}$ M$_{\odot}$, and the scalelength radius, $\epsilon^{\rm s}_{\rm nuc}=0.03$~kpc.

\subsubsection{The bulge}
The stellar bulge is the dominant component within $\sim 1$~kpc \citep{2002A&A...384..112L}. In this case, we use a Hernquist profile \citep{1990ApJ...356..359H} with a scalelength, $\epsilon^{\rm s}_{\rm bul}$:
\begin{equation}
\Phi_{\rm bul}=-\frac{C}{r+\epsilon^{\rm s}_{\rm bul}}\ ,
\end{equation}
where $C$ is a constant defined as $C=G\,M_{\rm bul}$, with $M_{\rm bul}$ its total mass.

\subsubsection{The disc}
To model a  stellar disc with a double exponential density profile we follow the procedure described by \citet{2015MNRAS.448.2934S}. The idea behind this method is to approximate an exponential profile by the superposition of three different Miyamoto \& Nagai (MN) profiles \citep{1975PASJ...27..533M}. In our case, the mass distribution of the resulting models deviates from the radial mass distribution of a pure exponential disc by $< 1$~per cent out to four disc scalelengths, and by $< 6$ per cent out to ten disc scalelengths. \citet{2015MNRAS.448.2934S} provides a user-friendly online web-form that computes the best-fitting parameters for an exponential disc\footnote{\url{http://astronomy.swin.edu.au/~cflynn/expmaker.php}},
\begin{equation}
\rho(R, z) = \rho_{0} \exp(-R/\epsilon^{\rm s}_{\rm disc})\exp(-|z|/\epsilon^{\rm h}_{\rm disc})\ ,
\end{equation}
with $\rho(R, z)$ the axysimmetric density, $R=\sqrt{r^2-z^2}$ the projected galactocentric distance, $\rho_0$ the central density and when the desired total mass $M_{\rm disc}$, scalelength $\epsilon^{\rm s}_{\rm disc}$ and scaleheight $\epsilon^{\rm h}_{\rm disc}$ are provided.

Let us remind the reader that the potential of a single MN disc obeys the following expression:
\begin{equation}
\label{eq:discMN}
\Phi^{\rm MN}_{\rm disc}=-\frac{D}{\sqrt{R^2+\left[\epsilon^{\rm s\, MN}_{\rm disc}+\sqrt{z^2+\left(\epsilon^{\rm h}_{\rm disc}\right)^2}\right]^2}}\ ,
\end{equation}
where $\epsilon^{\rm s\, MN}_{\rm disc}$ and $\epsilon^{\rm h}_{\rm disc}$ are the scalelength and scaleheight of the MN disc, respectively (it should be noticed that $\epsilon^{\rm h}_{\rm disc}$ is the same for the exponential and the three MN discs). Furthermore, $D$ is a constant defined as $D=G\,M^{\rm MN}_{\rm disc}$ with $M^{\rm MN}_{\rm disc}$, its mass. 

The experiments performed in this work consider the double exponential approximation described above.

\subsubsection{The dark matter halo}
\label{subsubsec:dmcomp}
\citet{1996ApJ...462..563N,1997ApJ...490..493N} introduced a universal spherical density profile (NFW profile) that provides a reasonable fitting to the mass distribution of DM haloes of galaxies over a very wide range of mass and redshift. It has been shown since, however, that  DM haloes are not spherical as assumed by this potential. In the absence of baryons, fully cosmological simulations have shown that DM haloes are strongly triaxial, with their shape varying  as a function of galactocentric distance \citep[see e.g.][]{2006MNRAS.367.1781A,2011MNRAS.416.1377V}. To model this behaviour, \citetalias{2015MNRAS.453.2830M} uses a triaxial extension of the NFW profile \citep[introduced in][]{2008MNRAS.385..236V}, 
\begin{equation}
\label{eq:nfw_tri}
\Phi_{\rm DM}=-\frac{A}{r_p}\ln \left( 1+\frac{r_p}{r_s}\right)\ ,
\end{equation}
where $A$ is a constant defined as: 
\begin{equation}
A=\frac{G\,M_{200}}{\ln\left(1+c_{\rm NFW}\right)-c_{\rm NFW}/\left(1+c_{\rm NFW}\right)}\ ,
\end{equation}
with $M_{200}$ being the virial mass of the DM halo and $c_{\rm NFW}$ the concentration parameter; $r_s=r_{200}/c_{\rm NFW}$ is a  scale radius with $r_{200}$ the virial radius. The triaxiality of this potential is introduced through $r_p$,
\begin{equation}
r_p=\frac{(r_s+r)r_i}{r_s+r_i}\ ,
\end{equation}
where $r_i$ is an ellipsoidal radius for the inner parts defined as:
\begin{equation}
r_i=\sqrt{\left(\frac{x}{a}\right)^2 + \left(\frac{y}{b}\right)^2 +
\left(\frac{z}{c}\right)^2}\ .
\end{equation}
The quantities $b/a$ and $c/a$ represent the intermediate-to-major and the minor-to-major principal axes ratios and are defined such that $a^{2} + b^{2} + c^{2} = 3$. Note that the potential shape changes from ellipsoidal in the inner regions to near spherical at the scale radius, $r_{s}$. Thus, for $r \ll r_{\rm s}$, $r_{\rm p} \simeq r_{\rm i}$ and for $r \gg r_{s}$, $r_{p} \simeq r$ \citep{2008MNRAS.385..236V}.

\begin{figure}
\begin{center}
\begin{tabular}{c}
\hspace{-5mm}\includegraphics[width=1\linewidth]{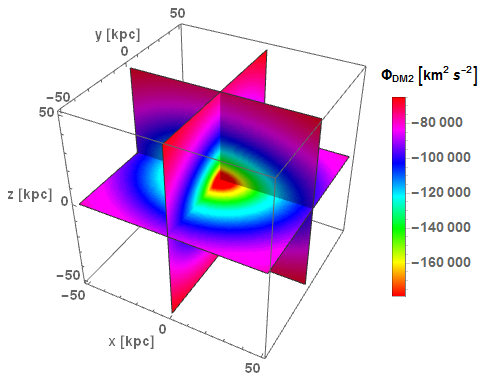}\\
\hspace{-5mm}\includegraphics[width=1\linewidth]{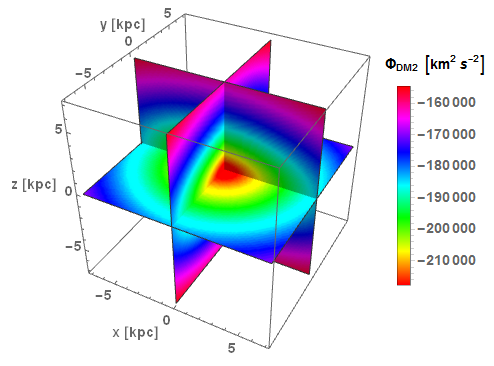}
\end{tabular}
\end{center}
\caption{Isopotentials for the outer (top panel) and inner parts (bottom panel) of our \textquotedblleft bi-triaxial'' extension of the NFW profile. The potential becomes more oblate as the centre is approached.}
\label{fig:isopot}
\end{figure}

The addition of baryons, however, significantly alters the DM phase space distribution. As a result of the central accumulation of baryons, DM haloes in  cosmological hydrodynamic simulations are found to be more oblate than triaxial in the inner parts \citep[e.g.][]{2014MNRAS.437.1750M,2017MNRAS.467..179G}. To account for this, we introduce a \textquotedblleft bi-triaxial'' extension of the NFW profile by defining a new parameter: 
\begin{equation}
r'_p=\frac{(r_s+r_o)r_i}{r_s+r_i}\ ,
\label{rpp}
\end{equation}
where a second ellipsoidal radius, $r_o$, is defined as:
\begin{equation}
r_o=\sqrt{\left(\frac{x}{a'}\right)^2 + \left(\frac{y}{b'}\right)^2 +
\left(\frac{z}{c'}\right)^2}\ ,
\end{equation}
with $a'^2+b'^2+c'^2=3$, while $(b/a,c/a)$ and $(b'/a',c'/a')$ denote the principal axes ratios of the DM halo in the inner and outer regions, respectively, with a smooth transition taking place at $r_{s}$. Thus, the new potential is simply $\Phi_{\rm DM2}=\Phi_{\rm DM}(r'_p)$. Fig.~\ref{fig:isopot} shows an example of isopotentials in the outer and inner parts of $\Phi_{\rm DM2}$. The transition from mildly triaxial to a more oblate shape can clearly be seen as the centre is approached \citep[compare Fig.~\ref{fig:isopot} with Fig. 10 of][]{2008MNRAS.385..236V}. Note, however, that our models have less exaggerated (and more realistic) axes ratios (see Table~\ref{table:t2} in Section~\ref{subsec:relevance-dist}) than those considered in the example, and the transition is not so evident.

\subsection{Cosmologically motivated initial conditions}
\label{subsec:meth-init}
To investigate the efficiency of chaotic mixing associated with the galactic potential described in Section~\ref{subsec:meth-pot}, Eq.~\eqref{eq:mw},  we first need to model the distribution of tracer particles in phase space. Rather than stochastically sampling the associated phase space distribution, we select particles from the different simulated galaxies as described below.

To characterize  how the efficiency of chaotic mixing varies as a function of galactocentric radius, we first select stellar particles located within a $15^{\circ}$ wedge, whose axis coincides with the disc semi-major axis. For these experiments we will focus only on the simulation dubbed Aq-C4. To increase the numerical resolutions we include stellar particles located within diametrically opposed wedges. Within these wedges, with vertex in the galactic center, we consider star particles with  $0.5 \lid r \lid 50.5$~kpc, where $r$ represents the galactocentric distance. Note that here we consider  stellar particles born both in-situ and in accreted satellites. 


In addition, and to compare with the results presented in \citetalias{2015MNRAS.453.2830M}, we also select all DM particles that are located within a sphere of $2.5$~kpc radius, centred at $8$~kpc along the semi-major axis of the DM halo in the Aq-C4 simulation. Note that the disc scalelength in this simulation is similar to that of the MW. Thus, this region can be regarded as a solar neighbourhood-like volume. The larger number of DM particles, with respect to their stellar counterparts, allow us to characterize more robustly the efficiency of chaotic mixing within this relatively small volume.

In Section \ref{subsubsec:relevance-6Au} we analyse the {\it Auriga} simulations to characterize how the stochasticity inherent to the process of galaxy formation can affect the fraction of chaotic orbits in solar neighbourhood analogs. Stellar particles located within a sphere of $5$~kpc radius are selected on each simulation. Note that the final scalelengths of our simulated galactic discs show a great diversity. Thus, in order to select regions with the same density contrast, the spheres are centred at galactocentric distances of $2.65\times\epsilon^{\rm s}_{\rm Au-i}$, where $2.65$ is the ratio between the Sun's galactocentric distance ($\sim 8$~kpc) and the scalelength of the MW disc ($\sim 3$~kpc), and $\epsilon^{\rm s}_{\rm Au-i}$ being the $\epsilon^{\rm s}_{{\rm disc}}$ for the simulated galaxy in simulation Au-i.

\subsection{Chaos Indicator: The Orthogonal Fast Lyapunov Indicator}
\label{subsec:meth-ind-OFLI}

In this work we use the Orthogonal Fast Lyapunov Indicator, OFLI \citep[][]{2002CeMDA..83..205F}, to quantify and characterize the fraction of chaotic orbits within different phase space volumes. Here we  briefly describe the method and refer the reader to \citetalias[][and references therein,]{2015MNRAS.453.2830M} for further
details\footnote{Note that this study is supported by similar results based on other CI, the MEGNO \citep[see for instance][]{2000A&AS..147..205C,2003PhyD..182..151C,2005ApJ...619.1084G,2012CeMDA.112...75C,2016LNP...915..93C}. The orbital classification obtained with the approximate galactic potential, described in Section~\ref{subsec:meth-pot}, is thus robust. Nonetheless, for the sake of brevity, the results based on the MEGNO are not included. For a thorough discussion on the advantages and disadvantages of the most popular CIs found in the literature we refer the reader to \citet{2011IJNLM..46...23M,2011CeMDA.111..285M,2012IJBC...2230033D,2013MNRAS.429.2700M,2016LNP..915.....S} and references therein.}.

The basic idea behind the OFLI is to track the time evolution of the distortion of an initially infinitesimal local phase space volume surrounding any given orbital initial condition. The rate at which 
this volume expands along the direction of maximum distortion can be used to identify chaos. In practice, we follow the time evolution of a unit
deviation vector $\mathbf{\hat{w}}(t)$, evolving on an $N$-dimensional Hamiltonian $\mathcal{H}$ along a given solution of the equations of motion (i.e. the orbit)  $\gamma(t)$. The vector $\mathbf{\hat{w}}(t)$ is initially chosen normal to the energy surface \citep[in order to avoid spurious structures, see][]{2016LNP...915..55B} and, as it evolves, we  take its orthogonal component to the flow, $\hat{w}(t)^{\perp}\in\mathbb{R}$. Its largest value ($\sup$, or minimum upper bound) between an initial time $t_0$ and a stopping time $t_f$ is retained. The OFLI is then defined as:
\begin{equation}
\mathrm{OFLI}^{\gamma}(t_f)=\sup_{t_0<t<t_f}\left[\hat{w}(t)^{\perp}\right]\ ,
\end{equation}
for the orbit $\gamma$. For both chaotic and non-periodic regular orbits, the value of the  $\mathrm{OFLI}^{\gamma}$ tends to
infinity as time increases. However,  on a logarithmic time-scale, the $\mathrm{OFLI}^{\gamma}$ presents and exponential growth  for chaotic orbits , while it is linear for resonant and non-resonant regular orbits (with different rates). In the case of periodic orbits, it  oscillates around a constant value \citep[for further details we refer the reader to][]{2002CeMDA..83..205F}. 

From now on, we integrate the orbits and compute the preferred CI using the {\sc lp-vicode} code \citep[see][]{2014A&C.....5...19C}. The numerical  integrator conserves energy to an accuracy of one part in $10^{-12}$ or less for all the experiments throughout the paper. 

\subsection{Orbital classification}
\label{orbitalclassification}
As discussed in \citetalias{2015MNRAS.453.2830M}, the local spatial density of a star moving on a chaotic orbit decreases exponentially with time. Stellar streams moving on these orbits experience a rapid phase space mixing process, thus eroding signatures of past accretion events. In contrast, the local density of a star moving on a regular orbit falls with time as a power law, with an exponent less than or equal to $3$ \citep[a significantly lower rate,][]{1999MNRAS.307..495H,2008MNRAS.385..236V,2013MNRAS.436.3602G}. The chances of finding stellar streams in the solar neighbourhood are thus higher if they are evolving on regular orbits.

In \citetalias{2015MNRAS.453.2830M} we presented an analysis that highlighted the very strong connection between the time evolution of the local (stream) density around a given particle and the time evolution of the corresponding OFLI. More precisely, we showed that if the OFLI grows linearly (i.e. regular behaviour), then the associated local density decreases as a power law, with index less than or equal to $3$. An exponential growth of the OFLI, instead, reflects an exponential decay of the corresponding local density. 

To characterize the impact of chaotic mixing on the different local phase space distributions, we proceed as in \citetalias{2015MNRAS.453.2830M}. We examine two central aspects associated with chaotic mixing efficiency: {\it (i)} the distribution of chaos onset times, i.e. the time at which a given orbit starts to showcase its chaotic behaviour, and {\it (ii)} the chaotic mixing diffusion rate, a mechanism that can lead to a significant drift in the integrals of motion space. 
To tackle {\it (i)},  we compute the time evolution of the OFLI of stellar and DM particles located within different local volumes (see Section~\ref{subsec:meth-init}), by integrated the orbits for a maximum period 
of 1000~Gyr. We use such a long timespan to identify very sticky orbits reliably (see below). We then classify as chaotic orbits those that show chaos onset times  smaller than 10~Gyr; i.e. approximately within a Hubble time. Orbits that showcase their chaotic behaviour on time-scales larger than a Hubble time are classified as sticky orbits. Note that the later should not be associated with the concept of weakly chaotic orbits. Sticky orbits are orbits that behave as regular for long periods of time, with their associated local stream densities evolving as a 
power-law until they reach the chaotic sea. Instead, weakly chaotic orbits behave like chaotic orbits right from the beginning, with local 
stream densities evolving exponentially, albeit with a very small exponent.

Orbits that never showcase chaotic behaviour are classified as regular. These threshold-dependent definitions are arbitrary. However, based on them it is possible to isolate orbits which are likely to showcase chaotic behaviour within relevant and physical periods of 
time from those in which chaos is clearly irrelevant (see \citetalias{2015MNRAS.453.2830M} for further details).
To address {\it (ii)}, we quantify the diffusion of pseudo-integrals of motion for large ensembles of initially nearby test particles in phase space.

\section{The actual relevance of chaos in multicomponent triaxial potentials}
\label{sec:relevance}
The OFLI allow us to robustly characterize the time evolution of the local (stream) density around any stellar particle (as shown in \citetalias{2015MNRAS.453.2830M}). In what follows, we use 
this tool to quantify the fraction of particles, located within different volumes, that are moving on regular, sticky and chaotic orbits. A large fraction of chaotic orbits would indicate that substructure in phase-space, specially those associated with the oldest accretion events, may have been efficiently erased due to chaotic mixing. This is especially relevant for the inner galactic regions, such as the solar neighbourhood, due to the much shorter dynamical time-scales associated with the corresponding orbits.

\subsection{The distribution of chaos onset times as a function of galactocentric distance}
\label{subsec:relevance-dist}
In this section we focus on the stellar particles located within $15^{\circ}$ wedges (see Section~\ref{subsec:meth-init}), extracted from the simulation Aq-C4. 
The corresponding distribution of initial conditions is dissected in bins of different galactocentric distances. Each bin covers a different (non-overlapping) galactocentric distance range of 5~kpc and contains at least of the order of 500 stellar particles.

In order to compute the OFLI, we integrate the equations of motion together with the first variational equations (see the Appendix in \citetalias{2015MNRAS.453.2830M}), assuming an analytic and static MW potential of the form given by Eq.~\eqref{eq:mw}, i.e.: 
\begin{equation}
\Phi_{\rm MW}^{\rm C4} = \Phi_{\rm nuc}+\Phi_{\rm bul}+\Phi_{\rm disc}+\Phi_{\rm DM2}\ ,
\label{eq:aqc4a}
\end{equation}
denoted as model C4 (see Section~\ref{subsec:meth-pot} for further descriptions of each component). The values of the parameters are given in Tables~\ref{table:t1} and~\ref{table:t2}. To describe the 
shape of this potential, we use the triaxiality parameter \citep[][]{1991ApJ...383..112F},
\begin{equation}
    T=\frac{1-(b/a)^2}{1-(c/a)^2}\ .
\end{equation}
The shape of the DM halo  can be characterized as mainly oblate for values of $0\lid T < 0.333$, strongly triaxial for $0.333\lid T\lid 0.666$, and mainly prolate for $0.666 < T\lid 1$. The principal axes ratios in the inner parts ($r\la r_s$) of the corresponding DM halo are computed using DM particles located within the first 10~kpc. Then, the triaxiality of the Aq-C4 halo in the inner regions is $T_{\rm inner} \sim 0.227$ (an oblate shape). For the outer parts ($r\gg r_s$) the principal axes ratios are computed using DM particles located within 40 and 70~kpc. The triaxiality parameter here is $T_{\rm outer} \sim 0.681$ (a mildly prolate shape).

\begin{table*}
\centering
\caption{Parameters of the components for the C4 and the six {\it Auriga} MW models. The first column labels the model. From left to right, the columns give: the mass, $M_{\rm bul}$, and the scalelength radius, $\epsilon^{\rm s}_{\rm bul}$, of the bulge; the mass, $M_{\rm disc}$, the scalelength radius, $\epsilon^{\rm s}_{\rm disc}$, and the scaleheight radius, $\epsilon^{\rm h}_{\rm disc}$,  of the disc; the distance from the Galactic centre where the solar neighbourhood-like sphere is located, $R_{\rm sph}$ ($2.65\times\epsilon^{\rm s}_{\rm disc}$); the intermediate-to-major, $b/a$, and the minor-to-major, $c/a$, principal axes ratios in the inner parts as well as $b'/a'$ and $c'/a'$, the corresponding principal axes ratios in the outer parts; the triaxiality parameter for the inner ($T_{\rm inner}$) and the outer ($T_{\rm outer}$) regions.}
\label{table:t2}
\begin{tabular}{@{}lcccccccccccc} \hline Name & $M_{\rm bul}$ & $\epsilon^{\rm s}_{\rm bul}$ & $M_{\rm disc}$ & $\epsilon^{\rm s}_{\rm disc}$ & $\epsilon^{\rm h}_{\rm disc}$ & $R_{\rm sph}$ & $b/a$ & $c/a$ & $b'/a'$ & $c'/a'$ & $T_{\rm inner}$ & $T_{\rm outer}$\\
& [$10^{10}$~M$_{\odot}$] & [kpc] & [$10^{10}$~M$_{\odot}$] & [kpc] & [kpc] & [kpc] &&&&&&\\
\hline 
C4 & 0.47 & 0.84 & 5.96 & 3.12 & 0.3  & 8 & 0.99 & 0.94 & 0.92 & 0.88 & 0.227 & 0.681\\ 
Au-3 & 4.10 & 1.51 & 4.29 & 7.50 & 0.3  & 19.875 & 0.998 & 0.926 & 0.976 & 0.935 & 0.022 & 0.379\\
Au-6 & 1.37 & 1.30 & 3.22 & 4.53 & 0.3  & 12.005 & 0.996 & 0.922 & 0.953 & 0.899 & 0.055 & 0.477\\
Au-15 & 0.79 & 0.90 & 2.74 & 4 & 0.3  & 10.06 & 0.996 & 0.93 & 0.976 & 0.94 & 0.057 & 0.419\\ 
Au-16 & 2.20 & 1.56 & 3.57 & 7.84 & 0.3 & 20.776 & 0.999 & 0.944 & 0.999 & 0.949 & 0.023 & 0.021\\
Au-19 & 2.02 & 1.02 & 2.88 & 4.31 & 0.3  & 11.422 & 0.989 & 0.931 & 0.964 & 0.907 & 0.160 & 0.399\\
Au-21 & 3.48 & 1.36 & 3.86 & 4.93 & 0.3  & 13.065 & 0.995 & 0.943 & 0.99 & 0.961 & 0.083 & 0.251\\
\hline
\end{tabular}
\end{table*}

The orbits of the stellar particles are integrated over a 1000~Gyr timespan, with a timestep of  $1$ Myr\footnote{An integration timestep of $10^{-2}$ Myr for a total integration time of $10$~Gyr was also used in order to check numerical convergence: no changes were found in the results.}. The orbits of the particles are then classified according to the shape of their OFLI time evolution curves following the procedure presented in \citetalias{2015MNRAS.453.2830M} and revisited below. 

\begin{figure}
\begin{center}
\begin{tabular}{c}
\hspace{-5mm}\includegraphics[width=1\linewidth]{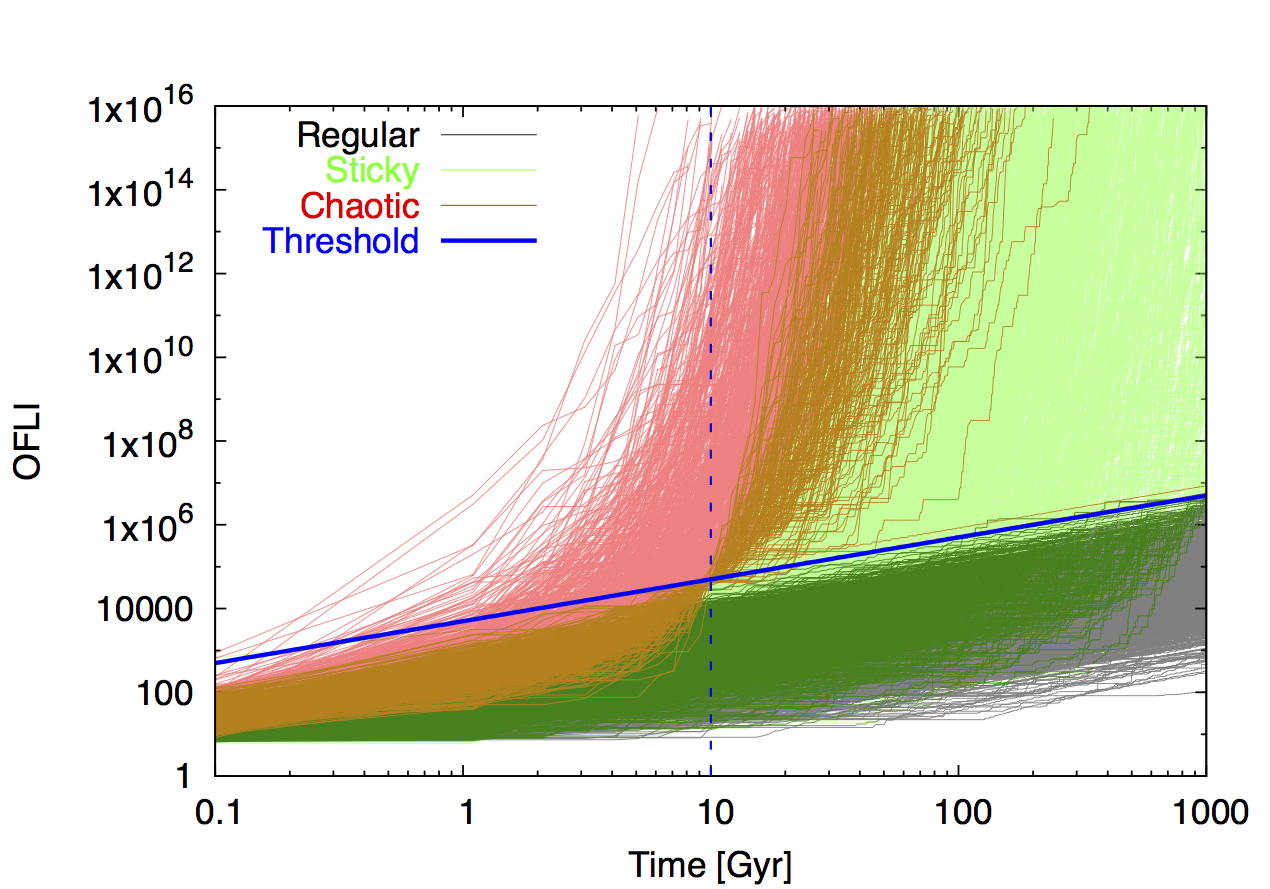}\\
\end{tabular}
\end{center}
\caption{Time evolution of the OFLI for 4410 DM particles considered for the MW model C4 and within an interval of time long enough to identify very sticky orbits (1000~Gyr). The upper limit used as a threshold for regular motion is shown with a blue solid line. The 10~Gyr threshold used to isolate sticky from chaotic orbits is shown with a vertical dashed-blue line. The three orbital components, i.e.  regular,  sticky and  chaotic, can be clearly distinguished by using the OFLI with both simple thresholds. Notice the logarithmic time-scale.}
\label{fig:3}
\end{figure}

As an example, and to demonstrate our method, we first show in Fig.~\ref{fig:3} the time evolution of the OFLI for a subset of 4410 DM particles located within a solar neighbourhood-like volume (the same subset will be use in next section)\footnote{We advise the readers affected with the common form of red-green color blindness to convert the paper in a grayscale format to distinguish the chaotic from the sticky components in the figures thorough the manuscript.}. It is clear that, for some orbits, the OFLI quickly diverges exponentially, whereas for others it continues to grow slowly following a power-law in time. For large samples of orbits, such as those shown in Fig.~\ref{fig:3}, individually inspecting each curve to estimate the chaos onset times (time at which the OFLI starts to diverge exponentially) becomes unfeasible. 
We address this by introducing a time evolving threshold value, which essentially is an upper limit for the typical linear behaviour of the OFLI seen in regular orbits. This threshold, indicated with a blue solid line in Fig.~\ref{fig:3}, evolves linearly with time and envelopes all the curves that present a linear behaviour. 
Particles are classified as either sticky or chaotic as soon as their corresponding OFLI crosses this threshold. Note that  threshold crossings within the first Gyr of evolution are neglected as this period corresponds to the typical transient stage of the indicator. Chaotic orbits are defined as those which cross the threshold  within the first 10~Gyr of their evolution. This 10~Gyr {\it barrier} is depicted in Fig.~\ref{fig:3} by a vertical dashed blue line. In a handful of cases we find that, even though the OFLI crossed the threshold at an early stage, it later continued to evolve linearly with time.  Thus, the fraction of chaotic orbits within each volume may 
be slightly overestimated. In other words, our procedure provides conservative estimates. 

\begin{figure}
\begin{center}
\begin{tabular}{c}
\hspace{-5mm}\includegraphics[width=1\linewidth]{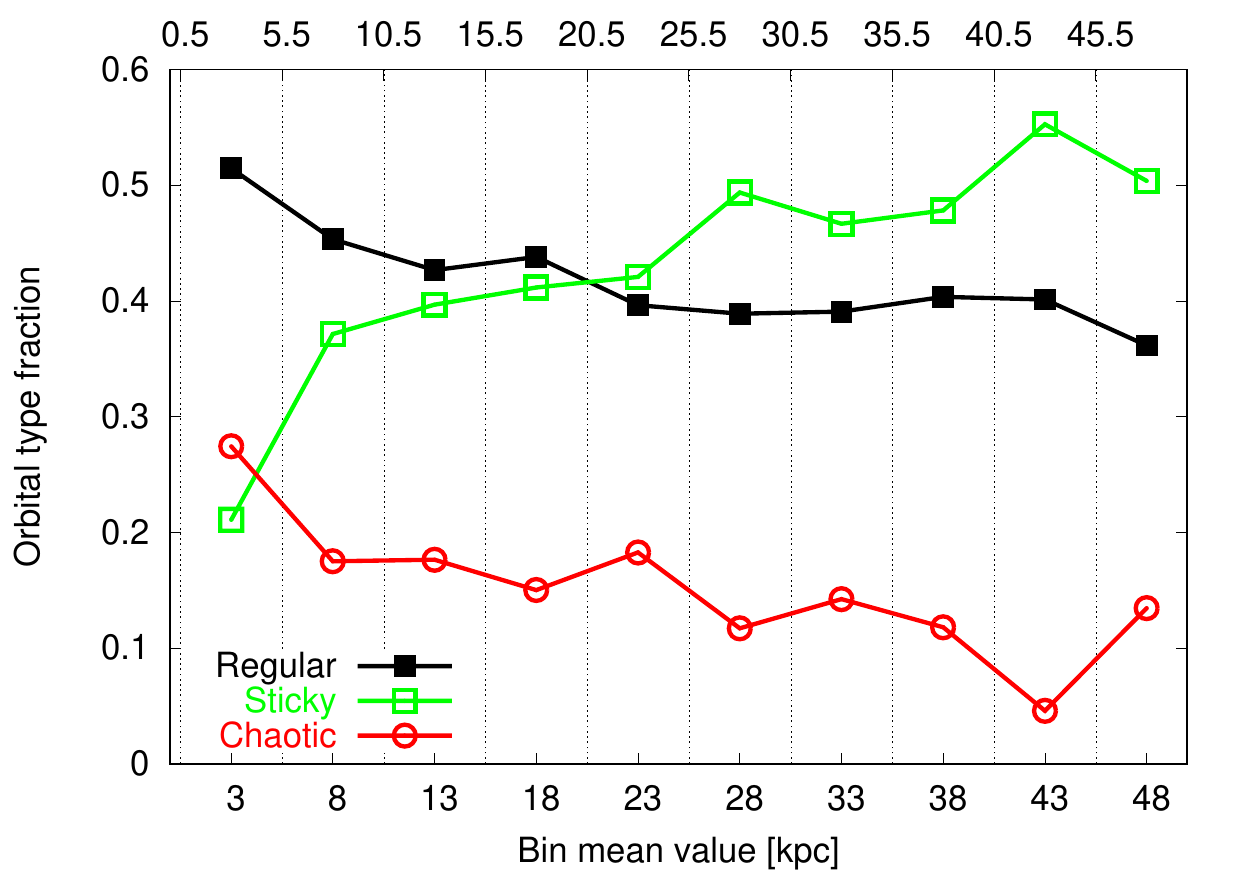}\\
\end{tabular}
\end{center}
\caption{Orbital type distribution as a function of the bin mean galactocentric distance value for MW model C4. Stellar particles on chaotic orbits remain bounded under $\sim27.45$~per cent at all galactocentric distances and below $\sim17.51$~per cent if the innermost bin is removed.}
\label{fig:adeterminar}
\end{figure}

The results of this procedure are presented in Fig.~\ref{fig:adeterminar}, where we show the fraction of regular (black solid squares), sticky (green open squares) and chaotic (red open dots) orbits, as a function of the mean galactocentric distance of the corresponding bin. We find that the fraction of regular orbits shows a very mild decrease as we move from the inner to the outer galactic regions, with orbital fractions in between $\sim37$ and $\sim53$~per cent. A more significant galactocentric distance dependence is shown by sticky and chaotic orbits, with orbital fractions varying from $20$ to $55$~per cent and $30$ to $5$~per cent, respectively. Note that sticky and  chaotic orbits show approximately a specular behaviour. Due to the longer dynamical time-scales associated with the outer galactic regions, clearly the fraction of chaotic orbits that exhibits their chaotic behaviour within a Hubble time becomes 
progressively smaller. 

Mainly, we find that, at all galactocentric distances $\approx 70$ -- $95$~per cent of the orbits show a regular behaviour within a Hubble time (i.e. regular +  sticky orbits by our definitions), with associated local stream densities that decrease as a power law rather than the much faster exponential decay associated with chaos. At the location of the Sun (bin enclosed between 5.5 and 10.5~kpc, i.e. centered at 8~kpc), this fraction takes a value of $\gtrsim 80$~per cent. Comparison with the results presented in \citetalias{2015MNRAS.453.2830M} (where only $\la20$~per cent of orbits could experience chaotic mixing) suggest that considering a multicomponent potential, including a central super massive black hole, a bulge, an axisymmetric disc and a double triaxial DM halo, does not significantly enhance the relevance of chaos within a Hubble time. We will further explore this in the following sections. 

\subsection{The relevance of baryons}
\label{subsubsec:relevance-C45}
In the previous section we have shown that, under the particular set up used for the multicomponent Galactic potential, the fraction of orbits exhibiting chaotic behaviour within a Hubble time is small in a solar neighbourhood-like volume ($\la$ 20~per cent). In this section we will explore whether different configurations of the baryonic components affect this result. To increase the numerical resolution, in these experiments we will analyse the orbits of the $\approx 4400$ DM particles enclosed within a 2.5~kpc sphere centred 8~kpc from the Galactic center of the C4 model (see Sec. \ref{subsec:meth-init})  

The orbits of this subset of DM particles are first integrated in a Galactic potential that only considers the double triaxial DM halo, i.e.,
\begin{equation}
\Phi^{\rm C4}_{\rm MW}=\Phi_{\rm DM2}\ . 
\end{equation}

We will use this result as a reference to contrast against the results obtained when the different baryonic components of the potential are introduced. Recall that, as discussed in Section~\ref{subsec:meth-pot}, the fraction of chaotic orbits expected after integrating initial conditions that are not self-consistent with the Galactic potential are larger than what would be obtained from the corresponding self-consistent model\footnote{Note that this holds true as long as the number of isolating integrals of motion in the non self-consistent potential is the same as in the self-consistent case.}. 
The fraction of regular, sticky and chaotic orbits in this experiment are $76$, $23.4$ and $0.6$~per cent, respectively. It is evident that the fraction of chaotic orbits within a Hubble time is negligible for this potential. 

In \citetalias{2015MNRAS.453.2830M} we estimated the fraction of chaotic orbits within solar neighbourhood-like volumes extracted from the {\it Aquarius} DM-only simulations. In particular, for the DM-only version of the simulation analysed in this section, the fraction of regular, sticky and chaotic orbits found were $31.6$, $46.6$ and $21.8$~per cent, respectively. It is clear that the inclusion of baryons in the Aq-C4 simulation resulted in a significant reduction of the chaotic orbits within solar neighbourhood-like volumes. As can be seen from \citet[figure 8]{2016MNRAS.458.1559Z}, where the shape of the main DM halo of the DM-only and hydro simulations are compared as a function of galactocentric distance, the addition of baryons strongly reduces the triaxiality within the inner galactic regions. 

We now integrate the same subset of orbits, now including in the Galactic potential the two main baryonic components, 
i.e.,  
\begin{equation}
\Phi^{\rm C4}_{\rm MW}=\Phi_{\rm bul}+\Phi_{\rm disc}+\Phi_{\rm DM2}\ . 
\label{eq:aqc4b-3}
\end{equation}
The fractions of regular, sticky and chaotic orbits are  $50.5$, $34.1$ and $15.4$~per cent, respectively. Note that the fraction of chaotic orbits has significantly increased with the inclusion of these two components. This result shows that, while the asymmetry of the DM halo inner parts is the source of chaos for stellar halo particles in this solar neighbourhood-like volume, the bulge-disc pair plays a significant role in amplifying the occurrence of chaotic motion. As we show in what follows, this is due to a significant enhancement of the asymmetry between the inner and outer Galactic potential. 

\begin{figure}
\begin{center}
\begin{tabular}{c}
\hspace{-5mm}\includegraphics[width=1\linewidth]{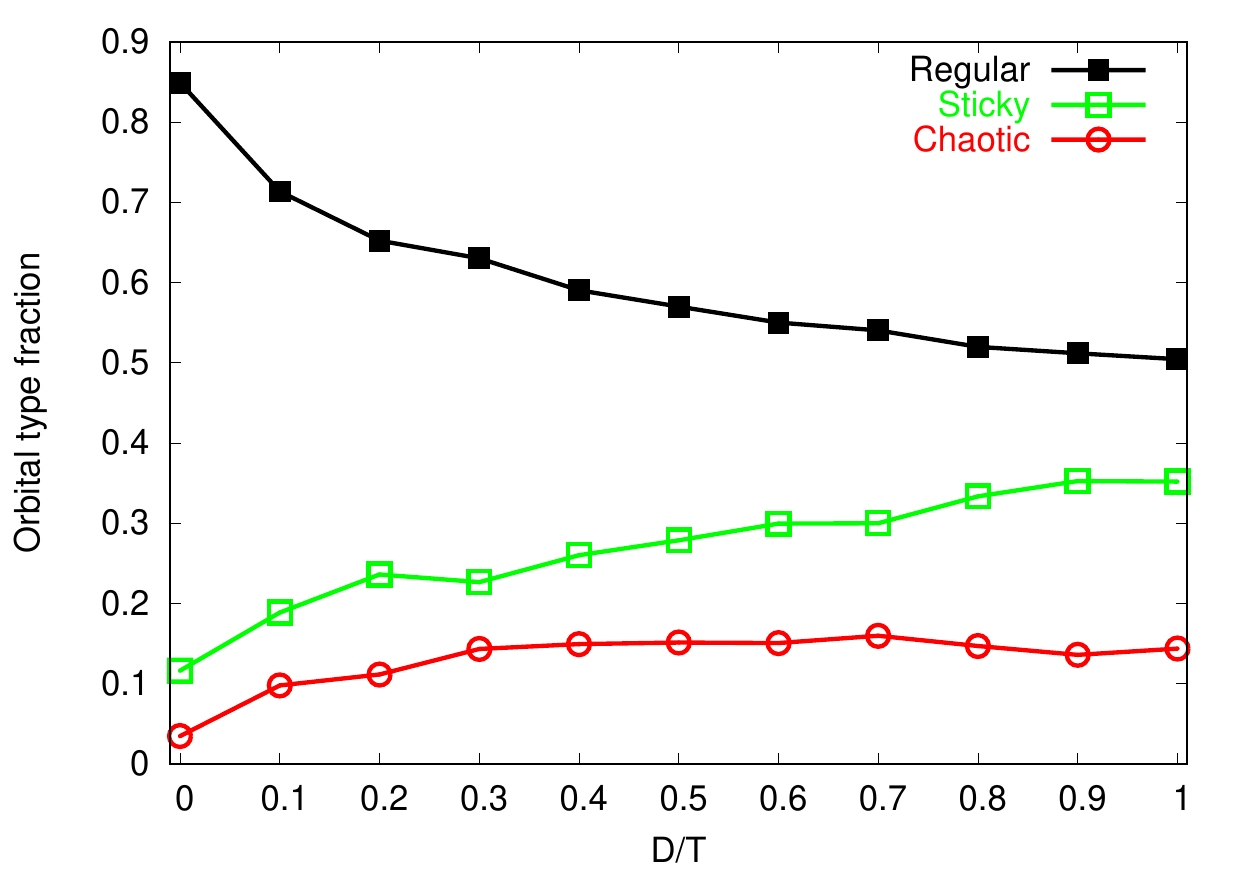}\\
\end{tabular}
\end{center}
\caption{Orbital type distribution as a function of the disc to total baryonic mass fraction ($D/T$). The fraction of DM particles on chaotic orbits is always smaller than $\sim16$~per cent.}
\label{fig:bd_frac}
\end{figure}

To study the impact of these chaos amplifiers, we repeat the latter experiment, now varying the masses of both baryonic components while keeping the total mass of the pair constant. The results are shown in Fig.~\ref{fig:bd_frac}, where $D/T$ is the disc to total baryonic mass fraction. 

Comparison with the result obtained considering the potential given by Eq.~\eqref{eq:aqc4b-3}, shows that associating all the baryonic mass to the galactic bulge ($D/T=0$) results in a reduction of the fraction of sticky and chaotic orbits; $11.6$ and $3.5$~per cent, respectively. The addition of this massive spherical bulge component, which dominates the potential in the inner Galactic region, reduces the impact of the inner triaxial shape of the underlying DM halo on the orbital distribution. 

Interestingly, as the values of $D/T$ increase, both chaotic and, more strongly, sticky orbital fractions increase. While the fraction of chaotic orbits reaches a maximum of $\sim 16$~per cent for values of $D/T \gtrsim 0.3$, the fraction of sticky orbits continues to grow to values of $\sim35$ percent at $D/T=1$ (all the baryonic mass is associated to the disc). As opposed to the spherical bulge, the disc strongly amplifies the effects of this mild oblateness of the inner triaxial DM halo potential. The minimum fraction of regular orbits is found for  values of $D/T = 1$, i.e. $50.4$~per cent. None the less, it is important to highlight that, despite the redistribution of the baryonic mass in the pair bulge-disc, the fraction of orbits that exhibits a chaotic behaviour within a Hubble time is always smaller than $\sim16$~per cent.   

In the next section we characterize the impact that different galactic formation histories may have on our chaotic orbital fraction estimates.

\subsection{The impact of formation history}
\label{subsubsec:relevance-6Au}
In the previous section we showed that, even though the inclusion of the baryonic component in our simulated Galaxy results in a reduction of the triaxiality of the inner DM halo, the addition of the disc enhances the   asymmetry  between the inner ($r \la r_{s}$) and the outer overall Galactic potential $(r > r_{s})$. As a consequence, the fraction of chaotic orbits remains consistent with that obtained from  strongly triaxial DM  potentials associated with DM-only simulations.   

Our results were based on a single numerical model, namely C4. Thus, in this section we will explore whether differences in shapes and masses of the different galactic components, originating as a consequence of different formation histories, have an effect on our previous results. 

As discussed in Section~\ref{subsec:meth-sim}, here we consider a subset of six simulations from the {\it Auriga project}. Recall that our analytic Galactic potentials do not account for the effect of Galactic bars. Thus, these simulations  were selected to not show strong bars at $z = 0$.  

The Galactic potential of each {\it Auriga} galaxy is modelled considering Eq~\ref{eq:mw}.  The values of the parameters that
describe the potentials are given in Tables~\ref{table:t1} and~\ref{table:t2}. With these parameter sets, a good agreement between the analytic and the numerical velocity curves of the models is obtained. Comparison between the triaxiality of the inner and outer DM haloes reveals a significant diversity in the asymmetric shape of this Galactic component.  

From each simulated Galaxy, stellar particles located within a sphere of 5~kpc radius centred at a distance of $2.65\times\epsilon^{\rm s}_{\rm Au-i}$ are selected (see Sec.~\ref{subsec:meth-init}). Since we are interested in studying the efficiency of chaotic mixing in erasing local signatures of accretion events, in this section we will only consider accreted stellar particles (i.e. stellar particles that were born within the potential wells of accreted satellite galaxies). Note however that our results are not significantly modified when in-situ stellar populations are taken into account. As before, orbits are integrated for 1000~Gyr, with an integration timestep of  $1$ Myr.

\begin{figure}
\begin{center}
\begin{tabular}{c}
\hspace{-5mm}\includegraphics[width=1\linewidth]{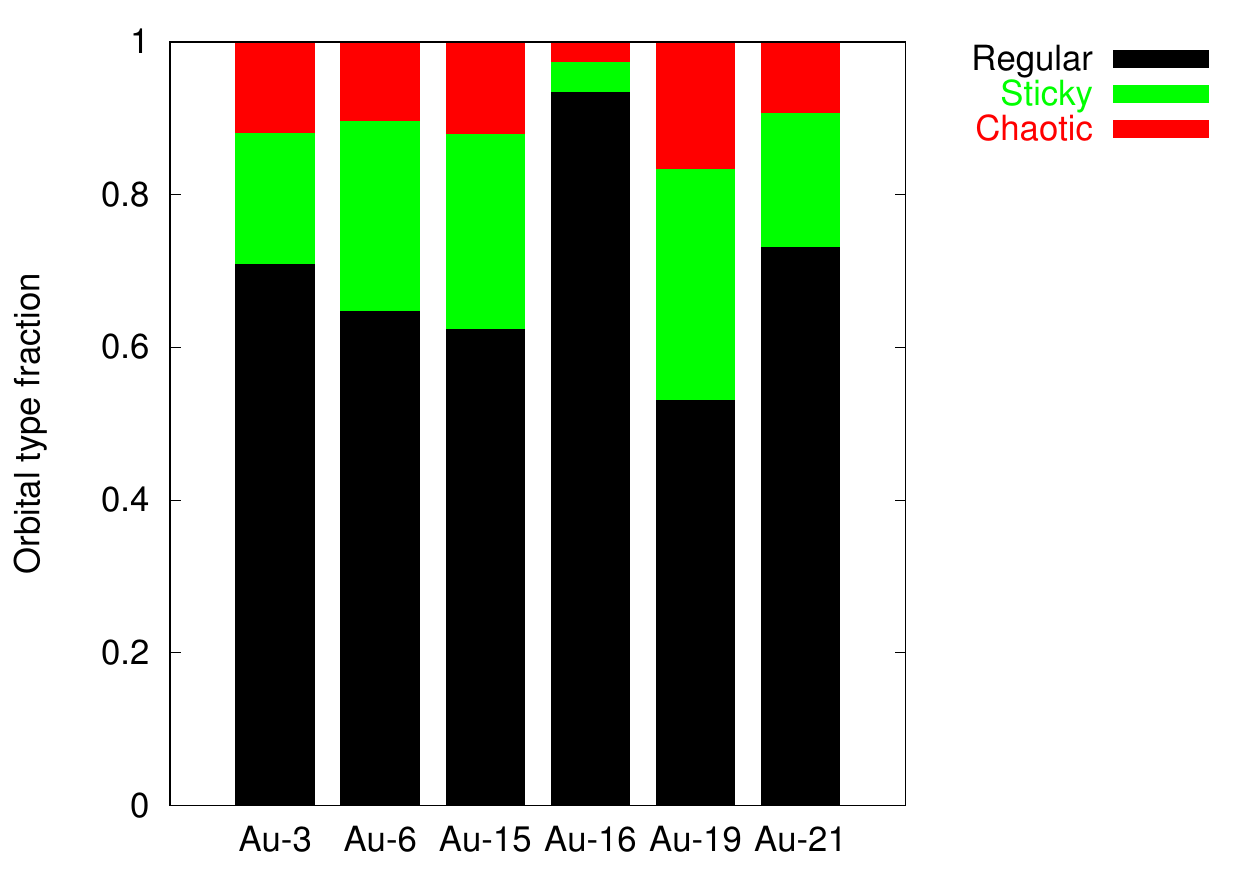}\\
\end{tabular}
\end{center}
\caption{Orbital type distribution for the six {\it Auriga} models. The fraction of accreted stellar particles on chaotic orbits is highest for model Au-19, $\sim16.6$~per cent.}
\label{fig:auriga_frac}
\end{figure}

The results of this analysis are summarized in Fig.~\ref{fig:auriga_frac} where we show with a normalized histogram and for each {\it Auriga} Galaxy model, the fraction of regular (black), sticky (green) and chaotic (red) orbits. The model with the smallest fraction of regular orbits is Au-19 ($\sim 53.2$~per cent). Interestingly, this model contains the most triaxial DM halo among the {\it Auriga} galaxies, with inner and outer triaxiality parameters (defined using the inner and outer principal axial ratios) of $T_{\rm inner} = 0.160$  and $T_{\rm outer} = 0.399$, respectively. On the other hand, the {\it Auriga} model with clearly the largest fraction of regular orbits is Au-16. Its associated DM halo potential has a nearly perfect oblate shape, with triaxiality parameter values in the inner and outer regions of $T_{\rm inner} \sim T_{\rm outer} \sim 0.02$. 

\begin{figure}
\begin{center}
\begin{tabular}{c}
\hspace{-5mm}\includegraphics[width=1\linewidth]{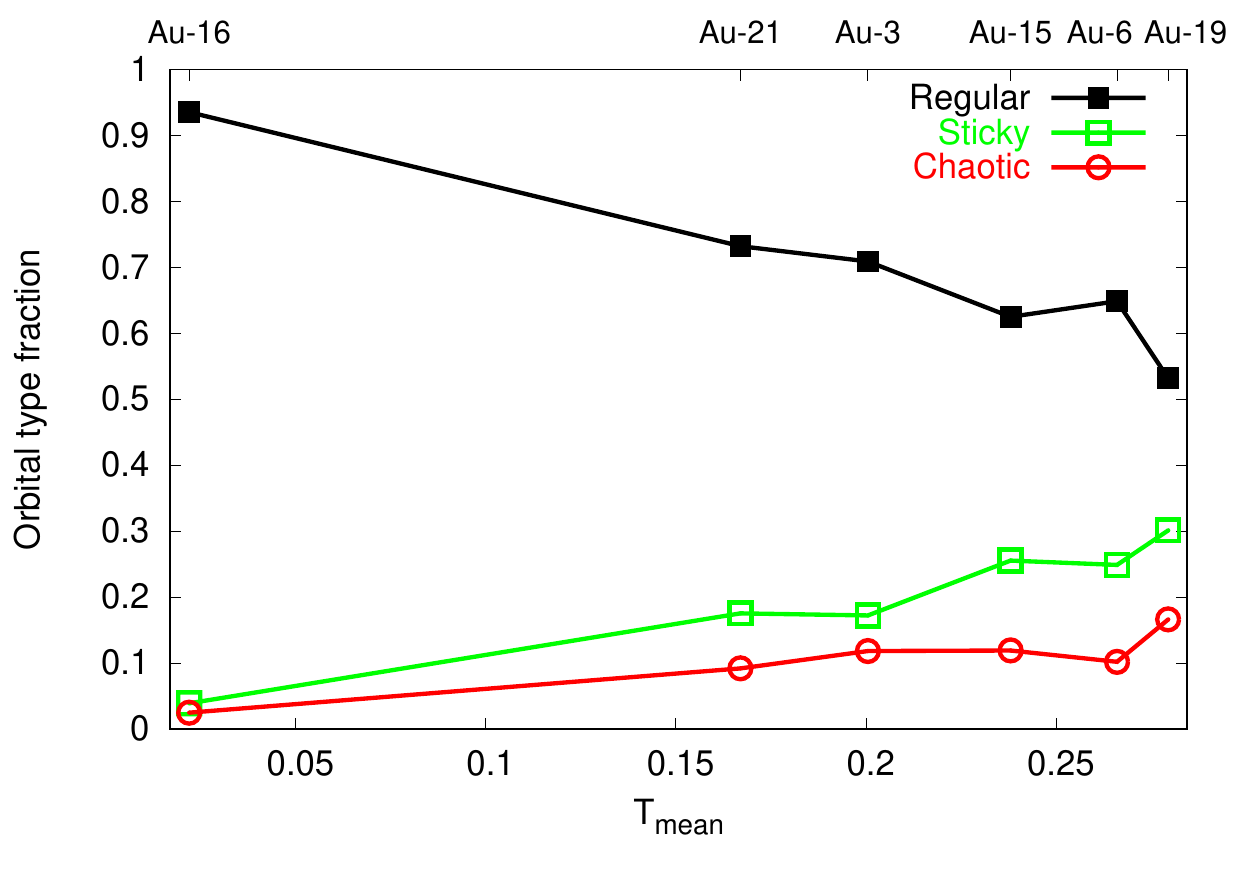}\\
\end{tabular}
\end{center}
\caption{Orbital type distribution as a function of an average triaxiality parameter ($T_{\rm mean}$). The fraction of stellar particles on chaotic orbits is always smaller than $17$~per cent.}
\label{fig:tri_frac}
\end{figure}

As expected, the fraction of regular orbits in each potential shows a dependence on the degree of asymmetry of the corresponding DM halo. This is more clearly seen in Fig.~\ref{fig:tri_frac}, where we show the fraction of regular (black solid squares), sticky (green open squares) and chaotic (red open dots) orbits in each {\it Auriga} model, as a function of an average triaxiality parameter, $T_{\rm mean} = (T_{\rm inner} + T_{\rm outer}) / 2$. Notice how the fraction of regular orbits
steadily decreases as $T_{\rm mean}$ increases. This highlights that once the bulge-disc pair is taken into account, for values of $D/T \gtrsim 0.3$, the dominant factor determining the fraction of regular orbits is the overall triaxiality of the underlying DM halo. \emph{None the less, in all cases, and independently of the properties of the analysed potential models, we find that the fraction of orbits exhibiting a chaotic behaviour within a Hubble time is smaller than $17$~per cent. }    

\section{Global dynamics and diffusion}
\label{sec:diffusion}

As shown in the previous section, a small but non-negligible fraction of stellar halo particles in solar neighbourhood-like volumes could indeed exhibit chaotic mixing. This fraction is comparable to the one found in \citetalias{2015MNRAS.453.2830M}, but now taking into account the contribution from both the DM and baryonic components to the overall galactic potentials. In what follows we will discuss the extent to which such mixing can erase  kinematic signatures of early accretion events within galactic regions such as the solar neighbourhood and physically relevant periods of time.
In this direction, we give a theoretical framework and approximate the potential model as a near-integrable one, i.e. a fully integrable
one plus a \textquotedblleft small'' perturbation in order to gain some insight on the expected result. Thus, by chaotic diffusion, roughly speaking, we mean the time variation of the 
prime integrals of the integrable potential when it is acted upon by a small non-integrable perturbation.

\subsection{Approximating the galactic potential within solar neighbourhood-like regions}
The potential $\Phi_{\mathrm{DM2}}$ given in Eq.~\eqref{eq:nfw_tri} in terms of the \textquotedblleft bi-triaxial'' radius defined in Eq.~\eqref{rpp} admits, for $r'_p<r_s$
(both quantities introduced in Section~\ref{subsubsec:dmcomp}),
the power series expansion:
\begin{equation}
\Phi_{\mathrm{DM2}}=-\frac{A}{r'_s}\sum_{n=1}^{\infty}\frac{{(-1)}^{n+1}}{n}
\left(\frac{r'_p}{r_s}\right)^{n-1}\ ,
\label{expansion}
\end{equation}
so it is analytic everywhere, and the condition $r'_p<r_s$ implies that $r < r_s$, which
holds for local volumes around the Sun.

Under the above assumption, the ratio $r'_p/r_s$ could be expanded as a power series and,  retaining terms up to $r_i^2/r_s^2$ and $r_i\,r_o/r_s^2$, it can be written as: 
\begin{equation}
\frac{r'_p}{r_s}\approx\frac{r_i}{r_s} \left(1+\frac{r_o-r_i}{r_s}\right)\ .
\end{equation}
Again, the relationship between these radii follows from  their definitions given in Section~\ref{subsubsec:dmcomp}.

Taking into account the values of the ratios of the principal axes describing the ellipsoidal inner and outer regions of the DM halo (MW model C4 in Table~\ref{table:t2}), we introduce the small parameters:
\begin{align}
\varepsilon_1 = \frac{a^2}{b^2}-1, &\quad \varepsilon'_1 = \frac{a'^2}{b'^2}-1\ ,\label{epsilon}\\ 
\varepsilon_2 = \frac{a^2}{c^2}-1, &\quad \varepsilon'_2 = \frac{a'^2}{c'^2}-1\ .\nonumber
\end{align}

Recalling that $a'\approx a\approx 1$ (1.07 and 1.02 respectively), introducing spherical coordinates $(r,\vartheta,\varphi)$,
retaining terms up to $r'_p/r_s$ in Eq.~\eqref{expansion} and neglecting those of second order in the parameters defined in Eq.~\eqref{epsilon}, the quadrupolar approximation to Eq.~\eqref{expansion} takes the form:
\begin{multline}
\Phi_{\mathrm{DM2}}\approx \phi_{\mathrm{DM2}}^0(r) + V(r)\,\{\alpha_1(r)\cos2\vartheta+\alpha_2(r)\,[\,\cos2\varphi-\\
0.5 \cos2(\vartheta+\varphi)- 0.5 \cos2(\vartheta-\varphi)\,]\}\ ,
\end{multline}
where 
\begin{eqnarray}
\Phi_{\rm DM2}^0(r)&=&\frac{Ar}{2r^2_s}\left[1+\frac{\varepsilon_1}{8}+\frac{\varepsilon_2}{4}+
\frac{r}{4r_s}\left(\frac{\delta\varepsilon_1}{2}+\delta\varepsilon_2\right)\right],\\ V(r)&=&\frac{Ar}{8r_s^2}\ ,\\
\alpha_1(r)&=&\varepsilon_2-\frac{\varepsilon_1}{2}+\frac{r}{r_s}\left(\delta\varepsilon_2-\frac{\delta\varepsilon_1}{2}\right)\ ,\\
\alpha_2(r)&=&-\frac{1}{2}\left(\varepsilon_1+\frac{r}{r_s}\delta\varepsilon_1\right)\ ,
\end{eqnarray}
with $\delta\varepsilon_i=\varepsilon'_i-\varepsilon_i$, the amplitudes $\alpha_s$ are then assumed to be small.

As far as our analysis concerned, the contribution 
of the disc component, Eq.~\eqref{eq:discMN},
to the overall potential is essentially spherical due to the smallness of the $z$-values around the position of the Sun. Then, neglecting terms of $\mathcal{O}(z^2/r^2)$, it can be well approximated by the expression:
\begin{equation}
\Phi_{\rm disc}^0(r)=-\frac{D}{\sqrt{r^2+\left(\epsilon^{\rm s}_{\rm disc}+\epsilon^{\rm h}_{\rm disc}\right)^2}}\ .
\end{equation}
As we have already mentioned (Section~\ref{subsec:meth-pot}), the exponential disc is approximated with the superposition of three different MN models, so the above approximation still holds in case of an exponential profile.

Summarizing, the central part of the Galactic potential can be written as:
\begin{equation}
\Phi_{\rm MW}^0=\Phi_{\rm nuc}+\Phi_{\rm bul}+\Phi_{\rm disc}^0+\Phi_{\rm DM2}^0\ ,
\end{equation}
which yields the following expression for the total (approximated) potential:
\begin{multline}
\Phi_{\rm MW}(r,\vartheta, \varphi) \approx \Phi_{\rm MW}^0(r)+ V(r)\{\alpha_1(r)\cos2\vartheta+\alpha_2(r)\,[\cos2\varphi-\\
0.5 \cos2(\vartheta+\varphi)-0.5 \cos2(\vartheta+\varphi)]\}\ .
\label{phi_app}
\end{multline}

In Eq.~\eqref{phi_app} the angular dependence of the potential, at this order, only comes from the DM halo.

In \citetalias{2015MNRAS.453.2830M} we have already provided a theoretical background for chaotic diffusion, 
so here we restrict the discussion to this model. The Hamiltonian associated to Eq.~\eqref{phi_app} can be written as:
\begin{equation}
\mathcal{H}(\mathbf{p},\mathbf{r})=\mathcal{H}_0(\mathbf{p},r,\vartheta)+\hat{\Phi}_1(\mathbf{r})\ ,
\end{equation}
with
\begin{equation}
\mathcal{H}_0(\mathbf{p},r,\theta)=\frac{p_r^2}{2}+\frac{p_{\vartheta}^2}{2r^2}+\frac{p_{\varphi}^2}{2r^2\sin^2\vartheta}+\Phi_{\rm MW}^0(r)\ ,
\end{equation}
and
\begin{multline}
\hat{\Phi}_1(\mathbf{r}) = V(r)\{\alpha_1(r)\cos2\vartheta+\alpha_2(r)\,[\cos2\varphi-\\
0.5\cos2(\vartheta+\varphi)-0.5 \cos2(\vartheta+\varphi)]\}\ ,
\end{multline}
where
\begin{equation}
p_r=\dot{r},\qquad p_{\vartheta}=r^2\dot{\vartheta}\ ,\qquad
p_{\varphi}=r^2\dot{\varphi}\sin^2\vartheta\ .
\end{equation}
Therefore, $\mathcal{H}_0$ is an integrable Hamiltonian, since:
\begin{equation}
\mathcal{H}_0=\mathrm{E}_0\ ,\qquad
L_z= p_{\varphi}\ ,\qquad
L^2=p_{\vartheta}^2+p_{\varphi}^2\csc^2\vartheta\ ,
\end{equation}
are the three global integrals, while $\hat{\Phi}_1$ can be considered as a {small} perturbation. The terms in $\hat{\Phi}_1$ that depend on $(\vartheta, \varphi)$ break the spherical symmetry leading to variations in the modulus of the total angular momentum and its $z$-component:
\begin{eqnarray}
\frac{dL^2}{dt}&=&[L^2,\mathcal{H}]=
-2p_{\vartheta}\frac{\partial     \hat{\Phi}_1}{\partial\vartheta}-\frac{2p_{\varphi}}{\sin^2\vartheta}\frac{\partial \hat{\Phi}_1}{\partial\varphi}\ ,\\
\frac{dL_{z}}{dt}&=&[L_z,\mathcal{H}]=
-\frac{\partial\hat{\Phi}_1}{\partial\varphi}\ ,
\end{eqnarray}
which are of order $\alpha_s$. These small variations of $|L|$ and $L_z$ within chaotic
domains would lead to chaotic diffusion. For instance, in those regions where stickiness is strong, the
changes $\Delta |L|$, $\Delta L_z$ over a given timespan $T$ would be very small and thus,  though the dynamics is chaotic, almost stability can be assumed for time-scales $\tau\sim T$. Instead, 
in other domains of phase space, large values of $\Delta |L|$ and $\Delta L_z$ could
be observed over the same timespan,  diffusion becoming significantly faster so that  chaotic mixing 
would be efficient over $\tau\la T$. In other words, chaos is a necessary but not sufficient condition for diffusion. Therefore, diffusion experiments would be required to obtain  reliable information about the stable/unstable 
character of the motion within chaotic domains, as shown, for instance, in \citetalias{2015MNRAS.453.2830M} and in \cite{2016MNRAS.460.1094M}, for the case of planetary dynamics. Notice that the approximation to the galactic potential given by Eq.~\eqref{phi_app} is derived and used just for the  theoretical discussion regarding chaotic diffusion delivered in the next section.

\subsection{The actual relevance of chaotic diffusion}
\label{subsec2:diffusion}
\begin{figure}
\begin{center}
\hspace{-5mm}\includegraphics[width=1\linewidth]{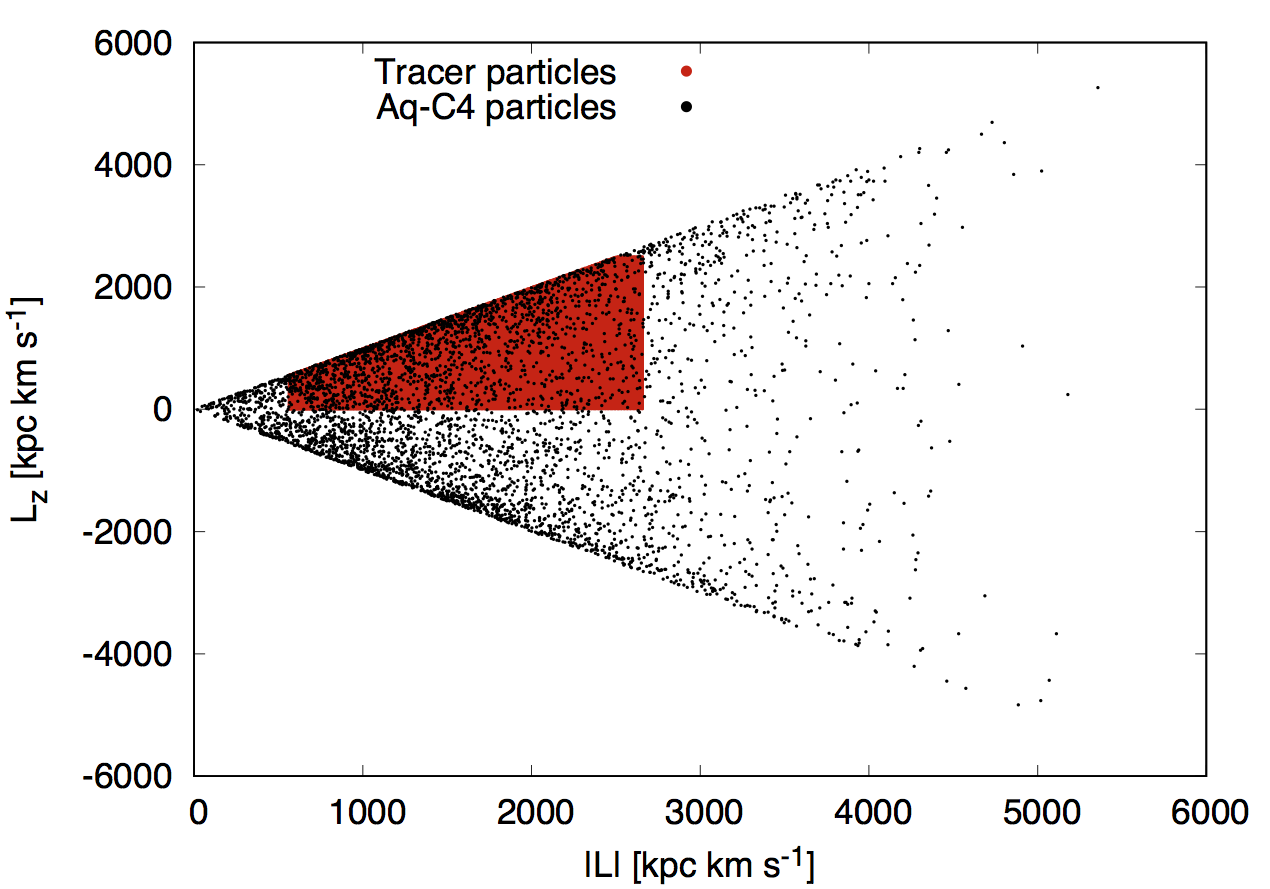}
\end{center}
\caption{Ranges in $|L|$ and $L_z$ for the whole set of $1171$ stellar particles (black dots) of the Aq-C4 simulation. In red, we highlight the areas of the plane considered in the experiments.}
\label{fig:angular-momentum}
\end{figure}

In this direction, we accomplish a global picture of the dynamics in the angular momentum space by computing a dynamical indicator for a large set of  initial conditions around the position vector of the Sun $\mathbf{x}_{\odot}$ and on a given energy surface. For this purpose, and adopting $(x_\odot,y_\odot,z_\odot)=(8,0,0)$~kpc, we take the mean value of the energy distribution of the $1171$ stellar particles located within a $15^{\circ}$ wedge, oriented along the disc semi-major axis and with galactocentric distances between 5.5 and 10.5~kpc (second bin, centered at 8~kpc in Fig.~\ref{fig:adeterminar}) in the Aq-C4 simulation (see also Section~\ref{subsec:meth-init} for further details on the initial distribution of the conditions), namely $\langle \mathrm{E}\rangle = h\simeq -164803$ km$^2$ s$^{-2}$ (such energy surface is computed with our analytic and static representation of the galactic potential). Then we sample a domain in the $(|L|,L_z)$ plane with a grid chosen such that, in both dimensions, nearly $80$~per cent of the corresponding stellar particles are encompassed (notice that actually, for the sake of symmetry, only the  $\approx 40$~per cent in the $L_z$-direction needs to be accounted for). Fig.~\ref{fig:angular-momentum} displays in red the region on the $(|L|,L_z)$ plane considered for the current dynamical study, and also shows, as black dots, the values corresponding to the Aq-C4 stellar particles.

\begin{figure*}
\begin{center}
\begin{tabular}{cc}
\hspace{-5mm}\includegraphics[width=0.5\linewidth]{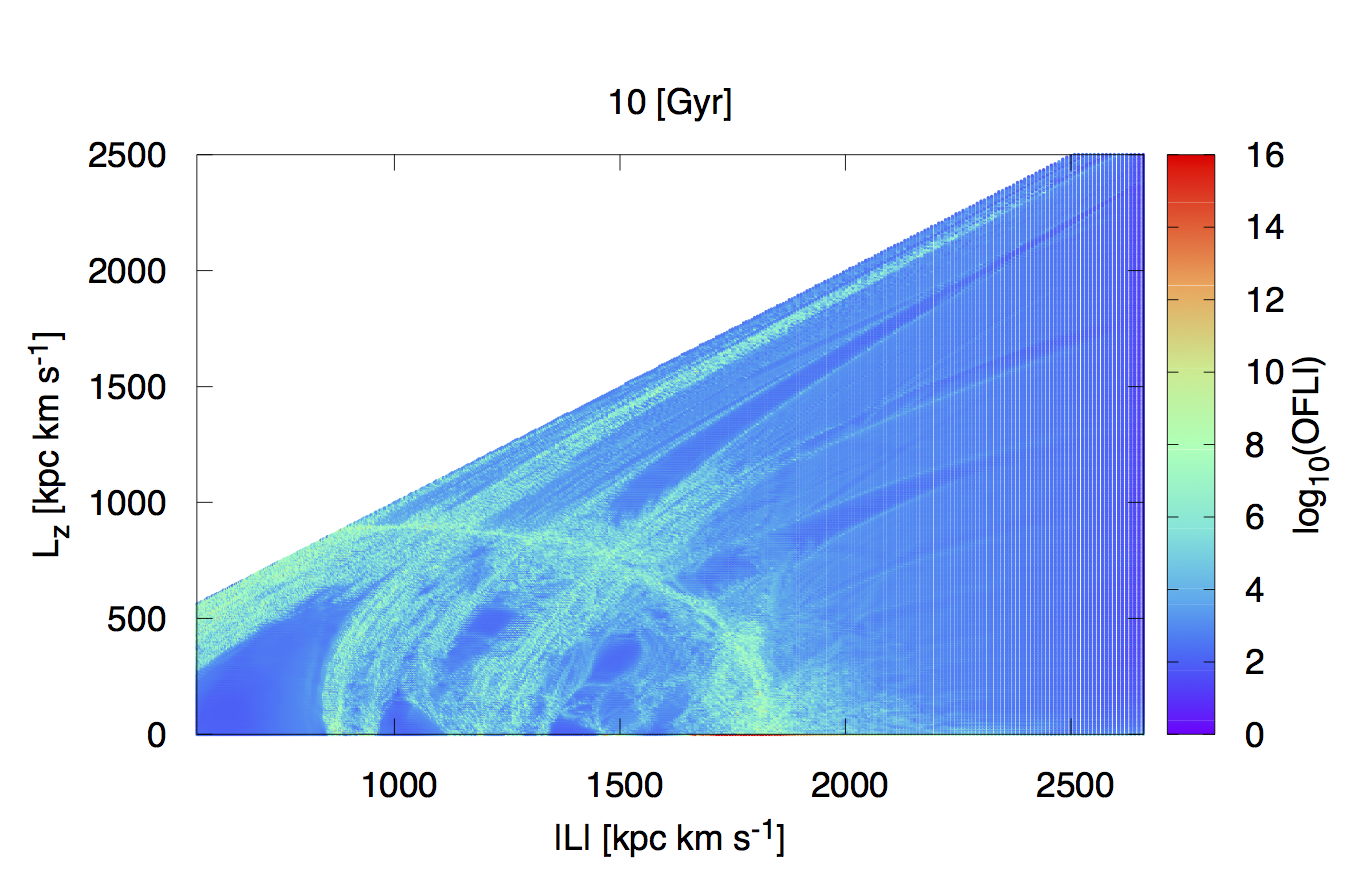}&
\includegraphics[width=0.5\linewidth]{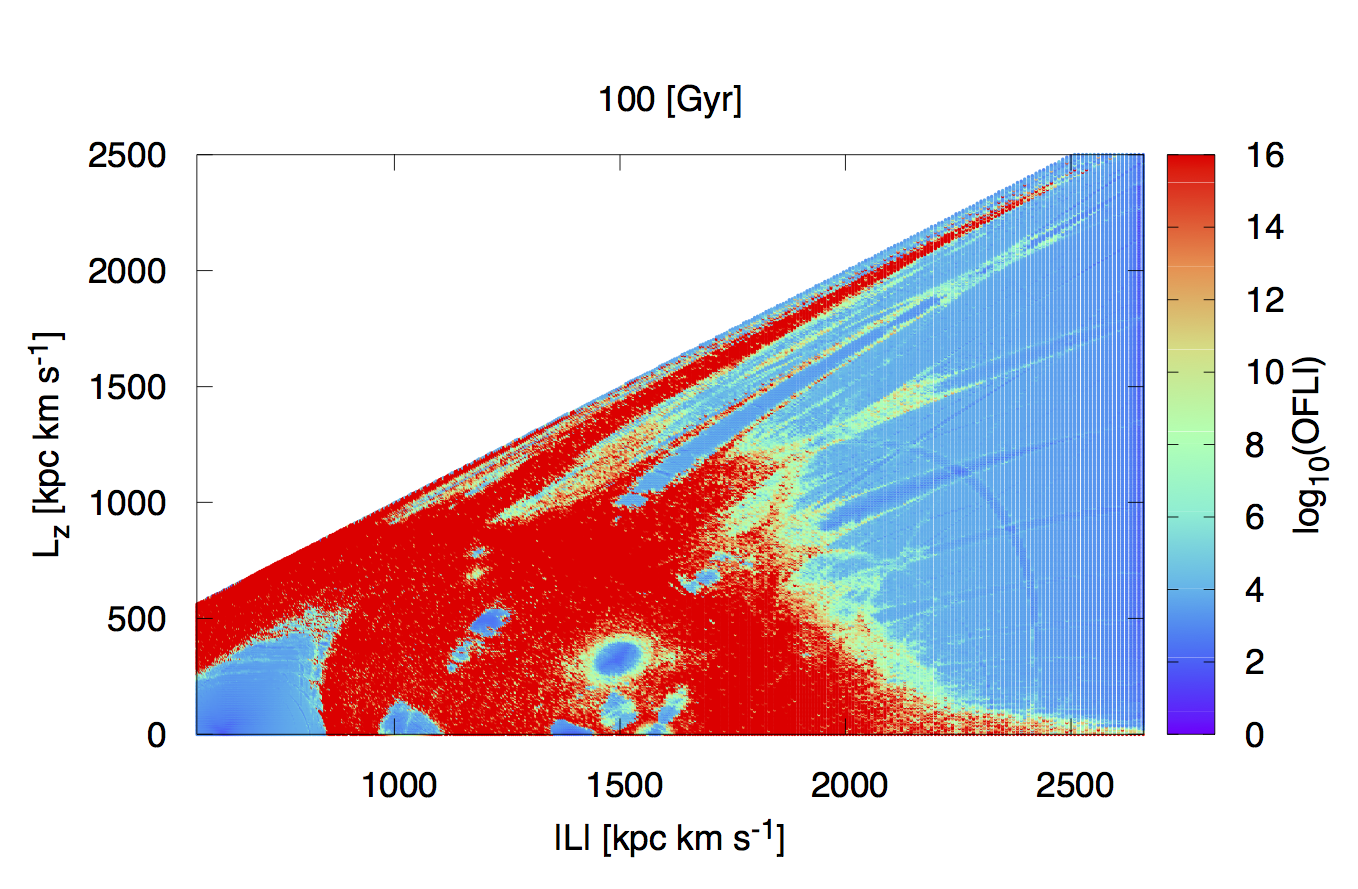}
\end{tabular}
\end{center}
\caption{OFLI maps for $10$ (left panel) and $100$ (right panel) Gyr for the C4 MW model for $(x_\odot, y_\odot,z_\odot)=(8,0,0)$ and $\hat{h}\simeq -164803$ km$^2$ s$^{-2}$. The solid black lines in the colour bars show the threshold used to distinguish regular from chaotic motion. Warm colours indicate chaotic motion while cool colours represent regular motion. The Arnold web is mostly unveiled and, as we can see from the warm colours on the right panel, it covers a considerable domain in phase space.}
\label{fig:404}
\end{figure*}

The left panel of Fig.~\ref{fig:404} displays an OFLI  map 
for a grid of $121224$ equidistant initial conditions in the plane $(\log(L^2),L_z)$ which   
have been integrated over a timespan of $10$~Gyr in order to obtain the concomitant value of the chaos indicator. The solid black line in the colour bar  shows the threshold used to distinguish regular from chaotic motion. In general, cool colours  represent regular motion whereas warmer ones indicate chaotic motion.

The resulting global dynamical portrait for this time-scale reveals the prevailing regular character of the motion in the angular momentum space. In fact, just a few invariant manifolds and narrow resonances are detected. Let us recall that the most relevant invariant manifolds separate different orbital families, the large resonance domains. Indeed, the region with smaller values
of $|L|$ and $L_z$ corresponds to the box family while the tube orbital family has $|L|\gtrsim 1750\,\mathrm{kpc\,km\,s^{-1}}, L_z\gtrsim 1000\, \mathrm{kpc\,km\,s^{-1}}$. The light blue
arc arising from $|L|\approx 1750\,\mathrm{kpc\,km\,s^{-1}},\, L_z = 0$ corresponds to the separatrix (actually the chaotic layer), that separates both orbital
families. Meanwhile, rather small high-order resonances show up as thin channels all over the angular momentum space. This
web of resonances is known as the Arnold web.

In sum, notice should be taken that the phase space is mainly covered by regular orbits for the considered timespan. Consequently, chaos is almost irrelevant after an evolution of 10~Gyr, even though the perturbation is not  negligible for the C4 MW model, being $\alpha_1\sim 0.1,\,\alpha_2\sim -0.03$ for local solar neighbourhood-like volumes. In this case, as already mentioned, the fraction of phase space corresponding to chaotic motion is small but non-negligible and amounts to $\simeq 25.57$~per cent. 

In order to detect any diffusive phenomena or chaotic mixing in the present model, a larger time-scale should be considered. Therefore, though without direct physical significance, we obtained the OFLI map corresponding to 100~Gyr, which is displayed in the right panel of Fig.~\ref{fig:404}. Such a map discloses chaotic motion that still appeared as regular at 10~Gyr, mainly due to stickiness. Indeed, the map reveals that the thin chaotic layer separating box and tube
families already discussed, now at 100~Gyr appears wider with large OFLI values, depicted in red. Moreover
other resonances also show up as highly chaotic and the Arnold web is seen to occupy a considerable region in phase space which amounts to almost $60$~per cent of the integrated orbits. The presence of a connected chaotic region of noticeable size would forecast a secular variation of the unperturbed integrals $(|L|, L_z)$, which would lead to the uprising of fast diffusion. 

Let us say that we are also interested in determining the time-scale for chaotic diffusion to take place. Therefore, following a similar approach to that presented in \citetalias{2015MNRAS.453.2830M}, we investigate diffusion over the ($|L|,\, L_z$) plane, 
for a given energy surface, $\hat{h}$, within a small sphere in configuration space, $|\mathbf{x}-\mathbf{x}_{\odot}|<\delta$. 
In this way we reduce the motion to an almost two dimensional section defined by:
\begin{equation}
\mathcal{S}=\left\{(|L|,\, L_z)\,:\, |\mathbf{x}-\mathbf{x}_{\odot}|<\delta,\,\, \mathcal{H} = \hat{h}\right\}\ .
\end{equation}

For our diffusion studies, we take  ensembles of $N_p=10^6$ tracer particles sampled uniformly in boxes of size $\sim10^{-6}$ in both $|L|$ and $L_z$. The centres of these boxes, whose  highly chaotic nature has been revealed by the OFLI indicator are listed in Table~\ref{table:ens}. We integrate the equations of motion for each initial condition taking into account the full potential given by Eq.~\eqref{eq:aqc4a}
over some timespan $T$, and every time the orbits of the ensemble intersect $\mathcal{S}$, we retain the corresponding values of $|L|(t)$ and 
 $\, L_z(t)$. For ensembles located in stable regions both unperturbed integrals slightly vary, being  $|\Delta |L|(t)|,\, |\Delta L_z(t)|\ll 1$, so that practically no evolution in the angular momentum plane should be  observed. In fact, the small variations in  $|L|,\, L_z$ arise as a consequence of the system being no longer spherical. For ensembles immersed  in chaotic domains instead, if no barriers to diffusion are present, both unperturbed integrals are expected to change with time and the trajectory over $\mathcal{S}$ would provide an indication of actual diffusion.
 
 \begin{table}
\centering
\caption{Ensembles of $10^6$ initial conditions sampled uniformly in boxes of size $\sim10^{-6}$ in both $|L|$ and $L_z$, whose  centres, given in the table, correspond to chaotic orbits. The units in $L$ and $L_z$ are $\mathrm{kpc\,km\,s^{-1}}$.}
\label{table:ens}
\begin{tabular}{@{}cll} \hline Ensemble & $|L|$ & $L_z$\\\hline 
(i) &  $562$  & 500  \\
(ii) & $1122$ & 900 \\
(iii) & $1820$ & 50 \\
(iv) & $861$ & 50 \\
(v) & $1413$ & 700 \\
\hline
\end{tabular}
\end{table}

Further, let us remark that the number of intersections of a given trajectory with  $\mathcal{S}$ strongly depends on the stability of the motion. In fact, in the case of stable regular motion, 
since the orbit lies in a three dimensional torus,
$\mathcal{S}$ is a slice of it and thus many crossings would occur. On the other hand, in the case of an unstable chaotic orbit, no tori structure exists and thus only a few intersections with  $\mathcal{S}$ are expected. So much that in the considered ensembles, which correspond to highly chaotic motion, no crossings are observed during the first $20$~Gyr. 

Moreover, taking into account the sticky character of most orbits, long  timespans should be considered. Indeed, such stickiness could vary for slightly different values of the model parameters, thus leading to rather different results. Therefore, in order to overcome the possible effect of sticky phenomena,  long-term diffusion experiments have been carried out, which are described straight away. 

\begin{figure*}
\begin{center}
\begin{tabular}{cc}
\hspace{-5mm}\includegraphics[width=0.5\linewidth]{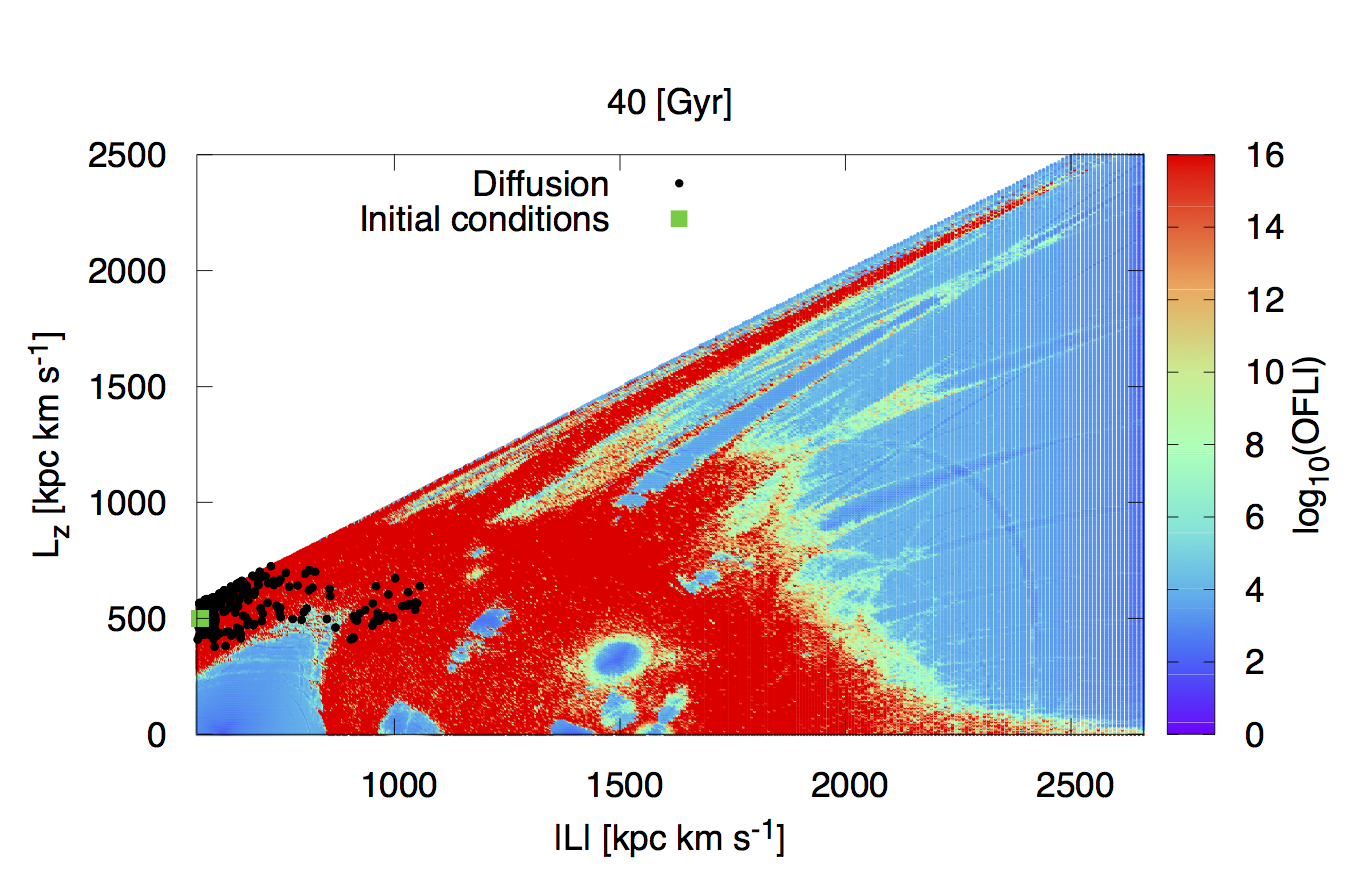}&
\includegraphics[width=0.5\linewidth]{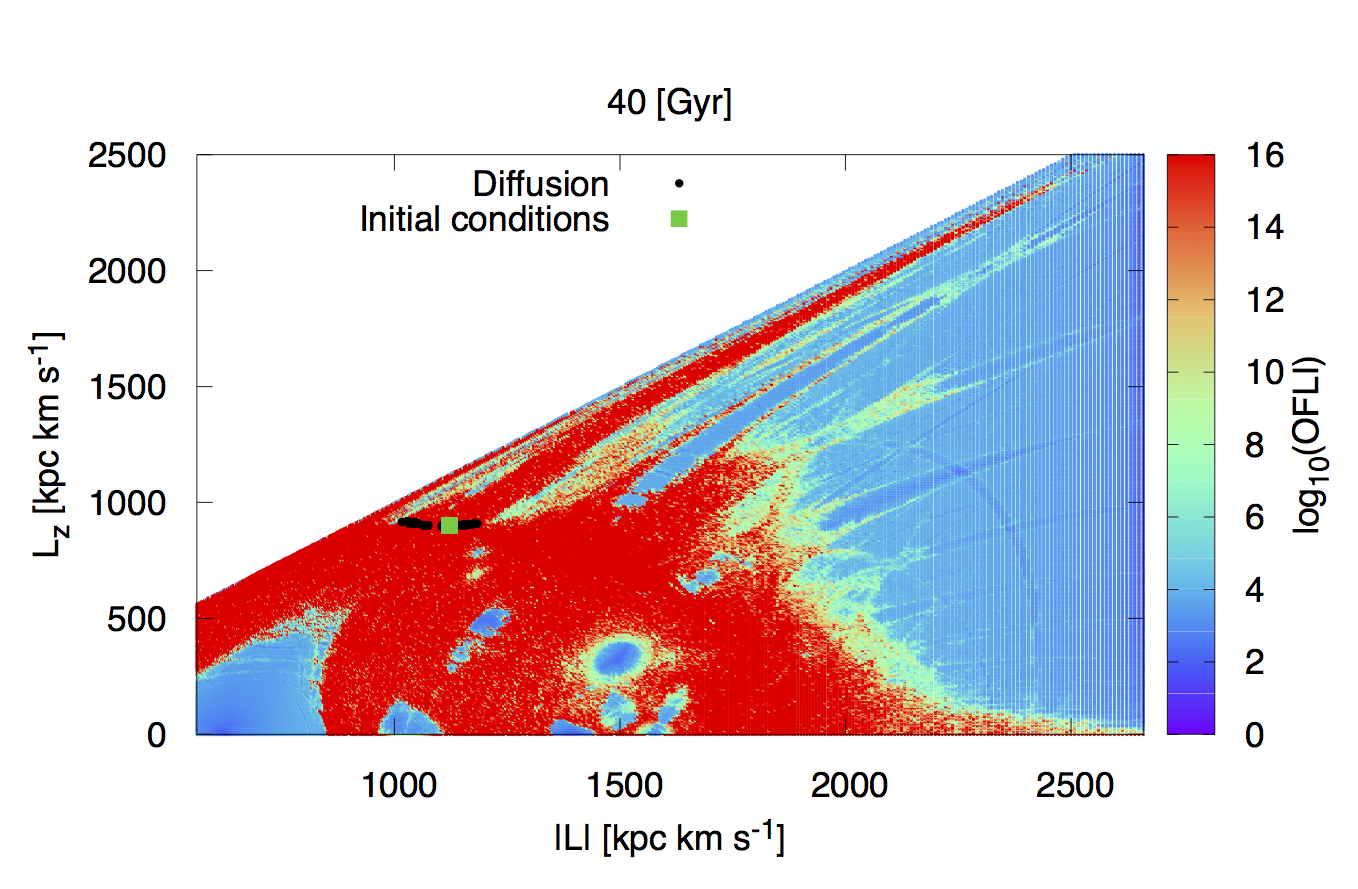}\\
\hspace{-5mm}\includegraphics[width=0.5\linewidth]{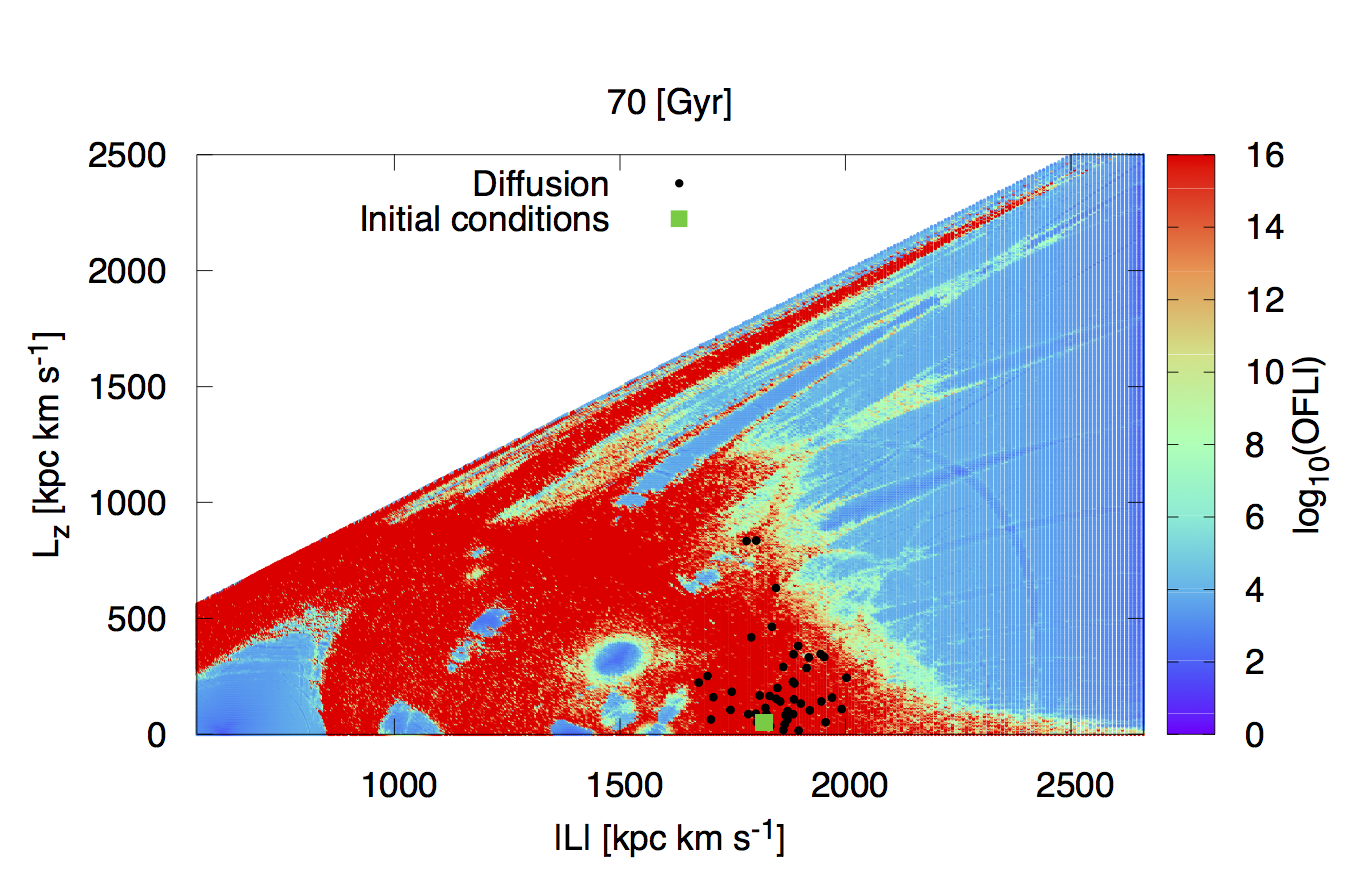}&
\includegraphics[width=0.5\linewidth]{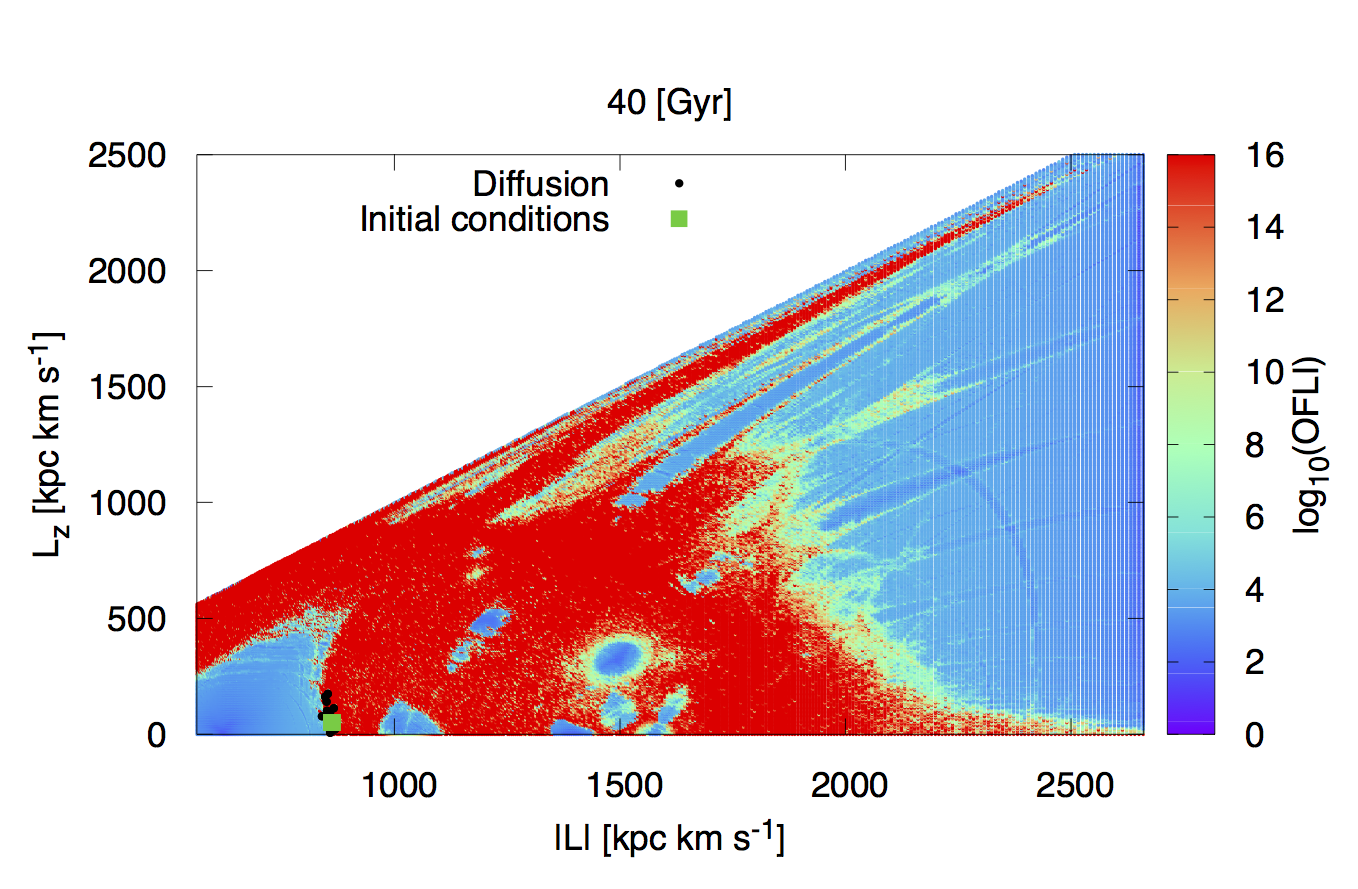}\\
\hspace{-5mm}\includegraphics[width=0.5\linewidth]{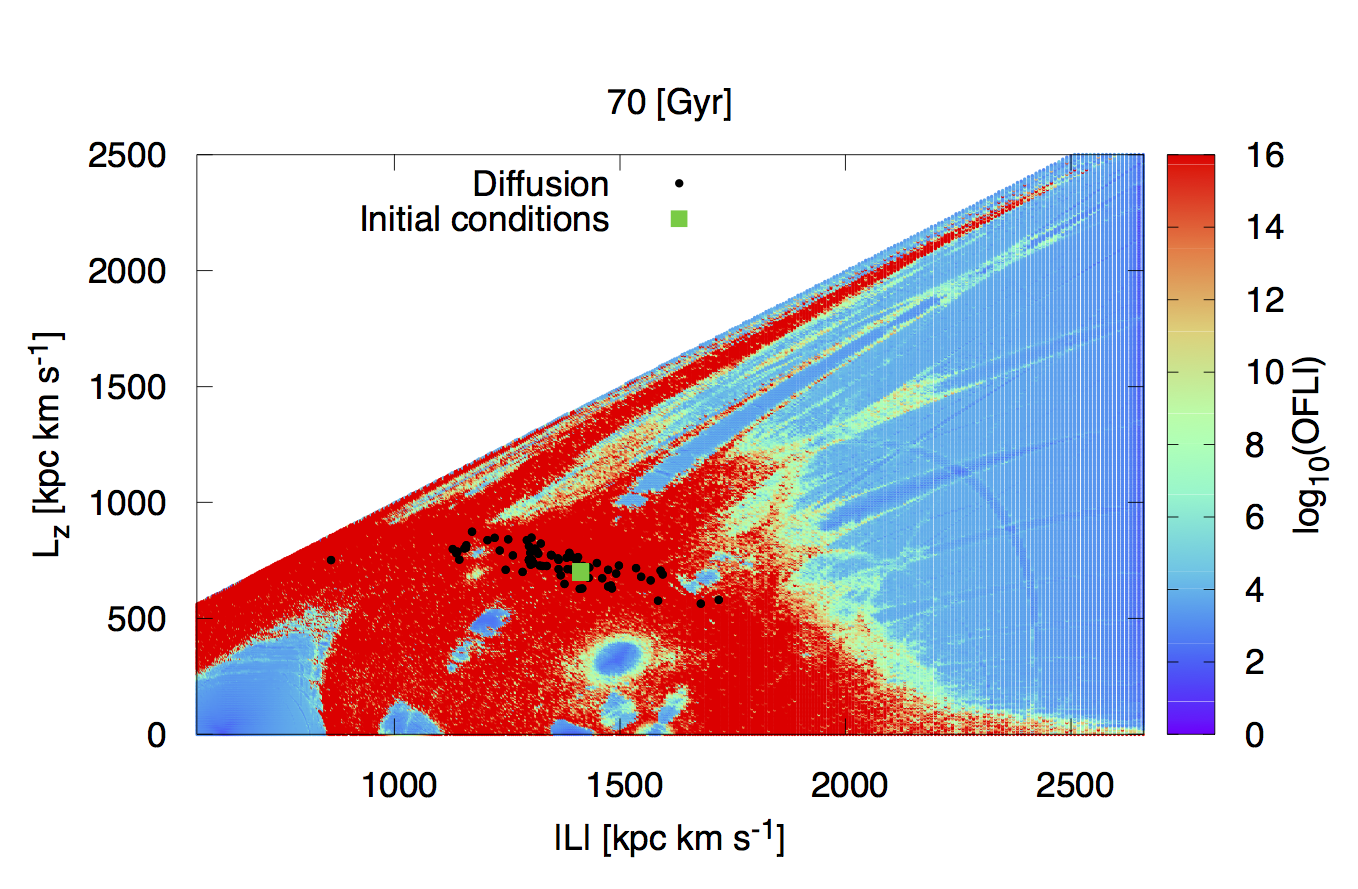}&
\end{tabular}
\end{center}
\caption{Top left panel: diffusion over $40$~Gyr for ensemble (i) of initial conditions (depicted in green) superimposed on the Arnold web. Top right panel: diffusion over $40$~Gyr  for ensemble (ii) of initial conditions (also depicted in green) superimposed on the Arnold web. The unperturbed integrals remain confined to a rather small domain, revealing that  diffusion turns out to be mostly inefficient. Middle left panel: long-term diffusion over $70$~Gyr for ensemble (iii) of initial conditions (in green) overplotted on the Arnold web. Middle right panel: drift of the unperturbed integrals over $40$~Gyr for ensemble (iv) of initial conditions (in green) superimposed on the Arnold web. Bottom panel: long-term diffusion over $70$~Gyr for ensemble (v) of initial conditions (in green) overplotted on the Arnold web.}
\label{fig:reg1_40}
\end{figure*}

The top left panel of Fig.~\ref{fig:reg1_40} shows how ensemble (i) evolves with time in action space. Diffusion is seen to proceed along the stochastic layer separating box from tube orbits. Let us point out the geometrical resemblance of the observed diffusion with the one that would be expected from the Arnold's theoretical conjecture, which forecasts that diffusion would proceed through phase space along the chaotic layers of the full resonance web. However, and since the perturbation is not sufficiently small, the detected diffusion should be interpreted as a consequence of the resonances' overlap. Even though fast diffusion could take place in such a scenario, this event does not occur at all, as follows from our numerical experiments.

To stress this fact we turn to the time evolution of ensemble (ii) shown in the top right panel of the same figure during $40$~Gyr. Therein, we recognize that the unperturbed integrals remain confined to a very small domain, even for a rather large timespan, diffusion neatly spreading over the unperturbed separatrix discriminating box from tube orbits. This still applies  when far larger time-scales are considered.

The wandering onto the resonance web of the unperturbed actions, $|L|$ and $L_z$ corresponding to the third ensemble is displayed in the middle left panel of Fig.~\ref{fig:reg1_40}. In this particular case, only three intersections with  $\mathcal{S}$ are observed up to $40$~Gyr so that a larger  timespan needs to be covered. Indeed, after $70$~Gyr we notice that diffusion advances along the outermost edge of the stochastic layer, near the bottom of the figure, and climbs up over the entangled assemblage of stable/unstable manifolds associated to different high order resonances.

For ensemble (iv) instead, already at $40$~Gyr some variation of the unperturbed integrals is seen to occur, as the middle right panel of Fig.~\ref{fig:reg1_40} displays. For an even larger time-scale, say $70$~Gyr, the ramble in action space breaks through the innermost region of the resonance interweave. 

The bottom panel of Fig.~\ref{fig:reg1_40} shows how diffusion proceeds for the ensemble (v). The successive intersections of the trajectories with the section $\mathcal{S}$ in action space adroitly diffuse along the layer discriminating box from tube orbits,  also after a rather long time-scale since up to $\sim 40$~Gyr no crossings take place.

We should note 
that the considered ensembles, except for the one denoted by (iv), were picked up very close to the main unstable region, that is the chaotic layer that separates box from tube families. 
From the above results, it turns out that the largest variation of the integrals corresponds to ensemble (v),
being $\Delta |L|\lesssim 800\, \mathrm{kpc\,km\,s^{-1}}$ over $\tau = 70$~Gyr so that a mean rate of variation could be estimated as
$\Delta |{L}|/\tau \lesssim 11.5 \,\mathrm{kpc\,km\,Gyr^{-1}\,s^{-1}}$, which is actually rather small \citep[for instance, on the left panel of Fig.~7 in][a resolved stream is shown with a typical extension of more than $500\, \mathrm{kpc\,km\,s^{-1}}$ in $L_z$]{2013MNRAS.436.3602G}.

\section{Discussion and conclusions}
\label{sec:discussion}
Stellar streams are the living records of galactic accretion events. Therefore, their identification as kinematically cold substructures is of key importance for galactic archaeology. Much effort has been devoted to locate such fossil signatures in the outer stellar halo, where typical dynamical time-scales are long enough to preserve this structure in a spatially coherent fashion. Several streams have indeed been identified and studied in great detail within these regions \citep{1994Natur.370..194I,2001ApJ...547L.133I,2001ApJ...551..294I,2001ApJ...548L.165O,2003MNRAS.340L..21I,2003ApJ...599.1082M,2006ApJ...637L..29B,2006ApJ...642L.137B,2007ApJ...658..337B,2007ApJ...668L.123M}\footnote{For a recent and very complete list of stellar streams in the Galactic halo we refer the reader to \citet[table 4.1]{2016ASSL..420...87G}.}. On the other hand, in the inner stellar halo, and particularly around the solar neighbourhood \citep[where information about the most ancient accretion events is expected to be stored,][]{2000MNRAS.319..657H,2008ApJ...689..936J,2010MNRAS.408..935G}, identifications of stellar streams are far less numerous, even though
theoretical models predict hundreds of them  \citep{1999MNRAS.307..495H,1999Natur.402...53H,2003MNRAS.339..834H,2006MNRAS.365.1309H}. Furthermore, the extragalactic origin of some of these substructures has not been proved conclusively yet \citep[see][for a full discussion and references therein]{2016ASSL..420..113S}. The ammount of substructure present in the solar neighbourhood's phase space distribution is subject to several factors. It has been often argued that the 
low identification rate may be mainly due to the lack of an accurate and large enough full phase space stellar catalog. Within 2018 {\it Gaia}--DR2 will be released and a robust quantification of substructure within the extended solar neighbourhood will become feasible for the first time. 

In addition to the previous astrometric limitations, another relevant factor playing a role for the quantification of stellar streams are the active sources of chaos, which can trigger chaotic mixing within relevant time-scales. As discussed in \citetalias{2015MNRAS.453.2830M}, thanks to the asymmetric nature of the underlying gravitational potential, a fraction of the local stellar streams is expected to be evolving on chaotic orbits. Chaos, in the Lyapunov sense, indicates exponential divergence of initially nearby orbits in phase space. Dynamical time-scales in the inner regions of the Galaxy are  
relatively short. Thus, a group of initially close-by stars in phase space, evolving on chaotic orbits, would experience a 
very rapid mixing. More importantly, regions filled with chaotic orbits can foster  chaotic diffusion, which effectively erases the `dynamical memory' imprinted in all phase space and results in a 
smooth distribution function. The detection of stellar streams  could be seriously threatened if such chaotic orbit
are indeed very common \citep{2003ApJ...592L..63G,2015ApJ...799...28P,2016MNRAS.460..497H,2016MNRAS.455.1079P,2016ApJ...824..104P,2017MNRAS.470...60E}.

In \citetalias{2015MNRAS.453.2830M} we explored whether chaos could indeed be playing a significant role in eroding substructure in the solar neighborhood phase space distribution. The experiments carried out in that work
strongly suggested that this would not be the case. Only a very small fraction of the orbits within solar neighbourhood-like volumes exhibit chaotic behaviour within a Hubble time. Diffusion did not have enough time, even in those cases. However, 
that study was based on dark matter only simulations,
which completely neglected the role
of the baryonic component.

In this second paper we re-examined the problem using a significantly more realistic set up to model the Galactic stellar halo.
We used a suite of seven state of the art fully cosmological hydrodynamic zoom-in simulations of the formation of Milky Way-like galaxies (Section~\ref{subsec:meth-sim}), to extract values of the parameters that describe our analytic potential models and to sample realistic phase space distributions of different volumes. The Galactic potential was  modelled  with a new analytic and static representation: a multicomponent model that accounts for the effect of both the baryonic and dark matter components (Section~\ref{subsec:meth-pot}). We integrated the equations of motion, coupled with the first variational equations, using the different sets of cosmologically motivated initial conditions (Section~\ref{subsec:meth-init}) and computed, for each orbit, the Orthogonal Fast Lyapunov Indicator, OFLI (Section~\ref{subsec:meth-ind-OFLI}). This chaos indicator allowed us to robustly classify the orbits of our stellar and dark matter particles into three different components: regular, sticky and chaotic (Section~\ref{orbitalclassification}). Their distinction is of pivotal importance due to the fact that the time evolution of the rate at which the local (stream) density around such a given particle decreases is completely different. In case of regular orbits, they are associated with a rate that follows a power law in time, while chaotic orbits have an exponential one \citep[][and \citetalias{2015MNRAS.453.2830M}]{1999MNRAS.307..495H,2008MNRAS.385..236V,2013MNRAS.436.3602G}. Sticky orbits, on the other hand, are not so easily defined. They behave as regular orbits for a given period of time to change their orbital character afterwards. Following \citetalias{2015MNRAS.453.2830M}, we used an arbitrary but physically relevant period of time  threshold to differentiate between sticky and chaotic orbits: 10~Gyr (roughly a Hubble time).

Our results show that, at all galactocentric distances, $\approx 70 - 95$~per cent of the orbits considered show a regular 
behaviour within a Hubble time. In particular, around the location of the Sun this fraction takes an average value of $\gtrsim 80$~per cent (Section~\ref{subsec:relevance-dist} for details). This holds true independently of the way the total baryonic mass is re-distributed within the  bulge-disc pair (Section~\ref{subsubsec:relevance-C45}) and, more importantly, the galactic formation history (Section~\ref{subsubsec:relevance-6Au}).  The lowest percentages of chaotic orbits is obtained for models Au-16 and Au-21 ($\sim 2$ and $\sim9$~per cent, respectively), where the shape of the dark matter haloes is oblate across all their extension. 

We performed a detailed study of the efficiency of chaotic diffusion based on first order perturbation theory. The numerical experiments presented in Section~\ref{subsec2:diffusion} showed that diffusion, the most critical mixing process, has a time-scale that by far surpasses the Hubble time. As we find from our most diffusive experiments, the largest measure of the relative diffusion rate barely amounts to $\sim 0.01 \mathrm{Gyr}^{-1}$.

Comparison with the results presented in \citetalias{2015MNRAS.453.2830M} suggests that considering a multicomponent representation of the galactic 
potential does not significantly enhance the relevance of chaos or chaotic diffusion in local halo stars  within a Hubble time. 
Instead, we find evidence that there is a direct connection between the amount of chaos found in the local stellar halo and the triaxiality of the underlying dark matter halo. It remains to be studied whether an accurate estimation of the amount of chaotic motion in halo stars could be used to constrain the shape of the underling dark matter halo potential.

Our results reinforce the idea that chaotic mixing is not a significant factor in erasing local signatures of accretion events, which is in very good agreement with previous theoretical predictions. However, fundamental caveats still persist and should be addressed in follow-up works. For instance, our models are a superposition of not only smooth but also static potentials, and substructure \citep[such as dark matter subhaloes:][]{2002MNRAS.332..915I,2009ApJ...705L.223C,2011ApJ...731...58Y,2015ApJ...808...15C,2015MNRAS.450.1136E,2015MNRAS.454.3542E,2016MNRAS.463..102E,2016ApJ...818..194N,2017MNRAS.470...60E} as well as time-dependence \citep[][and references therein]{2013JPhA...46y4017M,2014MNRAS.438.2201M,2016MNRAS.460..497H,2016MNRAS.458.3578M,2016MNRAS.461.3835M,2016ApJ...824..104P,2017MNRAS.470...60E,2017NatAs...1..633P} could enhance the efficiency of diffusion in phase space \citep{2013MNRAS.433.2576P}. It is worth noticing that sources of noise, such as scattering by short-scale irregularities, or periodic driving given by external coupling can, indeed, enhance the diffusion rate of sticky orbits \citep[see][and references therein]{1997ApJ...480..155H,2000MNRAS.311..719K,2000MNRAS.319...43S,2003ApJ...585..244K,2003MNRAS.345..727K}. Nevertheless, within the local sphere, previous studies that focus on evaluating the degree of substructure in solar neighbourhood like volumes, considering the evolution of the Galactic potential in a cosmological context, have suggested that this variation may not be responsible for any major substructure erosion \citep[e.g.][]{2013MNRAS.436.3602G}. Furthermore, as previously discussed in \citetalias{2015MNRAS.453.2830M}, it is unlikely that the inner parts of the Galactic potential have changed significantly during the last $\sim 8$~Gyr. The validity of these assumptions 
will be explored in detail in our 
forthcoming work.

\section*{Acknowledgements}

This work was started during a brief research visit to the Max-Planck-Institut f\"ur Astrophysik. NPM wish to thank their hospitality, particularly to Simon White and the people from the Galaxy Formation Group. We wish to thank David Campbell and Adrian Jenkins for generating the initial conditions and selecting the sample of the {\it Auriga} galaxies. We are grateful to the anonymous referee for detailed comments provided, which help us to greatly improve the manuscript. NPM, PMC and CMG were supported with grants from the Consejo Nacional de Investigaciones Cient\'ificas y T\'ecnicas (CONICET) de la Rep\'ublica Argentina and the Universidad Nacional de R\'io Negro (Sede Andina UNRN) and the Universidad Nacional de La Plata (FCAG UNLP). RG, RP, CMS and VS acknowledge support by the European Research Council under ERC-StG grant EXAGAL-308037, the SFB-881 'The Milky Way System' of the German Science Foundation and by the Klaus Tschira Foundation.



\bibliographystyle{mnras}
\bibliography{nfpc2017} 

\begin{thebibliography}{}
\makeatletter
\relax
\def\mn@urlcharsother{\let\do\@makeother \do\$\do\&\do\#\do\^\do\_\do\%\do\~}
\def\mn@doi{\begingroup\mn@urlcharsother \@ifnextchar [ {\mn@doi@}
  {\mn@doi@[]}}
\def\mn@doi@[#1]#2{\def\@tempa{#1}\ifx\@tempa\@empty \href
  {http://dx.doi.org/#2} {doi:#2}\else \href {http://dx.doi.org/#2} {#1}\fi
  \endgroup}
\def\mn@eprint#1#2{\mn@eprint@#1:#2::\@nil}
\def\mn@eprint@arXiv#1{\href {http://arxiv.org/abs/#1} {{\tt arXiv:#1}}}
\def\mn@eprint@dblp#1{\href {http://dblp.uni-trier.de/rec/bibtex/#1.xml}
  {dblp:#1}}
\def\mn@eprint@#1:#2:#3:#4\@nil{\def\@tempa {#1}\def\@tempb {#2}\def\@tempc
  {#3}\ifx \@tempc \@empty \let \@tempc \@tempb \let \@tempb \@tempa \fi \ifx
  \@tempb \@empty \def\@tempb {arXiv}\fi \@ifundefined
  {mn@eprint@\@tempb}{\@tempb:\@tempc}{\expandafter \expandafter \csname
  mn@eprint@\@tempb\endcsname \expandafter{\@tempc}}}

\bibitem[\protect\citeauthoryear{{Abadi}, {Navarro}, {Fardal}, {Babul}  \&
  {Steinmetz}}{{Abadi} et~al.}{2010}]{2010MNRAS.407..435A}
{Abadi} M.~G.,  {Navarro} J.~F.,  {Fardal} M.,  {Babul} A.,   {Steinmetz} M.,
  2010, \mn@doi [\mnras] {10.1111/j.1365-2966.2010.16912.x}, \href
  {http://adsabs.harvard.edu/abs/2010MNRAS.407..435A} {407, 435}

\bibitem[\protect\citeauthoryear{{Allgood}, {Flores}, {Primack}, {Kravtsov},
  {Wechsler}, {Faltenbacher}  \& {Bullock}}{{Allgood}
  et~al.}{2006}]{2006MNRAS.367.1781A}
{Allgood} B.,  {Flores} R.~A.,  {Primack} J.~R.,  {Kravtsov} A.~V.,  {Wechsler}
  R.~H.,  {Faltenbacher} A.,   {Bullock} J.~S.,  2006, \mn@doi [\mnras]
  {10.1111/j.1365-2966.2006.10094.x}, \href
  {http://adsabs.harvard.edu/abs/2006MNRAS.367.1781A} {367, 1781}

\bibitem[\protect\citeauthoryear{{Barrio}}{{Barrio}}{2016}]{2016LNP...915..55B}
{Barrio} R.,  2016, in {Skokos} C.,  {Gottwald} G.,   {Laskar} J.,  eds,
  Lecture Notes in Physics Vol. 915, Chaos Detection and Predictability. p.~55,
  \mn@doi{10.1007/978-3-662-48410-4_3}

\bibitem[\protect\citeauthoryear{{Belokurov}, {Evans}, {Irwin}, {Hewett}  \&
  {Wilkinson}}{{Belokurov} et~al.}{2006a}]{2006ApJ...637L..29B}
{Belokurov} V.,  {Evans} N.~W.,  {Irwin} M.~J.,  {Hewett} P.~C.,   {Wilkinson}
  M.~I.,  2006a, \mn@doi [\apjl] {10.1086/500362}, \href
  {http://adsabs.harvard.edu/abs/2006ApJ...637L..29B} {637, L29}

\bibitem[\protect\citeauthoryear{{Belokurov} et~al.,}{{Belokurov}
  et~al.}{2006b}]{2006ApJ...642L.137B}
{Belokurov} V.,  et~al., 2006b, \mn@doi [\apjl] {10.1086/504797}, \href
  {http://adsabs.harvard.edu/abs/2006ApJ...642L.137B} {642, L137}

\bibitem[\protect\citeauthoryear{{Belokurov} et~al.,}{{Belokurov}
  et~al.}{2007}]{2007ApJ...658..337B}
{Belokurov} V.,  et~al., 2007, \mn@doi [\apj] {10.1086/511302}, \href
  {http://adsabs.harvard.edu/abs/2007ApJ...658..337B} {658, 337}

\bibitem[\protect\citeauthoryear{{Binney} \& {Tremaine}}{{Binney} \&
  {Tremaine}}{1987}]{1987gady.book.....B}
{Binney} J.,  {Tremaine} S.,  1987, {Galactic dynamics}

\bibitem[\protect\citeauthoryear{{Bland-Hawthorn} \&
  {Freeman}}{{Bland-Hawthorn} \& {Freeman}}{2014}]{2014SAAS...37....1B}
{Bland-Hawthorn} J.,  {Freeman} K.,  2014, \mn@doi [The Origin of the Galaxy
  and Local Group, Saas-Fee Advanced Course, Volume 37.~ISBN
  978-3-642-41719-1.~Springer-Verlag Berlin Heidelberg, 2014, p.~1]
  {10.1007/978-3-642-41720-7_1}, \href
  {http://adsabs.harvard.edu/abs/2014SAAS...37....1B} {37, 1}

\bibitem[\protect\citeauthoryear{{Bonaca}, {Conroy}, {Wetzel}, {Hopkins}  \&
  {Kere{\v s}}}{{Bonaca} et~al.}{2017}]{2017ApJ...845..101B}
{Bonaca} A.,  {Conroy} C.,  {Wetzel} A.,  {Hopkins} P.~F.,   {Kere{\v s}} D.,
  2017, \mn@doi [\apj] {10.3847/1538-4357/aa7d0c}, \href
  {http://adsabs.harvard.edu/abs/2017ApJ...845..101B} {845, 101}

\bibitem[\protect\citeauthoryear{{Bovy}}{{Bovy}}{2017}]{2017MNRAS.470.1360B}
{Bovy} J.,  2017, \mn@doi [\mnras] {10.1093/mnras/stx1277}, \href
  {http://adsabs.harvard.edu/abs/2017MNRAS.470.1360B} {470, 1360}

\bibitem[\protect\citeauthoryear{{Carlberg}}{{Carlberg}}{2009}]{2009ApJ...705L.223C}
{Carlberg} R.~G.,  2009, \mn@doi [\apjl] {10.1088/0004-637X/705/2/L223}, \href
  {http://adsabs.harvard.edu/abs/2009ApJ...705L.223C} {705, L223}

\bibitem[\protect\citeauthoryear{{Carlberg}}{{Carlberg}}{2015}]{2015ApJ...808...15C}
{Carlberg} R.~G.,  2015, \mn@doi [\apj] {10.1088/0004-637X/808/1/15}, \href
  {http://adsabs.harvard.edu/abs/2015ApJ...808...15C} {808, 15}

\bibitem[\protect\citeauthoryear{{Carpintero}, {Maffione}  \&
  {Darriba}}{{Carpintero} et~al.}{2014}]{2014A&C.....5...19C}
{Carpintero} D.~D.,  {Maffione} N.,   {Darriba} L.,  2014, \mn@doi [Astronomy
  and Computing] {10.1016/j.ascom.2014.04.001}, \href
  {http://adsabs.harvard.edu/abs/2014A%26C.....5...19C} {5, 19}

\bibitem[\protect\citeauthoryear{{Chakrabarty}}{{Chakrabarty}}{2007}]{2007A&A...467..145C}
{Chakrabarty} D.,  2007, \mn@doi [\aap] {10.1051/0004-6361:20066677}, \href
  {http://adsabs.harvard.edu/abs/2007A%26A...467..145C} {467, 145}

\bibitem[\protect\citeauthoryear{{Chakrabarty} \& {Sideris}}{{Chakrabarty} \&
  {Sideris}}{2008}]{2008A&A...488..161C}
{Chakrabarty} D.,  {Sideris} I.~V.,  2008, \mn@doi [\aap]
  {10.1051/0004-6361:20079316}, \href
  {http://adsabs.harvard.edu/abs/2008A%26A...488..161C} {488, 161}

\bibitem[\protect\citeauthoryear{{Cincotta} \& {Giordano}}{{Cincotta} \&
  {Giordano}}{2016}]{2016LNP...915..93C}
{Cincotta} P.~M.,  {Giordano} C.~M.,  2016, in {Skokos} C.,  {Gottwald} G.,
  {Laskar} J.,  eds,  Lecture Notes in Physics Vol. 915, Chaos Detection and
  Predictability. p.~93, \mn@doi{10.1007/978-3-662-48410-4_4}

\bibitem[\protect\citeauthoryear{{Cincotta} \& {Sim{\'o}}}{{Cincotta} \&
  {Sim{\'o}}}{2000}]{2000A&AS..147..205C}
{Cincotta} P.~M.,  {Sim{\'o}} C.,  2000, \mn@doi [\aaps] {10.1051/aas:2000108},
  \href {http://adsabs.harvard.edu/abs/2000A%26AS..147..205C} {147, 205}

\bibitem[\protect\citeauthoryear{{Cincotta}, {Giordano}  \&
  {Sim{\'o}}}{{Cincotta} et~al.}{2003}]{2003PhyD..182..151C}
{Cincotta} P.~M.,  {Giordano} C.~M.,   {Sim{\'o}} C.,  2003, \mn@doi [Physica D
  Nonlinear Phenomena] {10.1016/S0167-2789(03)00103-9}, \href
  {http://adsabs.harvard.edu/abs/2003PhyD..182..151C} {182, 151}

\bibitem[\protect\citeauthoryear{{Cincotta}, {Giordano}  \&
  {P{\'e}rez}}{{Cincotta} et~al.}{2006}]{2006A&A...455..499C}
{Cincotta} P.~M.,  {Giordano} C.~M.,   {P{\'e}rez} M.~J.,  2006, \mn@doi [\aap]
  {10.1051/0004-6361:20054721}, \href
  {http://adsabs.harvard.edu/abs/2006A%26A...455..499C} {455, 499}

\bibitem[\protect\citeauthoryear{{Cincotta}, {Efthymiopoulos}, {Giordano}  \&
  {Mestre}}{{Cincotta} et~al.}{2014}]{2014PhyD..266...49C}
{Cincotta} P.~M.,  {Efthymiopoulos} C.,  {Giordano} C.~M.,   {Mestre} M.~F.,
  2014, \mn@doi [Physica D Nonlinear Phenomena] {10.1016/j.physd.2013.10.005},
  \href {http://adsabs.harvard.edu/abs/2014PhyD..266...49C} {266, 49}

\bibitem[\protect\citeauthoryear{{Clutton-Brock}}{{Clutton-Brock}}{1973}]{1973Ap&SS..23...55C}
{Clutton-Brock} M.,  1973, \mn@doi [\apss] {10.1007/BF00647652}, \href
  {http://adsabs.harvard.edu/abs/1973Ap%26SS..23...55C} {23, 55}

\bibitem[\protect\citeauthoryear{{Comp{\`e}re}, {Lema{\^i}tre}  \&
  {Delsate}}{{Comp{\`e}re} et~al.}{2012}]{2012CeMDA.112...75C}
{Comp{\`e}re} A.,  {Lema{\^i}tre} A.,   {Delsate} N.,  2012, \mn@doi [Celestial
  Mechanics and Dynamical Astronomy] {10.1007/s10569-011-9384-9}, \href
  {http://adsabs.harvard.edu/abs/2012CeMDA.112...75C} {112, 75}

\bibitem[\protect\citeauthoryear{{Contopoulos}}{{Contopoulos}}{2002}]{2002ocda.book.....C}
{Contopoulos} G.,  2002, {Order and chaos in dynamical astronomy}

\bibitem[\protect\citeauthoryear{{Cooper} et~al.,}{{Cooper}
  et~al.}{2010}]{2010MNRAS.406..744C}
{Cooper} A.~P.,  et~al., 2010, \mn@doi [\mnras]
  {10.1111/j.1365-2966.2010.16740.x}, \href
  {http://adsabs.harvard.edu/abs/2010MNRAS.406..744C} {406, 744}

\bibitem[\protect\citeauthoryear{{Darriba}, {Maffione}, {Cincotta}  \&
  {Giordano}}{{Darriba} et~al.}{2012}]{2012IJBC...2230033D}
{Darriba} L.~A.,  {Maffione} N.~P.,  {Cincotta} P.~M.,   {Giordano} C.~M.,
  2012, \mn@doi [International Journal of Bifurcation and Chaos]
  {10.1142/S0218127412300339}, \href
  {http://adsabs.harvard.edu/abs/2012IJBC...2230033D} {22, 1230033}

\bibitem[\protect\citeauthoryear{{Debattista}, {Moore}, {Quinn}, {Kazantzidis},
  {Maas}, {Mayer}, {Read}  \& {Stadel}}{{Debattista}
  et~al.}{2008}]{2008ApJ...681.1076D}
{Debattista} V.~P.,  {Moore} B.,  {Quinn} T.,  {Kazantzidis} S.,  {Maas} R.,
  {Mayer} L.,  {Read} J.,   {Stadel} J.,  2008, \mn@doi [\apj]
  {10.1086/587977}, \href {http://adsabs.harvard.edu/abs/2008ApJ...681.1076D}
  {681, 1076}

\bibitem[\protect\citeauthoryear{{Dubinski}}{{Dubinski}}{1994}]{1994ApJ...431..617D}
{Dubinski} J.,  1994, \mn@doi [\apj] {10.1086/174512}, \href
  {http://adsabs.harvard.edu/abs/1994ApJ...431..617D} {431, 617}

\bibitem[\protect\citeauthoryear{{ESA}}{{ESA}}{1997}]{1997ESASP1200.....E}
{ESA} ed. 1997, {The HIPPARCOS and TYCHO catalogues. Astrometric and
  photometric star catalogues derived from the ESA HIPPARCOS Space Astrometry
  Mission}  ESA Special Publication Vol. 1200

\bibitem[\protect\citeauthoryear{{Efthymiopoulos}, {Voglis}  \&
  {Kalapotharakos}}{{Efthymiopoulos} et~al.}{2007}]{2007LNP...729..297E}
{Efthymiopoulos} C.,  {Voglis} N.,   {Kalapotharakos} C.,  2007, in {Benest}
  D.,  {Froeschle} C.,   {Lega} E.,  eds,  Lecture Notes in Physics, Berlin
  Springer Verlag Vol. 729, Lecture Notes in Physics, Berlin Springer Verlag.
  pp 297--389 (\mn@eprint {} {astro-ph/0610246}),
  \mn@doi{10.1007/978-3-540-72984-6_11}

\bibitem[\protect\citeauthoryear{{Erkal} \& {Belokurov}}{{Erkal} \&
  {Belokurov}}{2015a}]{2015MNRAS.450.1136E}
{Erkal} D.,  {Belokurov} V.,  2015a, \mn@doi [\mnras] {10.1093/mnras/stv655},
  \href {http://adsabs.harvard.edu/abs/2015MNRAS.450.1136E} {450, 1136}

\bibitem[\protect\citeauthoryear{{Erkal} \& {Belokurov}}{{Erkal} \&
  {Belokurov}}{2015b}]{2015MNRAS.454.3542E}
{Erkal} D.,  {Belokurov} V.,  2015b, \mn@doi [\mnras] {10.1093/mnras/stv2122},
  \href {http://adsabs.harvard.edu/abs/2015MNRAS.454.3542E} {454, 3542}

\bibitem[\protect\citeauthoryear{{Erkal}, {Belokurov}, {Bovy}  \&
  {Sanders}}{{Erkal} et~al.}{2016}]{2016MNRAS.463..102E}
{Erkal} D.,  {Belokurov} V.,  {Bovy} J.,   {Sanders} J.~L.,  2016, \mn@doi
  [\mnras] {10.1093/mnras/stw1957}, \href
  {http://adsabs.harvard.edu/abs/2016MNRAS.463..102E} {463, 102}

\bibitem[\protect\citeauthoryear{{Erkal}, {Koposov}  \& {Belokurov}}{{Erkal}
  et~al.}{2017}]{2017MNRAS.470...60E}
{Erkal} D.,  {Koposov} S.~E.,   {Belokurov} V.,  2017, \mn@doi [\mnras]
  {10.1093/mnras/stx1208}, \href
  {http://adsabs.harvard.edu/abs/2017MNRAS.470...60E} {470, 60}

\bibitem[\protect\citeauthoryear{{Fouchard}, {Lega}, {Froeschl{\'e}}  \&
  {Froeschl{\'e}}}{{Fouchard} et~al.}{2002}]{2002CeMDA..83..205F}
{Fouchard} M.,  {Lega} E.,  {Froeschl{\'e}} C.,   {Froeschl{\'e}} C.,  2002,
  Celestial Mechanics and Dynamical Astronomy, \href
  {http://adsabs.harvard.edu/abs/2002CeMDA..83..205F} {83, 205}

\bibitem[\protect\citeauthoryear{{Franx}, {Illingworth}  \& {de Zeeuw}}{{Franx}
  et~al.}{1991}]{1991ApJ...383..112F}
{Franx} M.,  {Illingworth} G.,   {de Zeeuw} T.,  1991, \mn@doi [\apj]
  {10.1086/170769}, \href {http://adsabs.harvard.edu/abs/1991ApJ...383..112F}
  {383, 112}

\bibitem[\protect\citeauthoryear{{Freeman} \& {Bland-Hawthorn}}{{Freeman} \&
  {Bland-Hawthorn}}{2002}]{2002ARA&A..40..487F}
{Freeman} K.,  {Bland-Hawthorn} J.,  2002, \mn@doi [\araa]
  {10.1146/annurev.astro.40.060401.093840}, \href
  {http://adsabs.harvard.edu/abs/2002ARA%26A..40..487F} {40, 487}

\bibitem[\protect\citeauthoryear{{Fux}}{{Fux}}{2001}]{2001A&A...373..511F}
{Fux} R.,  2001, \mn@doi [\aap] {10.1051/0004-6361:20010561}, \href
  {http://adsabs.harvard.edu/abs/2001A%26A...373..511F} {373, 511}

\bibitem[\protect\citeauthoryear{{Giordano} \& {Cincotta}}{{Giordano} \&
  {Cincotta}}{2004}]{2004A&A...423..745G}
{Giordano} C.~M.,  {Cincotta} P.~M.,  2004, \mn@doi [\aap]
  {10.1051/0004-6361:20040153}, \href
  {http://adsabs.harvard.edu/abs/2004A%26A...423..745G} {423, 745}

\bibitem[\protect\citeauthoryear{{Gnedin}, {Kravtsov}, {Klypin}  \&
  {Nagai}}{{Gnedin} et~al.}{2004}]{2004ApJ...616...16G}
{Gnedin} O.~Y.,  {Kravtsov} A.~V.,  {Klypin} A.~A.,   {Nagai} D.,  2004,
  \mn@doi [\apj] {10.1086/424914}, \href
  {http://adsabs.harvard.edu/abs/2004ApJ...616...16G} {616, 16}

\bibitem[\protect\citeauthoryear{{G{\'o}mez}, {Helmi}, {Brown}  \&
  {Li}}{{G{\'o}mez} et~al.}{2010}]{2010MNRAS.408..935G}
{G{\'o}mez} F.~A.,  {Helmi} A.,  {Brown} A.~G.~A.,   {Li} Y.-S.,  2010, \mn@doi
  [\mnras] {10.1111/j.1365-2966.2010.17225.x}, \href
  {http://adsabs.harvard.edu/abs/2010MNRAS.408..935G} {408, 935}

\bibitem[\protect\citeauthoryear{{G{\'o}mez}, {Minchev}, {Villalobos}, {O'Shea}
   \& {Williams}}{{G{\'o}mez} et~al.}{2012a}]{2012MNRAS.419.2163G}
{G{\'o}mez} F.~A.,  {Minchev} I.,  {Villalobos} {\'A}.,  {O'Shea} B.~W.,
  {Williams} M.~E.~K.,  2012a, \mn@doi [\mnras]
  {10.1111/j.1365-2966.2011.19867.x}, \href
  {http://adsabs.harvard.edu/abs/2012MNRAS.419.2163G} {419, 2163}

\bibitem[\protect\citeauthoryear{{G{\'o}mez} et~al.,}{{G{\'o}mez}
  et~al.}{2012b}]{2012MNRAS.423.3727G}
{G{\'o}mez} F.~A.,  et~al., 2012b, \mn@doi [\mnras]
  {10.1111/j.1365-2966.2012.21176.x}, \href
  {http://adsabs.harvard.edu/abs/2012MNRAS.423.3727G} {423, 3727}

\bibitem[\protect\citeauthoryear{{G{\'o}mez}, {Minchev}, {O'Shea}, {Beers},
  {Bullock}  \& {Purcell}}{{G{\'o}mez} et~al.}{2013a}]{2013MNRAS.429..159G}
{G{\'o}mez} F.~A.,  {Minchev} I.,  {O'Shea} B.~W.,  {Beers} T.~C.,  {Bullock}
  J.~S.,   {Purcell} C.~W.,  2013a, \mn@doi [\mnras] {10.1093/mnras/sts327},
  \href {http://adsabs.harvard.edu/abs/2013MNRAS.429..159G} {429, 159}

\bibitem[\protect\citeauthoryear{{G{\'o}mez}, {Helmi}, {Cooper}, {Frenk},
  {Navarro}  \& {White}}{{G{\'o}mez} et~al.}{2013b}]{2013MNRAS.436.3602G}
{G{\'o}mez} F.~A.,  {Helmi} A.,  {Cooper} A.~P.,  {Frenk} C.~S.,  {Navarro}
  J.~F.,   {White} S.~D.~M.,  2013b, \mn@doi [\mnras] {10.1093/mnras/stt1838},
  \href {http://adsabs.harvard.edu/abs/2013MNRAS.436.3602G} {436, 3602}

\bibitem[\protect\citeauthoryear{{G{\'o}mez}, {White}, {Marinacci}, {Slater},
  {Grand}, {Springel}  \& {Pakmor}}{{G{\'o}mez}
  et~al.}{2016}]{2016MNRAS.456.2779G}
{G{\'o}mez} F.~A.,  {White} S.~D.~M.,  {Marinacci} F.,  {Slater} C.~T.,
  {Grand} R.~J.~J.,  {Springel} V.,   {Pakmor} R.,  2016, \mn@doi [\mnras]
  {10.1093/mnras/stv2786}, \href
  {http://adsabs.harvard.edu/abs/2016MNRAS.456.2779G} {456, 2779}

\bibitem[\protect\citeauthoryear{{G{\'o}mez}, {White}, {Grand}, {Marinacci},
  {Springel}  \& {Pakmor}}{{G{\'o}mez} et~al.}{2017}]{2017MNRAS.465.3446G}
{G{\'o}mez} F.~A.,  {White} S.~D.~M.,  {Grand} R.~J.~J.,  {Marinacci} F.,
  {Springel} V.,   {Pakmor} R.,  2017, \mn@doi [\mnras]
  {10.1093/mnras/stw2957}, \href
  {http://adsabs.harvard.edu/abs/2017MNRAS.465.3446G} {465, 3446}

\bibitem[\protect\citeauthoryear{{Gould}}{{Gould}}{2003}]{2003ApJ...592L..63G}
{Gould} A.,  2003, \mn@doi [\apjl] {10.1086/377525}, \href
  {http://adsabs.harvard.edu/abs/2003ApJ...592L..63G} {592, L63}

\bibitem[\protect\citeauthoryear{{Go{\'z}dziewski}, {Konacki}  \&
  {Wolszczan}}{{Go{\'z}dziewski} et~al.}{2005}]{2005ApJ...619.1084G}
{Go{\'z}dziewski} K.,  {Konacki} M.,   {Wolszczan} A.,  2005, \mn@doi [\apj]
  {10.1086/426775}, \href {http://adsabs.harvard.edu/abs/2005ApJ...619.1084G}
  {619, 1084}

\bibitem[\protect\citeauthoryear{{Grand}, {Springel}, {G{\'o}mez}, {Marinacci},
  {Pakmor}, {Campbell}  \& {Jenkins}}{{Grand}
  et~al.}{2016}]{2016MNRAS.459..199G}
{Grand} R.~J.~J.,  {Springel} V.,  {G{\'o}mez} F.~A.,  {Marinacci} F.,
  {Pakmor} R.,  {Campbell} D.~J.~R.,   {Jenkins} A.,  2016, \mn@doi [\mnras]
  {10.1093/mnras/stw601}, \href
  {http://adsabs.harvard.edu/abs/2016MNRAS.459..199G} {459, 199}

\bibitem[\protect\citeauthoryear{{Grand} et~al.,}{{Grand}
  et~al.}{2017}]{2017MNRAS.467..179G}
{Grand} R.~J.~J.,  et~al., 2017, \mn@doi [\mnras] {10.1093/mnras/stx071}, \href
  {http://adsabs.harvard.edu/abs/2017MNRAS.467..179G} {467, 179}

\bibitem[\protect\citeauthoryear{{Grillmair} \& {Carlin}}{{Grillmair} \&
  {Carlin}}{2016}]{2016ASSL..420...87G}
{Grillmair} C.~J.,  {Carlin} J.~L.,  2016, in {Newberg} H.~J.,  {Carlin} J.~L.,
   eds,  Astrophysics and Space Science Library Vol. 420, Tidal Streams in the
  Local Group and Beyond. p.~87 (\mn@eprint {arXiv} {1603.08936}),
  \mn@doi{10.1007/978-3-319-19336-6_4}

\bibitem[\protect\citeauthoryear{{Gustafsson}, {Fairbairn}  \&
  {Sommer-Larsen}}{{Gustafsson} et~al.}{2006}]{2006PhRvD..74l3522G}
{Gustafsson} M.,  {Fairbairn} M.,   {Sommer-Larsen} J.,  2006, \mn@doi [\prd]
  {10.1103/PhysRevD.74.123522}, \href
  {http://adsabs.harvard.edu/abs/2006PhRvD..74l3522G} {74, 123522}

\bibitem[\protect\citeauthoryear{{Habib}, {Kandrup}  \& {Elaine Mahon}}{{Habib}
  et~al.}{1997}]{1997ApJ...480..155H}
{Habib} S.,  {Kandrup} H.~E.,   {Elaine Mahon} M.,  1997, \mn@doi [\apj]
  {10.1086/303935}, \href
  {https://ui.adsabs.harvard.edu/#abs/1997ApJ...480..155H} {480, 155}

\bibitem[\protect\citeauthoryear{{Hattori}, {Erkal}  \& {Sanders}}{{Hattori}
  et~al.}{2016}]{2016MNRAS.460..497H}
{Hattori} K.,  {Erkal} D.,   {Sanders} J.~L.,  2016, \mn@doi [\mnras]
  {10.1093/mnras/stw1006}, \href
  {http://adsabs.harvard.edu/abs/2016MNRAS.460..497H} {460, 497}

\bibitem[\protect\citeauthoryear{{Helmi}}{{Helmi}}{2008}]{2008A&ARv..15..145H}
{Helmi} A.,  2008, \mn@doi [\aapr] {10.1007/s00159-008-0009-6}, \href
  {http://adsabs.harvard.edu/abs/2008A%26ARv..15..145H} {15, 145}

\bibitem[\protect\citeauthoryear{{Helmi} \& {White}}{{Helmi} \&
  {White}}{1999}]{1999MNRAS.307..495H}
{Helmi} A.,  {White} S.~D.~M.,  1999, \mn@doi [\mnras]
  {10.1046/j.1365-8711.1999.02616.x}, \href
  {http://adsabs.harvard.edu/abs/1999MNRAS.307..495H} {307, 495}

\bibitem[\protect\citeauthoryear{{Helmi} \& {de Zeeuw}}{{Helmi} \& {de
  Zeeuw}}{2000}]{2000MNRAS.319..657H}
{Helmi} A.,  {de Zeeuw} P.~T.,  2000, \mn@doi [\mnras]
  {10.1046/j.1365-8711.2000.03895.x}, \href
  {http://adsabs.harvard.edu/abs/2000MNRAS.319..657H} {319, 657}

\bibitem[\protect\citeauthoryear{{Helmi}, {White}, {de Zeeuw}  \&
  {Zhao}}{{Helmi} et~al.}{1999}]{1999Natur.402...53H}
{Helmi} A.,  {White} S.~D.~M.,  {de Zeeuw} P.~T.,   {Zhao} H.,  1999, \mn@doi
  [\nat] {10.1038/46980}, \href
  {http://adsabs.harvard.edu/abs/1999Natur.402...53H} {402, 53}

\bibitem[\protect\citeauthoryear{{Helmi}, {White}  \& {Springel}}{{Helmi}
  et~al.}{2003}]{2003MNRAS.339..834H}
{Helmi} A.,  {White} S.~D.~M.,   {Springel} V.,  2003, \mn@doi [\mnras]
  {10.1046/j.1365-8711.2003.06227.x}, \href
  {http://adsabs.harvard.edu/abs/2003MNRAS.339..834H} {339, 834}

\bibitem[\protect\citeauthoryear{{Helmi}, {Navarro}, {Nordstr{\"o}m},
  {Holmberg}, {Abadi}  \& {Steinmetz}}{{Helmi}
  et~al.}{2006}]{2006MNRAS.365.1309H}
{Helmi} A.,  {Navarro} J.~F.,  {Nordstr{\"o}m} B.,  {Holmberg} J.,  {Abadi}
  M.~G.,   {Steinmetz} M.,  2006, \mn@doi [\mnras]
  {10.1111/j.1365-2966.2005.09818.x}, \href
  {http://adsabs.harvard.edu/abs/2006MNRAS.365.1309H} {365, 1309}

\bibitem[\protect\citeauthoryear{{Helmi}, {Veljanoski}, {Breddels}, {Tian}  \&
  {Sales}}{{Helmi} et~al.}{2017}]{2017A&A...598A..58H}
{Helmi} A.,  {Veljanoski} J.,  {Breddels} M.~A.,  {Tian} H.,   {Sales} L.~V.,
  2017, \mn@doi [\aap] {10.1051/0004-6361/201629990}, \href
  {http://adsabs.harvard.edu/abs/2017A%26A...598A..58H} {598, A58}

\bibitem[\protect\citeauthoryear{{Hernquist}}{{Hernquist}}{1990}]{1990ApJ...356..359H}
{Hernquist} L.,  1990, \mn@doi [\apj] {10.1086/168845}, \href
  {http://adsabs.harvard.edu/abs/1990ApJ...356..359H} {356, 359}

\bibitem[\protect\citeauthoryear{{Hernquist} \& {Ostriker}}{{Hernquist} \&
  {Ostriker}}{1992}]{1992ApJ...386..375H}
{Hernquist} L.,  {Ostriker} J.~P.,  1992, \mn@doi [\apj] {10.1086/171025},
  \href {http://adsabs.harvard.edu/abs/1992ApJ...386..375H} {386, 375}

\bibitem[\protect\citeauthoryear{{Ibata}, {Gilmore}  \& {Irwin}}{{Ibata}
  et~al.}{1994}]{1994Natur.370..194I}
{Ibata} R.~A.,  {Gilmore} G.,   {Irwin} M.~J.,  1994, \mn@doi [\nat]
  {10.1038/370194a0}, \href {http://adsabs.harvard.edu/abs/1994Natur.370..194I}
  {370, 194}

\bibitem[\protect\citeauthoryear{{Ibata}, {Irwin}, {Lewis}  \&
  {Stolte}}{{Ibata} et~al.}{2001a}]{2001ApJ...547L.133I}
{Ibata} R.,  {Irwin} M.,  {Lewis} G.~F.,   {Stolte} A.,  2001a, \mn@doi [\apjl]
  {10.1086/318894}, \href {http://adsabs.harvard.edu/abs/2001ApJ...547L.133I}
  {547, L133}

\bibitem[\protect\citeauthoryear{{Ibata}, {Lewis}, {Irwin}, {Totten}  \&
  {Quinn}}{{Ibata} et~al.}{2001b}]{2001ApJ...551..294I}
{Ibata} R.,  {Lewis} G.~F.,  {Irwin} M.,  {Totten} E.,   {Quinn} T.,  2001b,
  \mn@doi [\apj] {10.1086/320060}, \href
  {http://adsabs.harvard.edu/abs/2001ApJ...551..294I} {551, 294}

\bibitem[\protect\citeauthoryear{{Ibata}, {Lewis}, {Irwin}  \& {Quinn}}{{Ibata}
  et~al.}{2002}]{2002MNRAS.332..915I}
{Ibata} R.~A.,  {Lewis} G.~F.,  {Irwin} M.~J.,   {Quinn} T.,  2002, \mn@doi
  [\mnras] {10.1046/j.1365-8711.2002.05358.x}, \href
  {http://adsabs.harvard.edu/abs/2002MNRAS.332..915I} {332, 915}

\bibitem[\protect\citeauthoryear{{Ibata}, {Irwin}, {Lewis}, {Ferguson}  \&
  {Tanvir}}{{Ibata} et~al.}{2003}]{2003MNRAS.340L..21I}
{Ibata} R.~A.,  {Irwin} M.~J.,  {Lewis} G.~F.,  {Ferguson} A.~M.~N.,   {Tanvir}
  N.,  2003, \mn@doi [\mnras] {10.1046/j.1365-8711.2003.06545.x}, \href
  {http://adsabs.harvard.edu/abs/2003MNRAS.340L..21I} {340, L21}

\bibitem[\protect\citeauthoryear{{Iorio}, {Belokurov}, {Erkal}, {Koposov},
  {Nipoti}  \& {Fraternali}}{{Iorio} et~al.}{2017}]{2017arXiv170703833I}
{Iorio} G.,  {Belokurov} V.,  {Erkal} D.,  {Koposov} S.~E.,  {Nipoti} C.,
  {Fraternali} F.,  2017, preprint, \href
  {http://adsabs.harvard.edu/abs/2017arXiv170703833I} {} (\mn@eprint {arXiv}
  {1707.03833})

\bibitem[\protect\citeauthoryear{{Jing} \& {Suto}}{{Jing} \&
  {Suto}}{2002}]{2002ApJ...574..538J}
{Jing} Y.~P.,  {Suto} Y.,  2002, \mn@doi [\apj] {10.1086/341065}, \href
  {http://adsabs.harvard.edu/abs/2002ApJ...574..538J} {574, 538}

\bibitem[\protect\citeauthoryear{{Johnston}, {Bullock}, {Sharma}, {Font},
  {Robertson}  \& {Leitner}}{{Johnston} et~al.}{2008}]{2008ApJ...689..936J}
{Johnston} K.~V.,  {Bullock} J.~S.,  {Sharma} S.,  {Font} A.,  {Robertson}
  B.~E.,   {Leitner} S.~N.,  2008, \mn@doi [\apj] {10.1086/592228}, \href
  {http://adsabs.harvard.edu/abs/2008ApJ...689..936J} {689, 936}

\bibitem[\protect\citeauthoryear{{Kalapotharakos}, {Voglis}  \&
  {Contopoulos}}{{Kalapotharakos} et~al.}{2004}]{2004A&A...428..905K}
{Kalapotharakos} C.,  {Voglis} N.,   {Contopoulos} G.,  2004, \mn@doi [\aap]
  {10.1051/0004-6361:20041492}, \href
  {http://adsabs.harvard.edu/abs/2004A%26A...428..905K} {428, 905}

\bibitem[\protect\citeauthoryear{{Kalapotharakos}, {Efthymiopoulos}  \&
  {Voglis}}{{Kalapotharakos} et~al.}{2008}]{2008MNRAS.383..971K}
{Kalapotharakos} C.,  {Efthymiopoulos} C.,   {Voglis} N.,  2008, \mn@doi
  [\mnras] {10.1111/j.1365-2966.2007.12417.x}, \href
  {http://adsabs.harvard.edu/abs/2008MNRAS.383..971K} {383, 971}

\bibitem[\protect\citeauthoryear{{Kandrup} \& {Sideris}}{{Kandrup} \&
  {Sideris}}{2001}]{2001PhRvE..64e6209K}
{Kandrup} H.~E.,  {Sideris} I.~V.,  2001, \mn@doi [\pre]
  {10.1103/PhysRevE.64.056209}, \href
  {http://adsabs.harvard.edu/abs/2001PhRvE..64e6209K} {64, 056209}

\bibitem[\protect\citeauthoryear{{Kandrup} \& {Sideris}}{{Kandrup} \&
  {Sideris}}{2003}]{2003ApJ...585..244K}
{Kandrup} H.~E.,  {Sideris} I.~V.,  2003, \mn@doi [\apj] {10.1086/345948},
  \href {https://ui.adsabs.harvard.edu/#abs/2003ApJ...585..244K} {585, 244}

\bibitem[\protect\citeauthoryear{{Kandrup} \& {Siopis}}{{Kandrup} \&
  {Siopis}}{2003}]{2003MNRAS.345..727K}
{Kandrup} H.~E.,  {Siopis} C.,  2003, \mn@doi [\mnras]
  {10.1046/j.1365-8711.2003.06985.x}, \href
  {http://adsabs.harvard.edu/abs/2003MNRAS.345..727K} {345, 727}

\bibitem[\protect\citeauthoryear{{Kandrup}, {Pogorelov}  \&
  {Sideris}}{{Kandrup} et~al.}{2000}]{2000MNRAS.311..719K}
{Kandrup} H.~E.,  {Pogorelov} I.~V.,   {Sideris} I.~V.,  2000, \mn@doi [\mnras]
  {10.1046/j.1365-8711.2000.03097.x}, \href
  {https://ui.adsabs.harvard.edu/#abs/2000MNRAS.311..719K} {311, 719}

\bibitem[\protect\citeauthoryear{{Kushniruk}, {Schirmer}  \&
  {Bensby}}{{Kushniruk} et~al.}{2017}]{2017A&A...608A..73K}
{Kushniruk} I.,  {Schirmer} T.,   {Bensby} T.,  2017, \mn@doi [\aap]
  {10.1051/0004-6361/201731147}, \href
  {https://ui.adsabs.harvard.edu/#abs/2017A&A...608A..73K} {608, A73}

\bibitem[\protect\citeauthoryear{{Laporte}, {Johnston}, {G{\'o}mez},
  {Garavito-Camargo}  \& {Besla}}{{Laporte} et~al.}{2017}]{2017arXiv171002538L}
{Laporte} C.~F.~P.,  {Johnston} K.~V.,  {G{\'o}mez} F.~A.,  {Garavito-Camargo}
  N.,   {Besla} G.,  2017, preprint, \href
  {http://adsabs.harvard.edu/abs/2017arXiv171002538L} {} (\mn@eprint {arXiv}
  {1710.02538})

\bibitem[\protect\citeauthoryear{{Laporte}, {G{\'o}mez}, {Besla}, {Johnston}
  \& {Garavito-Camargo}}{{Laporte} et~al.}{2018}]{2018MNRAS.473.1218L}
{Laporte} C.~F.~P.,  {G{\'o}mez} F.~A.,  {Besla} G.,  {Johnston} K.~V.,
  {Garavito-Camargo} N.,  2018, \mn@doi [\mnras] {10.1093/mnras/stx2146}, \href
  {http://adsabs.harvard.edu/abs/2018MNRAS.473.1218L} {473, 1218}

\bibitem[\protect\citeauthoryear{{Launhardt}, {Zylka}  \& {Mezger}}{{Launhardt}
  et~al.}{2002}]{2002A&A...384..112L}
{Launhardt} R.,  {Zylka} R.,   {Mezger} P.~G.,  2002, \mn@doi [\aap]
  {10.1051/0004-6361:20020017}, \href
  {http://adsabs.harvard.edu/abs/2002A%26A...384..112L} {384, 112}

\bibitem[\protect\citeauthoryear{{Lindegren} et~al.,}{{Lindegren}
  et~al.}{2008}]{2008IAUS..248..529L}
{Lindegren} L.,  et~al., 2008, in {Jin} W.~J.,  {Platais} I.,   {Perryman}
  M.~A.~C.,  eds,  IAU Symposium Vol. 248, A Giant Step: from Milli- to
  Micro-arcsecond Astrometry. pp 529--530, \mn@doi{10.1017/S1743921308020061}

\bibitem[\protect\citeauthoryear{{Lindegren} et~al.,}{{Lindegren}
  et~al.}{2016}]{2016A&A...595A...4L}
{Lindegren} L.,  et~al., 2016, \mn@doi [\aap] {10.1051/0004-6361/201628714},
  \href {http://adsabs.harvard.edu/abs/2016A%26A...595A...4L} {595, A4}

\bibitem[\protect\citeauthoryear{{Lowing}, {Jenkins}, {Eke}  \&
  {Frenk}}{{Lowing} et~al.}{2011}]{2011MNRAS.416.2697L}
{Lowing} B.,  {Jenkins} A.,  {Eke} V.,   {Frenk} C.,  2011, \mn@doi [\mnras]
  {10.1111/j.1365-2966.2011.19222.x}, \href
  {http://adsabs.harvard.edu/abs/2011MNRAS.416.2697L} {416, 2697}

\bibitem[\protect\citeauthoryear{{Machado} \& {Manos}}{{Machado} \&
  {Manos}}{2016}]{2016MNRAS.458.3578M}
{Machado} R.~E.~G.,  {Manos} T.,  2016, \mn@doi [\mnras]
  {10.1093/mnras/stw572}, \href
  {http://adsabs.harvard.edu/abs/2016MNRAS.458.3578M} {458, 3578}

\bibitem[\protect\citeauthoryear{{Maffione}, {Giordano}  \&
  {Cincotta}}{{Maffione} et~al.}{2011a}]{2011IJNLM..46...23M}
{Maffione} N.~P.,  {Giordano} C.~M.,   {Cincotta} P.~M.,  2011a, \mn@doi
  [International Journal of Non Linear Mechanics]
  {10.1016/j.ijnonlinmec.2010.06.008}, \href
  {http://adsabs.harvard.edu/abs/2011IJNLM..46...23M} {46, 23}

\bibitem[\protect\citeauthoryear{{Maffione}, {Darriba}, {Cincotta}  \&
  {Giordano}}{{Maffione} et~al.}{2011b}]{2011CeMDA.111..285M}
{Maffione} N.~P.,  {Darriba} L.~A.,  {Cincotta} P.~M.,   {Giordano} C.~M.,
  2011b, \mn@doi [Celestial Mechanics and Dynamical Astronomy]
  {10.1007/s10569-011-9373-z}, \href
  {http://adsabs.harvard.edu/abs/2011CeMDA.111..285M} {111, 285}

\bibitem[\protect\citeauthoryear{{Maffione}, {Darriba}, {Cincotta}  \&
  {Giordano}}{{Maffione} et~al.}{2013}]{2013MNRAS.429.2700M}
{Maffione} N.~P.,  {Darriba} L.~A.,  {Cincotta} P.~M.,   {Giordano} C.~M.,
  2013, \mn@doi [\mnras] {10.1093/mnras/sts539}, \href
  {http://adsabs.harvard.edu/abs/2013MNRAS.429.2700M} {429, 2700}

\bibitem[\protect\citeauthoryear{{Maffione}, {G{\'o}mez}, {Cincotta},
  {Giordano}, {Cooper}  \& {O'Shea}}{{Maffione}
  et~al.}{2015}]{2015MNRAS.453.2830M}
{Maffione} N.~P.,  {G{\'o}mez} F.~A.,  {Cincotta} P.~M.,  {Giordano} C.~M.,
  {Cooper} A.~P.,   {O'Shea} B.~W.,  2015, \mn@doi [\mnras]
  {10.1093/mnras/stv1778}, \href
  {http://adsabs.harvard.edu/abs/2015MNRAS.453.2830M} {453, 2830}

\bibitem[\protect\citeauthoryear{{Majewski}, {Skrutskie}, {Weinberg}  \&
  {Ostheimer}}{{Majewski} et~al.}{2003}]{2003ApJ...599.1082M}
{Majewski} S.~R.,  {Skrutskie} M.~F.,  {Weinberg} M.~D.,   {Ostheimer} J.~C.,
  2003, \mn@doi [\apj] {10.1086/379504}, \href
  {http://adsabs.harvard.edu/abs/2003ApJ...599.1082M} {599, 1082}

\bibitem[\protect\citeauthoryear{{Manos} \& {Machado}}{{Manos} \&
  {Machado}}{2014}]{2014MNRAS.438.2201M}
{Manos} T.,  {Machado} R.~E.~G.,  2014, \mn@doi [\mnras]
  {10.1093/mnras/stt2355}, \href
  {http://adsabs.harvard.edu/abs/2014MNRAS.438.2201M} {438, 2201}

\bibitem[\protect\citeauthoryear{{Manos}, {Bountis}  \& {Skokos}}{{Manos}
  et~al.}{2013}]{2013JPhA...46y4017M}
{Manos} T.,  {Bountis} T.,   {Skokos} C.,  2013, \mn@doi [Journal of Physics A
  Mathematical General] {10.1088/1751-8113/46/25/254017}, \href
  {http://adsabs.harvard.edu/abs/2013JPhA...46y4017M} {46, 254017}

\bibitem[\protect\citeauthoryear{{Marinacci}, {Pakmor}  \&
  {Springel}}{{Marinacci} et~al.}{2014}]{2014MNRAS.437.1750M}
{Marinacci} F.,  {Pakmor} R.,   {Springel} V.,  2014, \mn@doi [\mnras]
  {10.1093/mnras/stt2003}, \href
  {http://adsabs.harvard.edu/abs/2014MNRAS.437.1750M} {437, 1750}

\bibitem[\protect\citeauthoryear{{Mart{\'{\i}}}, {Cincotta}  \&
  {Beaug{\'e}}}{{Mart{\'{\i}}} et~al.}{2016}]{2016MNRAS.460.1094M}
{Mart{\'{\i}}} J.~G.,  {Cincotta} P.~M.,   {Beaug{\'e}} C.,  2016, \mn@doi
  [\mnras] {10.1093/mnras/stw1035}, \href
  {http://adsabs.harvard.edu/abs/2016MNRAS.460.1094M} {460, 1094}

\bibitem[\protect\citeauthoryear{{Martin}, {Ibata}  \& {Irwin}}{{Martin}
  et~al.}{2007}]{2007ApJ...668L.123M}
{Martin} N.~F.,  {Ibata} R.~A.,   {Irwin} M.,  2007, \mn@doi [\apjl]
  {10.1086/522791}, \href {http://adsabs.harvard.edu/abs/2007ApJ...668L.123M}
  {668, L123}

\bibitem[\protect\citeauthoryear{{Matteucci}}{{Matteucci}}{2014}]{2014SAAS...37..145M}
{Matteucci} F.,  2014, \mn@doi [The Origin of the Galaxy and Local Group,
  Saas-Fee Advanced Course, Volume 37.~ISBN 978-3-642-41719-1.~Springer-Verlag
  Berlin Heidelberg, 2014, p.~145] {10.1007/978-3-642-41720-7_2}, \href
  {http://adsabs.harvard.edu/abs/2014SAAS...37..145M} {37, 145}

\bibitem[\protect\citeauthoryear{{Meiron}, {Li}, {Holley-Bockelmann}  \&
  {Spurzem}}{{Meiron} et~al.}{2014}]{2014ApJ...792...98M}
{Meiron} Y.,  {Li} B.,  {Holley-Bockelmann} K.,   {Spurzem} R.,  2014, \mn@doi
  [\apj] {10.1088/0004-637X/792/2/98}, \href
  {http://adsabs.harvard.edu/abs/2014ApJ...792...98M} {792, 98}

\bibitem[\protect\citeauthoryear{{Merritt} \& {Fridman}}{{Merritt} \&
  {Fridman}}{1996}]{1996ApJ...460..136M}
{Merritt} D.,  {Fridman} T.,  1996, \mn@doi [\apj] {10.1086/176957}, \href
  {http://adsabs.harvard.edu/abs/1996ApJ...460..136M} {460, 136}

\bibitem[\protect\citeauthoryear{{Merritt} \& {Valluri}}{{Merritt} \&
  {Valluri}}{1996}]{1996ApJ...471...82M}
{Merritt} D.,  {Valluri} M.,  1996, \mn@doi [\apj] {10.1086/177955}, \href
  {http://adsabs.harvard.edu/abs/1996ApJ...471...82M} {471, 82}

\bibitem[\protect\citeauthoryear{{Michalik}, {Lindegren}  \&
  {Hobbs}}{{Michalik} et~al.}{2015}]{2015A&A...574A.115M}
{Michalik} D.,  {Lindegren} L.,   {Hobbs} D.,  2015, \mn@doi [\aap]
  {10.1051/0004-6361/201425310}, \href
  {http://adsabs.harvard.edu/abs/2015A%26A...574A.115M} {574, A115}

\bibitem[\protect\citeauthoryear{{Minchev} et~al.,}{{Minchev}
  et~al.}{2014}]{2014ApJ...781L..20M}
{Minchev} I.,  et~al., 2014, \mn@doi [\apjl] {10.1088/2041-8205/781/1/L20},
  \href {http://adsabs.harvard.edu/abs/2014ApJ...781L..20M} {781, L20}

\bibitem[\protect\citeauthoryear{{Miyamoto} \& {Nagai}}{{Miyamoto} \&
  {Nagai}}{1975}]{1975PASJ...27..533M}
{Miyamoto} M.,  {Nagai} R.,  1975, \pasj, \href
  {http://adsabs.harvard.edu/abs/1975PASJ...27..533M} {27, 533}

\bibitem[\protect\citeauthoryear{{Monari}, {Famaey}, {Siebert}, {Grand},
  {Kawata}  \& {Boily}}{{Monari} et~al.}{2016}]{2016MNRAS.461.3835M}
{Monari} G.,  {Famaey} B.,  {Siebert} A.,  {Grand} R.~J.~J.,  {Kawata} D.,
  {Boily} C.,  2016, \mn@doi [\mnras] {10.1093/mnras/stw1564}, \href
  {http://adsabs.harvard.edu/abs/2016MNRAS.461.3835M} {461, 3835}

\bibitem[\protect\citeauthoryear{{Monari}, {Kawata}, {Hunt}  \&
  {Famaey}}{{Monari} et~al.}{2017}]{2017MNRAS.466L.113M}
{Monari} G.,  {Kawata} D.,  {Hunt} J.~A.~S.,   {Famaey} B.,  2017, \mn@doi
  [\mnras] {10.1093/mnrasl/slw238}, \href
  {http://adsabs.harvard.edu/abs/2017MNRAS.466L.113M} {466, L113}

\bibitem[\protect\citeauthoryear{{Muzzio}, {Carpintero}  \& {Wachlin}}{{Muzzio}
  et~al.}{2005}]{2005CeMDA..91..173M}
{Muzzio} J.~C.,  {Carpintero} D.~D.,   {Wachlin} F.~C.,  2005, \mn@doi
  [Celestial Mechanics and Dynamical Astronomy] {10.1007/s10569-005-1608-4},
  \href {http://adsabs.harvard.edu/abs/2005CeMDA..91..173M} {91, 173}

\bibitem[\protect\citeauthoryear{{Navarro}, {Frenk}  \& {White}}{{Navarro}
  et~al.}{1996}]{1996ApJ...462..563N}
{Navarro} J.~F.,  {Frenk} C.~S.,   {White} S.~D.~M.,  1996, \mn@doi [\apj]
  {10.1086/177173}, \href {http://adsabs.harvard.edu/abs/1996ApJ...462..563N}
  {462, 563}

\bibitem[\protect\citeauthoryear{{Navarro}, {Frenk}  \& {White}}{{Navarro}
  et~al.}{1997}]{1997ApJ...490..493N}
{Navarro} J.~F.,  {Frenk} C.~S.,   {White} S.~D.~M.,  1997, \mn@doi [\apj]
  {10.1086/304888}, \href {http://adsabs.harvard.edu/abs/1997ApJ...490..493N}
  {490, 493}

\bibitem[\protect\citeauthoryear{{Newberg} \& {Carlin}}{{Newberg} \&
  {Carlin}}{2016}]{2016ASSL..420.....N}
{Newberg} H.~J.,  {Carlin} J.~L.,  eds, 2016, {Tidal Streams in the Local Group
  and Beyond}  Astrophysics and Space Science Library Vol. 420,
  \mn@doi{10.1007/978-3-319-19336-6.
}

\bibitem[\protect\citeauthoryear{{Ngan}, {Carlberg}, {Bozek}, {Wyse}, {Szalay}
  \& {Madau}}{{Ngan} et~al.}{2016}]{2016ApJ...818..194N}
{Ngan} W.,  {Carlberg} R.~G.,  {Bozek} B.,  {Wyse} R.~F.~G.,  {Szalay} A.~S.,
  {Madau} P.,  2016, \mn@doi [\apj] {10.3847/0004-637X/818/2/194}, \href
  {http://adsabs.harvard.edu/abs/2016ApJ...818..194N} {818, 194}

\bibitem[\protect\citeauthoryear{{Odenkirchen} et~al.,}{{Odenkirchen}
  et~al.}{2001}]{2001ApJ...548L.165O}
{Odenkirchen} M.,  et~al., 2001, \mn@doi [\apjl] {10.1086/319095}, \href
  {http://adsabs.harvard.edu/abs/2001ApJ...548L.165O} {548, L165}

\bibitem[\protect\citeauthoryear{{Pakmor} \& {Springel}}{{Pakmor} \&
  {Springel}}{2013}]{2013MNRAS.432..176P}
{Pakmor} R.,  {Springel} V.,  2013, \mn@doi [\mnras] {10.1093/mnras/stt428},
  \href {http://adsabs.harvard.edu/abs/2013MNRAS.432..176P} {432, 176}

\bibitem[\protect\citeauthoryear{{Pakmor}, {Springel}, {Bauer}, {Mocz},
  {Munoz}, {Ohlmann}, {Schaal}  \& {Zhu}}{{Pakmor}
  et~al.}{2016}]{2016MNRAS.455.1134P}
{Pakmor} R.,  {Springel} V.,  {Bauer} A.,  {Mocz} P.,  {Munoz} D.~J.,
  {Ohlmann} S.~T.,  {Schaal} K.,   {Zhu} C.,  2016, \mn@doi [\mnras]
  {10.1093/mnras/stv2380}, \href
  {http://adsabs.harvard.edu/abs/2016MNRAS.455.1134P} {455, 1134}

\bibitem[\protect\citeauthoryear{{Pakmor} et~al.,}{{Pakmor}
  et~al.}{2017}]{2017MNRAS.469.3185P}
{Pakmor} R.,  et~al., 2017, \mn@doi [\mnras] {10.1093/mnras/stx1074}, \href
  {http://adsabs.harvard.edu/abs/2017MNRAS.469.3185P} {469, 3185}

\bibitem[\protect\citeauthoryear{{Pe{\~n}arrubia}}{{Pe{\~n}arrubia}}{2013}]{2013MNRAS.433.2576P}
{Pe{\~n}arrubia} J.,  2013, \mn@doi [\mnras] {10.1093/mnras/stt935}, \href
  {http://adsabs.harvard.edu/abs/2013MNRAS.433.2576P} {433, 2576}

\bibitem[\protect\citeauthoryear{{Pearson}, {K{\"u}pper}, {Johnston}  \&
  {Price-Whelan}}{{Pearson} et~al.}{2015}]{2015ApJ...799...28P}
{Pearson} S.,  {K{\"u}pper} A.~H.~W.,  {Johnston} K.~V.,   {Price-Whelan}
  A.~M.,  2015, \mn@doi [\apj] {10.1088/0004-637X/799/1/28}, \href
  {http://adsabs.harvard.edu/abs/2015ApJ...799...28P} {799, 28}

\bibitem[\protect\citeauthoryear{{Pearson}, {Price-Whelan}  \&
  {Johnston}}{{Pearson} et~al.}{2017}]{2017NatAs...1..633P}
{Pearson} S.,  {Price-Whelan} A.~M.,   {Johnston} K.~V.,  2017, \mn@doi [Nature
  Astronomy] {10.1038/s41550-017-0220-3}, \href
  {http://adsabs.harvard.edu/abs/2017NatAs...1..633P} {1, 633}

\bibitem[\protect\citeauthoryear{{Perryman} et~al.,}{{Perryman}
  et~al.}{2001}]{2001A&A...369..339P}
{Perryman} M.~A.~C.,  et~al., 2001, \mn@doi [\aap]
  {10.1051/0004-6361:20010085}, \href
  {http://adsabs.harvard.edu/abs/2001A%26A...369..339P} {369, 339}

\bibitem[\protect\citeauthoryear{{Plummer}}{{Plummer}}{1911}]{1911MNRAS..71..460P}
{Plummer} H.~C.,  1911, \mn@doi [\mnras] {10.1093/mnras/71.5.460}, \href
  {http://adsabs.harvard.edu/abs/1911MNRAS..71..460P} {71, 460}

\bibitem[\protect\citeauthoryear{{Poveda}, {Allen}  \& {Schuster}}{{Poveda}
  et~al.}{1992}]{1992IAUS..149..471P}
{Poveda} A.,  {Allen} C.,   {Schuster} W.,  1992, in {Barbuy} B.,  {Renzini}
  A.,  eds,  IAU Symposium Vol. 149, The Stellar Populations of Galaxies.
  p.~471

\bibitem[\protect\citeauthoryear{{Price-Whelan}, {Johnston}, {Valluri},
  {Pearson}, {K{\"u}pper}  \& {Hogg}}{{Price-Whelan}
  et~al.}{2016a}]{2016MNRAS.455.1079P}
{Price-Whelan} A.~M.,  {Johnston} K.~V.,  {Valluri} M.,  {Pearson} S.,
  {K{\"u}pper} A.~H.~W.,   {Hogg} D.~W.,  2016a, \mn@doi [\mnras]
  {10.1093/mnras/stv2383}, \href
  {http://adsabs.harvard.edu/abs/2016MNRAS.455.1079P} {455, 1079}

\bibitem[\protect\citeauthoryear{{Price-Whelan}, {Sesar}, {Johnston}  \&
  {Rix}}{{Price-Whelan} et~al.}{2016b}]{2016ApJ...824..104P}
{Price-Whelan} A.~M.,  {Sesar} B.,  {Johnston} K.~V.,   {Rix} H.-W.,  2016b,
  \mn@doi [\apj] {10.3847/0004-637X/824/2/104}, \href
  {http://adsabs.harvard.edu/abs/2016ApJ...824..104P} {824, 104}

\bibitem[\protect\citeauthoryear{{Quillen}}{{Quillen}}{2003}]{2003AJ....125..785Q}
{Quillen} A.~C.,  2003, \mn@doi [\aj] {10.1086/345725}, \href
  {http://adsabs.harvard.edu/abs/2003AJ....125..785Q} {125, 785}

\bibitem[\protect\citeauthoryear{{Quillen}, {Minchev}, {Bland-Hawthorn}  \&
  {Haywood}}{{Quillen} et~al.}{2009}]{2009MNRAS.397.1599Q}
{Quillen} A.~C.,  {Minchev} I.,  {Bland-Hawthorn} J.,   {Haywood} M.,  2009,
  \mn@doi [\mnras] {10.1111/j.1365-2966.2009.15054.x}, \href
  {http://adsabs.harvard.edu/abs/2009MNRAS.397.1599Q} {397, 1599}

\bibitem[\protect\citeauthoryear{{Sawala} et~al.,}{{Sawala}
  et~al.}{2016}]{2016MNRAS.457.1931S}
{Sawala} T.,  et~al., 2016, \mn@doi [\mnras] {10.1093/mnras/stw145}, \href
  {http://adsabs.harvard.edu/abs/2016MNRAS.457.1931S} {457, 1931}

\bibitem[\protect\citeauthoryear{{Schaye} et~al.,}{{Schaye}
  et~al.}{2015}]{2015MNRAS.446..521S}
{Schaye} J.,  et~al., 2015, \mn@doi [\mnras] {10.1093/mnras/stu2058}, \href
  {http://adsabs.harvard.edu/abs/2015MNRAS.446..521S} {446, 521}

\bibitem[\protect\citeauthoryear{{Sch{\"o}nrich} \& {Dehnen}}{{Sch{\"o}nrich}
  \& {Dehnen}}{2017}]{2017arXiv171206616S}
{Sch{\"o}nrich} R.,  {Dehnen} W.,  2017, preprint, \href
  {http://adsabs.harvard.edu/abs/2017arXiv171206616S} {} (\mn@eprint {arXiv}
  {1712.06616})

\bibitem[\protect\citeauthoryear{{Schuster} \& {Allen}}{{Schuster} \&
  {Allen}}{1997}]{1997A&A...319..796S}
{Schuster} W.~J.,  {Allen} C.,  1997, \aap, \href
  {http://adsabs.harvard.edu/abs/1997A%26A...319..796S} {319, 796}

\bibitem[\protect\citeauthoryear{{Schwarzschild}}{{Schwarzschild}}{1993}]{1993ApJ...409..563S}
{Schwarzschild} M.,  1993, \mn@doi [\apj] {10.1086/172687}, \href
  {http://adsabs.harvard.edu/abs/1993ApJ...409..563S} {409, 563}

\bibitem[\protect\citeauthoryear{{Shevchenko}}{{Shevchenko}}{2011}]{2011ApJ...733...39S}
{Shevchenko} I.~I.,  2011, \mn@doi [\apj] {10.1088/0004-637X/733/1/39}, \href
  {http://adsabs.harvard.edu/abs/2011ApJ...733...39S} {733, 39}

\bibitem[\protect\citeauthoryear{{Siopis} \& {Kandrup}}{{Siopis} \&
  {Kandrup}}{2000}]{2000MNRAS.319...43S}
{Siopis} C.,  {Kandrup} H.~E.,  2000, \mn@doi [\mnras]
  {10.1046/j.1365-8711.2000.03740.x}, \href
  {https://ui.adsabs.harvard.edu/#abs/2000MNRAS.319...43S} {319, 43}

\bibitem[\protect\citeauthoryear{{Skokos}, {Gottwald}  \& {Laskar}}{{Skokos}
  et~al.}{2016}]{2016LNP..915.....S}
{Skokos} C.,  {Gottwald} G.,   {Laskar} J.,  eds, 2016, {Chaos Detection and
  Predictability}  Lecture Notes in Physics Vol. 915,
  \mn@doi{10.1007/978-3-662-48410-4.
}

\bibitem[\protect\citeauthoryear{{Skrutskie} et~al.,}{{Skrutskie}
  et~al.}{2006}]{2006AJ....131.1163S}
{Skrutskie} M.~F.,  et~al., 2006, \mn@doi [\aj] {10.1086/498708}, \href
  {http://adsabs.harvard.edu/abs/2006AJ....131.1163S} {131, 1163}

\bibitem[\protect\citeauthoryear{{Smith}}{{Smith}}{2016}]{2016ASSL..420..113S}
{Smith} M.~C.,  2016, in {Newberg} H.~J.,  {Carlin} J.~L.,  eds,  Astrophysics
  and Space Science Library Vol. 420, Tidal Streams in the Local Group and
  Beyond. p.~113 (\mn@eprint {arXiv} {1603.02149}),
  \mn@doi{10.1007/978-3-319-19336-6_5}

\bibitem[\protect\citeauthoryear{{Smith}, {Flynn}, {Candlish}, {Fellhauer}  \&
  {Gibson}}{{Smith} et~al.}{2015}]{2015MNRAS.448.2934S}
{Smith} R.,  {Flynn} C.,  {Candlish} G.~N.,  {Fellhauer} M.,   {Gibson} B.~K.,
  2015, \mn@doi [\mnras] {10.1093/mnras/stv228}, \href
  {http://adsabs.harvard.edu/abs/2015MNRAS.448.2934S} {448, 2934}

\bibitem[\protect\citeauthoryear{{Springel}}{{Springel}}{2010}]{2010MNRAS.401..791S}
{Springel} V.,  2010, \mn@doi [\mnras] {10.1111/j.1365-2966.2009.15715.x},
  \href {http://adsabs.harvard.edu/abs/2010MNRAS.401..791S} {401, 791}

\bibitem[\protect\citeauthoryear{{Springel} et~al.,}{{Springel}
  et~al.}{2008a}]{2008MNRAS.391.1685S}
{Springel} V.,  et~al., 2008a, \mn@doi [\mnras]
  {10.1111/j.1365-2966.2008.14066.x}, \href
  {http://adsabs.harvard.edu/abs/2008MNRAS.391.1685S} {391, 1685}

\bibitem[\protect\citeauthoryear{{Springel} et~al.,}{{Springel}
  et~al.}{2008b}]{2008Natur.456...73S}
{Springel} V.,  et~al., 2008b, \mn@doi [\nat] {10.1038/nature07411}, \href
  {http://adsabs.harvard.edu/abs/2008Natur.456...73S} {456, 73}

\bibitem[\protect\citeauthoryear{{Tsiganis}, {Anastasiadis}  \&
  {Varvoglis}}{{Tsiganis} et~al.}{2000}]{2000CSF....11.2281T}
{Tsiganis} K.,  {Anastasiadis} A.,   {Varvoglis} H.,  2000, \mn@doi [Chaos
  Solitons and Fractals] {10.1016/S0960-0779(99)00147-2}, \href
  {http://adsabs.harvard.edu/abs/2000CSF....11.2281T} {11, 2281}

\bibitem[\protect\citeauthoryear{{Valluri} \& {Merritt}}{{Valluri} \&
  {Merritt}}{1998}]{1998ApJ...506..686V}
{Valluri} M.,  {Merritt} D.,  1998, \mn@doi [\apj] {10.1086/306269}, \href
  {http://adsabs.harvard.edu/abs/1998ApJ...506..686V} {506, 686}

\bibitem[\protect\citeauthoryear{{Valluri}, {Debattista}, {Quinn}, {Ro{\v
  s}kar}  \& {Wadsley}}{{Valluri} et~al.}{2012}]{2012MNRAS.419.1951V}
{Valluri} M.,  {Debattista} V.~P.,  {Quinn} T.~R.,  {Ro{\v s}kar} R.,
  {Wadsley} J.,  2012, \mn@doi [\mnras] {10.1111/j.1365-2966.2011.19853.x},
  \href {http://adsabs.harvard.edu/abs/2012MNRAS.419.1951V} {419, 1951}

\bibitem[\protect\citeauthoryear{{Valluri}, {Debattista}, {Stinson}, {Bailin},
  {Quinn}, {Couchman}  \& {Wadsley}}{{Valluri}
  et~al.}{2013}]{2013ApJ...767...93V}
{Valluri} M.,  {Debattista} V.~P.,  {Stinson} G.~S.,  {Bailin} J.,  {Quinn}
  T.~R.,  {Couchman} H.~M.~P.,   {Wadsley} J.,  2013, \mn@doi [\apj]
  {10.1088/0004-637X/767/1/93}, \href
  {http://adsabs.harvard.edu/abs/2013ApJ...767...93V} {767, 93}

\bibitem[\protect\citeauthoryear{{Vasiliev}}{{Vasiliev}}{2013}]{2013MNRAS.434.3174V}
{Vasiliev} E.,  2013, \mn@doi [\mnras] {10.1093/mnras/stt1235}, \href
  {http://adsabs.harvard.edu/abs/2013MNRAS.434.3174V} {434, 3174}

\bibitem[\protect\citeauthoryear{{Vera-Ciro}, {Sales}, {Helmi}, {Frenk},
  {Navarro}, {Springel}, {Vogelsberger}  \& {White}}{{Vera-Ciro}
  et~al.}{2011}]{2011MNRAS.416.1377V}
{Vera-Ciro} C.~A.,  {Sales} L.~V.,  {Helmi} A.,  {Frenk} C.~S.,  {Navarro}
  J.~F.,  {Springel} V.,  {Vogelsberger} M.,   {White} S.~D.~M.,  2011, \mn@doi
  [\mnras] {10.1111/j.1365-2966.2011.19134.x}, \href
  {http://adsabs.harvard.edu/abs/2011MNRAS.416.1377V} {416, 1377}

\bibitem[\protect\citeauthoryear{{Vogelsberger}, {White}, {Helmi}  \&
  {Springel}}{{Vogelsberger} et~al.}{2008}]{2008MNRAS.385..236V}
{Vogelsberger} M.,  {White} S.~D.~M.,  {Helmi} A.,   {Springel} V.,  2008,
  \mn@doi [\mnras] {10.1111/j.1365-2966.2007.12746.x}, \href
  {http://adsabs.harvard.edu/abs/2008MNRAS.385..236V} {385, 236}

\bibitem[\protect\citeauthoryear{{Vogelsberger}, {Genel}, {Sijacki}, {Torrey},
  {Springel}  \& {Hernquist}}{{Vogelsberger}
  et~al.}{2013}]{2013MNRAS.436.3031V}
{Vogelsberger} M.,  {Genel} S.,  {Sijacki} D.,  {Torrey} P.,  {Springel} V.,
  {Hernquist} L.,  2013, \mn@doi [\mnras] {10.1093/mnras/stt1789}, \href
  {http://adsabs.harvard.edu/abs/2013MNRAS.436.3031V} {436, 3031}

\bibitem[\protect\citeauthoryear{{Voglis}, {Kalapotharakos}  \&
  {Stavropoulos}}{{Voglis} et~al.}{2002}]{2002MNRAS.337..619V}
{Voglis} N.,  {Kalapotharakos} C.,   {Stavropoulos} I.,  2002, \mn@doi [\mnras]
  {10.1046/j.1365-8711.2002.05938.x}, \href
  {http://adsabs.harvard.edu/abs/2002MNRAS.337..619V} {337, 619}

\bibitem[\protect\citeauthoryear{{Weinberg}}{{Weinberg}}{1999}]{1999AJ....117..629W}
{Weinberg} M.~D.,  1999, \mn@doi [\aj] {10.1086/300669}, \href
  {http://adsabs.harvard.edu/abs/1999AJ....117..629W} {117, 629}

\bibitem[\protect\citeauthoryear{{Widrow} \& {Bonner}}{{Widrow} \&
  {Bonner}}{2015}]{2015MNRAS.450..266W}
{Widrow} L.~M.,  {Bonner} G.,  2015, \mn@doi [\mnras] {10.1093/mnras/stv574},
  \href {http://adsabs.harvard.edu/abs/2015MNRAS.450..266W} {450, 266}

\bibitem[\protect\citeauthoryear{{Yoon}, {Johnston}  \& {Hogg}}{{Yoon}
  et~al.}{2011}]{2011ApJ...731...58Y}
{Yoon} J.~H.,  {Johnston} K.~V.,   {Hogg} D.~W.,  2011, \mn@doi [\apj]
  {10.1088/0004-637X/731/1/58}, \href
  {http://adsabs.harvard.edu/abs/2011ApJ...731...58Y} {731, 58}

\bibitem[\protect\citeauthoryear{{York} et~al.,}{{York}
  et~al.}{2000}]{2000AJ....120.1579Y}
{York} D.~G.,  et~al., 2000, \mn@doi [\aj] {10.1086/301513}, \href
  {http://adsabs.harvard.edu/abs/2000AJ....120.1579Y} {120, 1579}

\bibitem[\protect\citeauthoryear{{Zhao}, {Chen}, {Shi}, {Liang}, {Hou}, {Chen},
  {Zhang}  \& {Li}}{{Zhao} et~al.}{2006}]{2006ChJAA...6..265Z}
{Zhao} G.,  {Chen} Y.-Q.,  {Shi} J.-R.,  {Liang} Y.-C.,  {Hou} J.-L.,  {Chen}
  L.,  {Zhang} H.-W.,   {Li} A.-G.,  2006, \mn@doi [\cjaa]
  {10.1088/1009-9271/6/3/01}, \href
  {http://adsabs.harvard.edu/abs/2006ChJAA...6..265Z} {6, 265}

\bibitem[\protect\citeauthoryear{{Zhu}, {Marinacci}, {Maji}, {Li}, {Springel}
  \& {Hernquist}}{{Zhu} et~al.}{2016}]{2016MNRAS.458.1559Z}
{Zhu} Q.,  {Marinacci} F.,  {Maji} M.,  {Li} Y.,  {Springel} V.,   {Hernquist}
  L.,  2016, \mn@doi [\mnras] {10.1093/mnras/stw374}, \href
  {http://adsabs.harvard.edu/abs/2016MNRAS.458.1559Z} {458, 1559}

\bibitem[\protect\citeauthoryear{{Zwitter} et~al.,}{{Zwitter}
  et~al.}{2008}]{2008AJ....136..421Z}
{Zwitter} T.,  et~al., 2008, \mn@doi [\aj] {10.1088/0004-6256/136/1/421}, \href
  {http://adsabs.harvard.edu/abs/2008AJ....136..421Z} {136, 421}

\makeatother
\end{thebibliography}





\bsp	
\label{lastpage}
\end{document}